\documentclass[11pt]{article}

\newcommand{\mABCD}{\textbf{mABCD}}
\newcommand{\Rr}{\mathbf{R}}
\usepackage{graphicx}
\usepackage{subcaption} 
\usepackage{multirow}

\usepackage[T1]{fontenc}
\usepackage[utf8]{inputenc}
\usepackage{lmodern}
\usepackage[style=apa, backend=biber]{biblatex}
\pdfoutput=1
\usepackage{arxiv}
\usepackage{hyperref}
\usepackage[normalem]{ulem}
\usepackage{amsthm}
\usepackage{amsmath}
\usepackage{amssymb}
\usepackage{mathrsfs}
\usepackage{mathtools}
\usepackage{graphicx}
\usepackage{enumerate}
\usepackage{comment}
\usepackage{algorithm}
\usepackage{algpseudocode}
\usepackage{color}
\usepackage{stmaryrd}
\usepackage{booktabs} 
\usepackage{fullpage}
\usepackage{appendix}
\usepackage[table]{xcolor}

\theoremstyle{definition}
 
\theoremstyle{remark}

\theoremstyle{remark}

\numberwithin{theorem}{section}

\providecommand{\R}{}
\providecommand{\Z}{}
\providecommand{\N}{}

\renewcommand{\R}{\mathbb{R}}
\renewcommand{\Z}{\mathbb{Z}}
\renewcommand{\N}{{\mathbb N}}

\newcommand{\E}[1]{{\mathbb E}\left[#1\right]}

\newcommand{\p}[1]{\mathbb{P}\left(#1\right)}

\newcommand{\abcdDist}{\ensuremath{\mathcal{A}}}

\newcommand{\round}[1]{\ensuremath{\left\lfloor #1 \right\rceil}}

\newcommand{\tpl}[3]{\ensuremath{\mathcal{P}\left(#1,#2,#3\right)}}

\expandafter\def\expandafter\normalsize\expandafter{%
    \normalsize%
    \setlength\abovedisplayskip{0pt}%
    \setlength\belowdisplayskip{8pt}%
    \setlength\abovedisplayshortskip{-4pt}%
    \setlength\belowdisplayshortskip{4pt}%
}

\newcommand{\revision}[1]{{#1}}

\title{\textbf{M}ultilayer \textbf{A}rtificial \textbf{B}enchmark \\ for \textbf{C}ommunity \textbf{D}etection (\mABCD)}

\author{
\L{}ukasz Krai\'{n}ski\thanks{Decision Analysis and Support Unit, SGH Warsaw School of Economics, Warsaw, Poland; email: \texttt{lkrain@sgh.waw.pl}; ORCID: \texttt{0000-0002-3664-468X}},
\hspace{.15cm}
Micha\l{} Czuba\thanks{Department of Artificial Intelligence, Wrocław University of Science and Technology, Wrocław, Poland; e-mail: \texttt{michal.czuba@pwr.edu.pl}; ORCID: \texttt{0000-0001-8652-3678}},
\hspace{.15cm}
Piotr Br\'odka\thanks{Department of Artificial Intelligence, Wrocław University of Science and Technology, Wrocław, Poland; e-mail: \texttt{piotr.brodka@pwr.edu.pl}; ORCID: \texttt{0000-0002-6474-0089}},
\hspace{.15cm}
Pawe\l{} Pra\l{}at\thanks{Department of Mathematics, Toronto Metropolitan University, Toronto, ON, Canada; e-mail: \texttt{pralat@torontomu.ca}; ORCID: \texttt{0000-0001-9176-8493}},
\hspace{.15cm}
Bogumi\l{} Kami\'{n}ski\thanks{Decision Analysis and Support Unit, SGH Warsaw School of Economics, Warsaw, Poland; email: \texttt{bkamins@sgh.waw.pl}; ORCID: \texttt{0000-0002-0678-282X}},
\hspace{.15cm}
Fran\c{c}ois Th\'{e}berge\thanks{Tutte Institute for Mathematics and Computing, Ottawa, ON, Canada; email: \texttt{theberge@ieee.org}; ORCID: \texttt{0000-0002-5499-3680}}
}

\begin{document}

\maketitle            

\begin{abstract}
\revision{One of the most persistent challenges in network science is the development of various synthetic graph models to support subsequent analyses. Among the most notable frameworks addressing this issue is} the \textbf{A}rtificial \textbf{B}enchmark for \textbf{C}ommunity \textbf{D}etection (\textbf{ABCD}) --- a random graph model with community structure and power-law distribution for both degrees and community sizes. The model generates graphs similar to the well-known \textbf{LFR} model but it is faster, more interpretable, and can be investigated analytically. In this paper, we use the underlying ingredients of \textbf{ABCD} and introduce its \revision{variant, \mABCD, thereby addressing the gap in models capable of generating multilayer networks. The uniqueness of the proposed approach lies in its flexibility at both levels of modelling: the internal structure of individual layers and the inter-layer dependencies, which together make the network a coherent structure rather than a collection of loosely coupled graphs. In addition to the conceptual description of the framework, we provide a comprehensive analysis of its efficient Julia implementation. Finally, we illustrate the applicability of \mABCD\ to one of the most prominent problems in the area of complex systems --- spreading phenomena analysis.}
\end{abstract}


\section{Introduction}\label{sec:intro} 
One of the most important features of real-world networks is their community structure, as it reveals the internal organization of nodes~\parencite{fortunato2010community,kaminski2021mining}. In social networks, communities may represent groups by interest; in citation networks, they correspond to related papers; in the Web graph, communities are formed by pages on related topics, etc. Identifying communities in a network is therefore valuable as this information helps us to understand the network structure better.

\revision{The need for synthetic graph models with community structure that resemble real-world networks arises for several reasons. Foremost, when modelling a given system is infeasible (for instance, due to data scarcity), such frameworks can provide a structure that globally approximates the analyzed environment. For example, in the field of epidemiology, the spread of diseases is commonly simulated on synthetic networks owing to the difficulty of obtaining ground-truth data at the human level~\parencite{watroba2023influence} --- only recently have some countries developed network models resembling the social structures of their populations~\parencite{panayiotou2025swedishstatenetwork}. Continuing this example, simulating disease dissemination using the model proposed by~\textcite{er-model} could be less accurate than doing so on a network generated by the stochastic block model~\parencite{holland1983stochastic}, as the latter yields community-like structures that better reflect social bonds. Another motivation for developing synthetic graph generators with community structure lies in the need to benchmark and tune clustering algorithms that are unsupervised by nature, given that only a few datasets exist with ground-truth communities identified and labelled~\parencite{magnani2021community}. Having such models enables experiments to be conducted in a controlled environment, thereby making the assessment of different community detection algorithms more reliable and reproducible. The generators proposed so far, although important milestones, have some limitations that can affect their applicability to different research scenarios (see Section~\ref{subsec:literature} for details). These limitations underscore the ongoing need for new models, tailored to both more advanced approaches to modelling complex systems and the increasing size of networks used in analysis.}

The \textbf{LFR} (\textbf{L}ancichinetti, \textbf{F}ortunato, \textbf{R}adicchi) model~\parencite{lancichinetti2008benchmark,lancichinetti2009benchmarks} is a highly popular model that generates networks with communities and, at the same time, allows for heterogeneity in the distributions of both node degrees and of community sizes. It became a standard and extensively used method for generating artificial networks. A similar synthetic network to \textbf{LFR}, the \textbf{A}rtificial \textbf{B}enchmark for \textbf{C}ommunity \textbf{D}etection (\textbf{ABCD})~\parencite{kaminski2021artificial} was recently introduced and implemented\footnote{\url{https://github.com/bkamins/ABCDGraphGenerator.jl/}}, including a fast implementation\footnote{\url{https://github.com/tolcz/ABCDeGraphGenerator.jl/}} that uses multiple threads (\textbf{ABCDe})~\parencite{kaminski2022abcde}. Undirected variants of \textbf{LFR} and \textbf{ABCD} produce graphs with comparable properties, but \textbf{ABCD}/\textbf{ABCDe} is faster than \textbf{LFR} and can be easily tuned to allow the user to make a smooth transition between the two extremes: pure (disjoint) communities and random graphs with no community structure. Moreover, it is easier to analyze theoretically --- for example, in~\parencite{kaminski2022modularity} various theoretical asymptotic properties of the \textbf{ABCD} model are investigated, including the modularity function that, despite some known issues such as the ``resolution limit'' reported in~\parencite{fortunato2007resolution}, is an important graph property of networks in the context of community detection. In~\parencite{self-similarityABCD}, some interesting and desired self-similar behaviour of the \textbf{ABCD} model is analyzed; namely, it is shown that the degree distribution of ground-truth communities is asymptotically the same as the degree distribution of the whole graph (appropriately normalized based on their sizes). Finally, the building blocks in the model are flexible and may be adjusted to satisfy different needs. Indeed, the original \textbf{ABCD} model was recently adjusted to include potential outliers (\textbf{ABCD+o})~\parencite{kaminski2023artificial}, overlapping communities (\textbf{ABCD+o$^2$})~\parencite{barrett2025artificial}, and extended to hypergraphs (\textbf{h--ABCD})~\parencite{kaminski2023hypergraph}\footnote{\url{https://github.com/bkamins/ABCDHypergraphGenerator.jl}}. The \textbf{ABCD} model is used by practitioners but, for the reasons mentioned above, it also gains recognition among scientists. For example, \cite{aref2022bayan} suggests to use \textbf{Adjusted Mutual Information} (\textbf{AMI}) between the partitions returned by various algorithms with the ground-truth partitions of synthetically generated random graphs, \textbf{ABCD} and \textbf{LFR}. In particular, they use both models to compare 30 community detection algorithms, mentioning that \emph{being directly comparable to \textbf{LFR}, \textbf{ABCD} offers additional benefits, including higher scalability and better control for adjusting an analogous mixing parameter}. In the context of this paper, the most important of the above features is that the \textbf{ABCD} model is flexible and can be easily extended, \revision{as we utilize that property to deliver a new model to} producing multilayer networks.

The study of complex networks has evolved significantly over the past two decades, driven by the growing need to model and analyze interconnected systems. Among the notable advancements in this field is the development of multilayer networks, a powerful framework that captures multiple types of relationships within a single network structure. Unlike traditional single-layer \revision{(a.k.a. monoplex)} networks where all edges represent the same type of interaction, multilayer networks allow nodes to participate in diverse types of relationships across different layers. For instance, in a multilayer social network, layers may represent distinct relations such as friendships, coworker connections, family ties, and online interactions via platforms like Facebook or LinkedIn~\parencite{kivela2014multilayer, dickison2016multilayer}. This flexibility makes multilayer networks uniquely suited to model the complexity of real-world systems. The concept of multilayer networks was introduced to address the limitations of traditional network models, particularly their inability to represent systems with heterogeneous interactions. Early research focused on formalizing the structure of multilayer networks, including their mathematical representation and classification. \cite{kivela2014multilayer} provided a seminal framework for multilayer networks, categorizing them into multiplex, interconnected, and temporal networks. This foundational work established the groundwork for studying the dynamics and structures of multilayer systems, sparking a surge of interest in the network science community.

As we already mentioned, one of the most important aspects of network analysis is community detection, which aims to identify groups of nodes that are densely connected within the network while being sparsely connected to nodes outside their group \parencite{deluca2023communitydetectionapproach}. In multilayer networks, this problem is significantly more complex because communities may exist consistently across layers, vary between layers, or even exhibit dependencies across layers~\parencite{magnani2021community}. The field of community detection in multilayer networks emerged shortly after the formalization of multilayer network structures, with pioneering studies exploring both theoretical and algorithmic approaches~\parencite{berlingerio2011finding, tang2012community, brodka2013introduction}. The survey by \cite{magnani2021community} provides an exhaustive overview of existing methods for community detection in multilayer networks, categorizing them into optimization-based methods (e.g., modularity maximization), statistical methods (e.g., stochastic block models), and heuristic approaches. These methods address diverse challenges, such as detecting overlapping communities, identifying temporal changes, and incorporating inter-layer dependencies. Despite these advancements, the problem of community retrieval in multilayer networks remains pertinent, and new approaches continue to emerge~\parencite{karimi2020multiplexcommunitydetection, qing2024bipartitemixedmembership}. One of the major challenges associated with this task is the evaluation of algorithms, primarily due to the scarcity of real-world multilayer datasets with ground-truth \revision{communities. In the aforementioned review study,~\textcite{magnani2021analysis} recalled only two multilayer networks with partially known partitions, which, however, do not arise from their geometry but from actor labels assigned according to affiliation (e.g., political party). Moreover, both networks were constructed such that their communities are similar (i.e., pillar-like) across layers, which makes them insufficiently representative for benchmarking diverse community detection algorithms.} 

In this paper, we use the underlying flexible ingredients of the \textbf{ABCD} family of models and introduce its variant for multilayer networks, \mABCD \footnote{\url{https://github.com/KrainskiL/MLNABCDGraphGenerator.jl}} \footnote{\revision{Please note that the name of the proposed method is intended to emphasize its connection to the \textbf{ABCD} model rather than to draw the reader’s attention to benchmarking different community detection methods.}}. It is a journal version of a short conceptual paper~\parencite{brodka2025multilayer} presented at the 20th Workshop on Modelling and Mining Networks (WAW2025). We significantly extend the previous work by improving, evaluating, and validating the model as well as setting \mABCD~in the context of related works. 
\revision{In summary, the main contributions of this work can be expressed in five key points:
\begin{itemize}
    \item The proposed model, \mABCD, is capable of generating synthetic, scale-free multilayer networks with community structure, where structural properties are shaped both within individual layers and across them.
    \item We introduce a latent, biscuit-like layer that supports the process of node assignment into communities and is shared across all layers during the generation process. This structure represents intrinsic actor properties and can be readily replaced with embeddings of real-world entities, thereby enhancing the model’s flexibility.
    \item The proposed model extends and strengthens the broader \textbf{ABCD} family of synthetic graph generators, contributing to its coherence and generality.
    \item \mABCD\ is an efficient model that outperforms its closest equivalent, multilayerGM, by one order of magnitude in time needed to generate small networks (around 1{,}000 nodes), with the improvement increasing up to two orders of magnitude for larger networks (around 33{,}000 nodes).
    \item We demonstrate a potential application of the model by addressing the problem of spread controllability in multilayer networks.
\end{itemize}}

The paper is structured as follows. We start with a gentle introduction to various aspects of multilayer networks (Section~\ref{subsec:related_work}), including an overview of existing synthetic models, summarizing approaches to measure correlations between layers. Then, we present design assumptions under which \mABCD\ was developed (Section~\ref{sec:design}). In Section~\ref{subsec:model}, we formally define the \mABCD\ model. Experiments highlighting properties of the model are discussed in Section~\ref{sec:properties}. We finish the paper with a demonstration of how one can use the \mABCD\ model to investigate properties of complex networks and algorithms that are running on them. As an example, we concentrate on \revision{information spread (Section~\ref{sec:spreading}), which constitutes a problem of greater complexity in the realm of multilayer networks compared to its monoplex counterpart}.

\section{Related Work}\label{subsec:related_work}

In this section, we introduce the standard notation used across the paper and properties of interest, especially from the perspective of multilayer networks~\parencite{dickison2016multilayer, kivela2014multilayer}. For a given $n \in \N = \{1, 2, \ldots \}$, we use $[n]$ to denote the set consisting of the first $n$ natural numbers, that is, $[n] = \{1, 2, \ldots, n\}$. We define a multilayer network as a quadruple $M = ([n],[\ell],V=[n] \times [\ell],E)$, where:
\begin{itemize}
 \item $[n]$ is a set of $n$ actors (for example, users of various social networking sites),
 \item $[\ell]$ is a set of layers (for example, different social networking platforms, such as LinkedIn, Facebook and Instagram, on which actors interact with each other),
 \item $V \subseteq [n] \times [\ell]$ a set of nodes (vertices); node $v = (a,\ell_i) \in V$ represents an actor $a$ in layer ${\ell}_i$,
 \item $E$ is a set of (undirected) edges between nodes; if $e = v_1 v_2 \in E$ with $v_1 = (a_1, {\ell}_1) \in V$ and $v_2=(a_2, {\ell}_2) \in V$, then ${\ell}_1={\ell}_2$, that is, edges occur only within layers.
\end{itemize}

\begin{figure}[ht!]
    \centering
    \includegraphics[width=.35\linewidth]{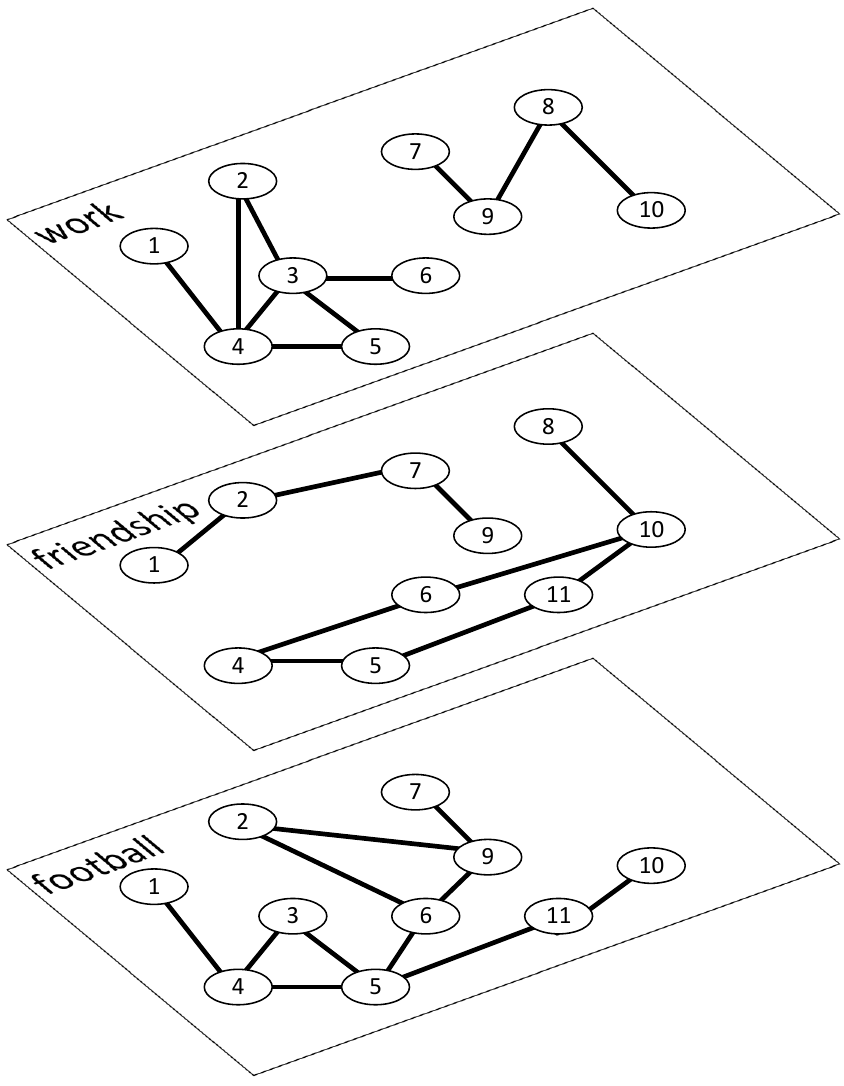}
    \caption{Example of a multilayer network.}
    \label{fig:toy_net}
\end{figure}

Figure~\ref{fig:toy_net} presents an example of a simple multilayer network consisting of eleven actors (represented as integers from $[11]$), three layers (professional, associated with the actors' interactions in a work environment, friendship relations between actors, and a layer representing actors playing football together), thirty nodes, and thirty edges.

Note that not every actor in the example above is present on all layers. For example, actor 8 does not exist in the football layer, meaning that it does not play football with other actors. For simplicity, in our model, we assume that each layer has exactly $n$ nodes associated with all actors. Actors that do not engage with a given layer (we will call them inactive) will be associated with isolated nodes (nodes of degree zero).

\subsection{Existing synthetic models for multilayer networks}\label{subsec:literature}

The first synthetic models for multilayer networks, with known community structure, were extensions of existing single-layer benchmarks. For example, in~\parencite{brodka2016method}, the \textbf{mLFR} model was introduced, as an extension of the original \textbf{LFR} benchmark, to generate multilayer networks with realistic (scale-free) degree and community size distributions. While the \textbf{mLFR} model preserves realistic degree and community size distributions, it assumes homogeneity in community generation across layers, limiting its applicability to networks where community structures vary between layers. Another drawback is its limited ability to model inter-layer correlations, a critical aspect of many real-world multilayer networks. Furthermore, the \textbf{mLFR} benchmark inherits the computational complexity of the original \textbf{LFR} model, making it resource-intensive for generating large networks with detailed configurations. Another example of early works presenting multilayer network generators is~\parencite{popiel2015muneg}, the MuNeG generator that was proposed as an extension of the Eldardiry model~\parencite{eldardiry2010multi-networkfusion}. It is governed by six scalar parameters: the number of nodes, layers, and groups; the probability that two nodes from the same (or different) group are connected; and community homophily. However, as an early contribution to the field of multilayer network generation, its functionality is limited. The model does not capture inter-layer relations, nor does it provide direct control over the distributions of degrees or communities.

As the body of literature on multilayer networks expanded, various generators addressing specific forms of multilayer networks emerged. For example, \parencite{tarvid2017stimulation} introduces a synthetic model designed to construct multilayer networks in which a link between two actors may exist in only one layer, thereby representing a social system composed of four distinct types of relationship: support, sympathy, band, and community. \cite{elsisy2022networkgenerator} also proposed a generator tailored to a specific problem. Specifically, they developed a framework for modelling covert social structures, such as criminal or terrorist networks. Their goal was to provide a model capable to represent both the organizational structure and the community partition of the considered system. These models, despite useful for specific problems, remain limited in terms of their applicability to more general scenarios.

A review of related work also reveals a broad influx of methods extending the Stochastic Block Model~\parencite{holland1983stochastic}. For instance, \cite{huang2019community} build upon this framework by assuming that a given community may appear in multiple layers, and propose a method for retrieving such exactly matching groups from the network. On the other hand, the generator proposed by~\cite{pamfil2019relating} emerges as an indirect outcome of efforts to generalize the modularity function to a multilayer setting. It attempts to address several limitations of extensions of the Stochastic Block Model, such as the restriction to identical communities across layers, but imposes an additional constraint: any inter-layer dependencies are induced solely through the multilayer partition and the coupling of nodes. Another significant contribution (also based on the Stochastic Block Model) came from \cite{bazzi2020framework}, who proposed a framework --- \textbf{multilayerGM} --- for generating networks with mesoscale structures, including communities. This is a very general and powerful model that enables researchers to control the properties of generated networks, such as the strength of inter-layer connections and the overlap of communities between layers. Because of its versatility (multilayer, temporal networks and more), the model might be challenging to use for a novice researcher (the need to define an inter-layer dependency tensor, implementing a new network model if one wants to use something else than the Degree Corrected Stochastic Block Model) and it suffers from high computational complexity (see Figure~\ref{fig:execution_time_multilayerGM_comparison} and Table~\ref{tab:multilayergm_mABCD_time}). Finally, it lacks the flexibility needed to reflect the various dependencies existing in real networks (for example, injecting degree sequences from real multilayer networks). As a result, despite its advanced capabilities, the model does not address all needs.

\cite{magnani2021community}, in the recent review paper on community detection in multiplex networks, have not used the model proposed by Bazzi. Instead, they proposed a simple model for generating multilayer networks with community structures designed specifically for various algorithm comparisons. This model, partially integrated into the multinet library~\parencite{magnani2021analysis}, simplifies the generation of multilayer networks and facilitates the benchmarking of various community detection algorithms. While this simplicity facilitates comparisons of different algorithms, it limits the model's ability to replicate the complexity of real-world networks, such as diversity in community sizes. Furthermore, the model offers only basic representations of inter-layer dependencies, which reduces its relevance for studying networks with varying edge correlations between layers. 

\section{Design Assumptions}\label{sec:design}

Before building the model, it is essential to address the question if and how degrees, edges, and partitions correlate between layers in real-world networks. Existing studies~\parencite{magnani2021community, brodka2018quantifying} have not provided a definitive answer to this question. To bridge this gap, we conducted an analysis of eight real-world networks from diverse domains and of different sizes. In the following section, we first present a methodology for computing these correlations. \revision{Then, we discuss obtained results and conclude with the assumptions underlying the design of \mABCD, which stem from this analysis}.

\subsection{Correlations between layers}\label{subsec:correlations}

Although there are no edges between nodes in different layers, in most real-world multilayered networks, layers are clearly \emph{not} independently generated. Each actor is associated with $\ell$ nodes, one in each layer, and there are some highly non-trivial correlations between edges across layers. For example, active users on one social media platform are often also active on another one~\parencite{gottfried2024americans}. This creates correlations between degree distributions across layers. Communities that are naturally formed in various layers often depend on the properties of the associated actors. For example, users interested in soccer might group together on Instagram and on Facebook. As a result, partitions of nodes into communities (associated with different layers) are often correlated. Finally, interactions between actors in one layer might increase their chances of interacting in another layer, yielding correlations at the level of edges.

Below, we summarize how we measure these three types of correlations. The first two measures are standard, and their description can be found in~\parencite{brodka2018quantifying, kaminski2021mining}.

\subsubsection*{Correlations between node degrees in various layers}

We will use \textbf{Kendall rank correlation coefficient} $\tau$~\parencite{kendall1938new} to measure correlations between sequences of node degrees in two different layers. It is a nonparametric measure of the ordinal association between two measured quantities: the similarity of the orderings of the data when ranked by each of the quantities (in our application, the degree sequences). The Kendall correlation between two variables ranges from $-1$ to $1$. It is large when observations have a similar rank between the two variables and is small when observations have a dissimilar rank. 

Specifically, we will use the ``tau-b'' statistic, which is adjusted to handle ties. If an actor is inactive in one of the two layers we compare against each other, then we simply ignore the two nodes corresponding to this actor. As a result, the degree sequences are always of the same length.

\subsubsection*{Correlations between partitions in various layers}

The \textbf{adjusted mutual information} (\textbf{AMI}), a variation of \textbf{mutual information} (\textbf{MI}), is a common way to compare partitions of the same set~\parencite{kaminski2021mining, vinh2009information}. Usually, one may want to compare the partitions returned by some clustering algorithms. In our present context, we may want to compare partitions into ground-truth communities from two different layers. The \textbf{AMI} takes a value of $1$ when the two partitions are identical and $0$ when the \textbf{MI} between two partitions equals the value expected due to chance alone. Actors, that are inactive in at least one of the two layers we compare against each other, are ignored so that a comparison of partitions is made on the same set of actors.

\subsubsection*{Correlations between edges in various layers}\label{subsubsec:correlations}

To measure correlations between edges in different layers, we define $\Rr$, a $\ell\times\ell$ matrix in which elements $r_{i,j} \in [0,1]$ ($i,j \in [\ell]$) capture correlation between edges present in layers $i$ and $j$. For any $i,j \in [\ell]$ with $i < j$, let
\begin{equation}\label{eq:Eij}
E_i^j = \{a_1a_2 : (a_1, i) (a_2, i) \in E \wedge  a_1, a_2 \in [n] \wedge a_1 \text{ and } a_2 \text{ are active in layers $i$ and $j$}\},
\end{equation}
be the set of edges that are present in layer $i$, involving actors that are also active in layer $j$. Note that in the definition of $E_i^j$, edges are defined over actors that are active in both layers, not nodes in layer $i$, so that we can perform set operations on edges between layers. Entries $r_{i,j}$ in $\Rr$ are computed using the following formula:
\begin{equation}\label{eq:rij}
    r_{i,j} = \frac{| E_i^j \cap E_j^i |}{\min\{|E_i^j|,|E_j^i|\}}.
\end{equation}
If $\min\{|E_i^j|,|E_j^i|\}=0$, then we leave $r_{i,j}$ undefined; in the implementation, \texttt{NaN} value is produced.

Note that the definition of $\Rr$ implies that $r_{i,i}=1$ for any $i \in [\ell]$ and $r_{i,j}=r_{j,i}$ for $1 \le i < j \le \ell$. The maximum value of $1$ is attained when edges in one of the layers form a subset of edges in the other layer. The minimum value of $0$ is attained when the two sets of edges in the corresponding layers are completely disjoint. As a result, $r_{ij}$ aims to capture correlations between individual edges, but it is not normalized as, for example, the Kendall rank correlation coefficient $\tau$. The coefficient $\tau$ ranges from $-1$ to $1$, corresponding to the two extremes, and $0$ corresponds to a neutral case. Graphs associated with the layers are sparse, but one layer might have substantially more edges than the other. Hence, $r_{ij}$ is convenient, but it does not have a natural interpretation as~$\tau$. Finally, let us mention that one can easily update the value of $r_{ij}$ when some small operations are applied to either layer $i$ or $j$. It will become handy when such operations must be performed on our synthetic model to converge to the desired correlation matrix $\Rr$.

\subsection{Examples of multilayer networks}\label{subsec:examples_networks}

The aspects of inter-layer dependencies, in the form introduced above, were derived from eight real-world networks, which varied in the number of actors, nodes, edges, layers, and in the domains they represented. Table~\ref{tab:networks_eda} presents a detailed overview of the characteristics of each network, offering insights into their structural properties.

\begin{table}[ht!]
    \centering
    {\footnotesize
        \caption{Real-world networks evaluated with respect to their inter-layer dependencies, with their basic parameters summarized.}
        \begin{tabular}{lrrrrp{8.1cm}}
        Name & Layers & Actors & Nodes & Edges & Note \\ \hline \hline
        arxiv & 13 & 14,065  & 26,796  & 59,026 & Coauthorship network obtained from articles published on the ``arXiv'' repository~\parencite{dedomenico2015arxiv}. Each layer represents a different arXiv category, e.g. Physics and Society or Social and Information Networks.\\
        aucs & 5 & 61 & 224 & 620 & A graph of interactions (friends on Facebook, leisure, work, co-authorship and lunch) between employees of \textbf{A}arhus \textbf{U}niversity, Department of \textbf{C}omputer \textbf{S}cience~\parencite{rossi2015aucs}. \\
        cannes & 3 & 438,537 & 659,951 & 974,743 & A network of interactions (retweets, mentions and replies) between Twitter (now X) users during the Cannes Film Festival in 2013~\parencite{omodei2015characterizing}. \\
        ckmp & 3 & 241 & 674 & 1,370 & A network depicting diffusion of innovations among physicians~\parencite{coleman1957ckmp}. Each layer was built based on the physician's answers to questions like ``Tell me the first names of your three friends whom you see most often socially".\\
        eutr-A & 37 & 417 & 2,034 & 3,588 & The European air transportation network~\parencite{cardillo2013eutransportation}. Each layer represents the connections of different airline operator in Europe.\\
        l2-course & 2 & 41 & 82 & 297 & A network of interactions (collaboration, friends) between U.S. students learning Arabic language during an intensive course in Jordan (1st month snapshot)~\parencite{paradowski2024peer}. \\
        lazega & 3 & 71 & 212 & 1,659 & A network of interactions (co-work, friendship and advice) between staff of a law corporation~\parencite{snijders2006lazega}. \\
        timik & 3 & 61,702 & 102,247 & 881,676 & A graph of interactions (text messages, online transactions, home visits) between users of the virtual world platform for teenagers~\parencite{jankowski2017timik}. 
        \end{tabular}
        \label{tab:networks_eda}
    }
\end{table}

\begin{figure}[ht!]
    \centering
    \begin{subfigure}{\textwidth}
        \centering
        \caption{Inter-layer correlations of degrees within each network.}
        \includegraphics[width=\textwidth]{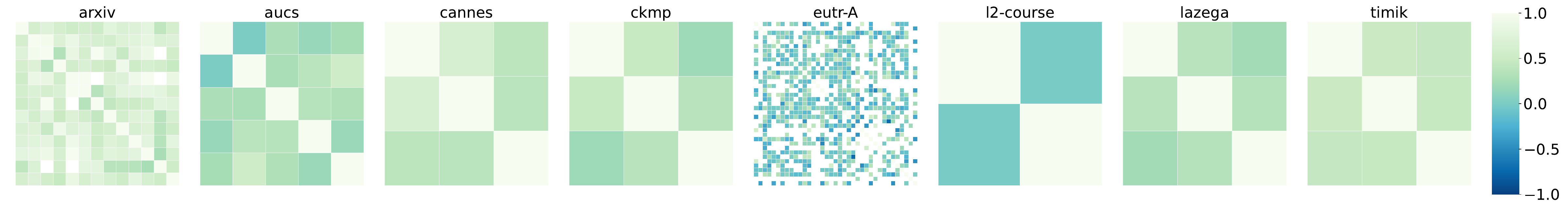}
        \label{fig:corr_degrees}
    \end{subfigure}
    \begin{subfigure}{\textwidth}
        \centering
        \caption{Inter-layer correlations of edges within each network.}
        \includegraphics[width=\textwidth]{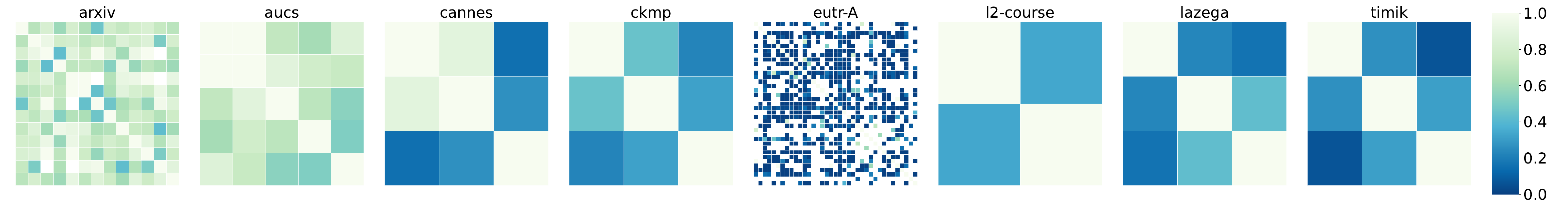}
        \label{fig:corr_edges}
    \end{subfigure}
    \begin{subfigure}{\textwidth}
        \centering
        \caption{Inter-layer correlations of partitions within each network.}
        \includegraphics[width=\textwidth]{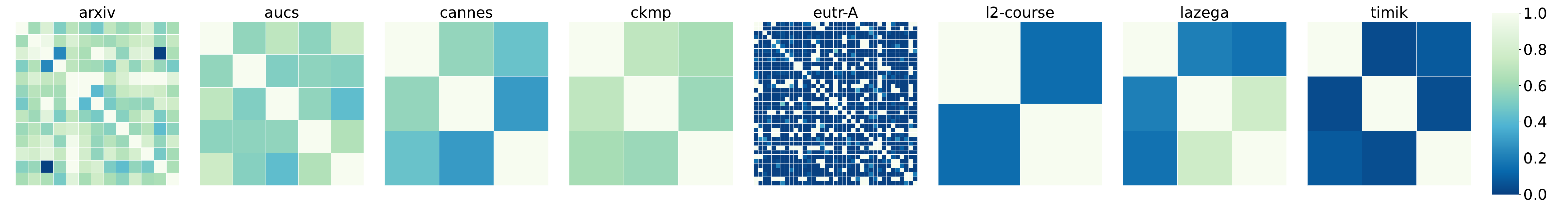}
        \label{fig:corr_partitions}
    \end{subfigure}
    \caption{Correlations between degrees, edges, and partitions presented as heatmaps for real-world networks: arxiv, aucs, cannes, ckmp, eutr-A, l2-course, lazega, and timik (layer names and explicit values have been removed for clarity). Note that there is no consistent trend across networks or within the given property.}
    \label{fig:real_graphs_correlation}
\end{figure}

Next, in Figure~\ref{fig:real_graphs_correlation}, we present correlation matrices between layers for degrees of nodes, partitions, and edges, according to the methodology described in Section~\ref{subsec:correlations}. Due to a lack of ground truth partitions, communities were identified by the outcome of the Louvain algorithm~\parencite{blondel2008fast}. After conducting a thorough analysis of the results, we observed a lack of consistent patterns in the examined real-world multilayer networks. Specifically, there was no universal relationship between degrees, edges, and community partitions across the layers. In some cases, these features were correlated, while in others, they appeared to be independent. Furthermore, within a single network, we found that while two layers might exhibit a correlation, this relationship did not necessarily extend to other layers. For instance, a node might have similar neighbours across two layers but exhibit a completely different connectivity pattern in a third layer. These observations highlight the inherent heterogeneity and complexity of real-world multilayer networks, emphasizing the need for a flexible and customizable synthetic network generation framework.

\subsection{Assumed functionalities}

In light of these findings, we assumed that the framework has to accommodate inter-layer diversities. Moreover, since the question of how precisely these relations should be modelled constitutes an issue that must be addressed individually, depending on the specific problem one is working on, we aimed to provide a tool that balances flexibility and accessibility. Thus, building the framework upon \textbf{ABCD} seemed a plausible approach, as it allows for extension in this direction.

On the one hand, \mABCD\ aims to provide advanced users with full control over critical properties of the network, such as the distributions of degrees or edges correlations. This level of customization ensures that users can generate networks tailored to their specific research needs, allowing for precise modelling of complex scenarios observed in real-world systems. On the other hand, we recognized the importance of making the framework accessible to less experienced users who may not be familiar with the intricacies of network science and programming. To address this, \mABCD\ was designed to include default settings, like generating networks with a power-law degree distribution, which is a common feature of many real-world systems. These default parameters provide a starting point for users while maintaining the ability to modify them as needed. Additionally, this is complemented by guidelines, default configurations, and examples provided at the associated GitHub repository.

In conclusion, by combining flexibility with ease of use, we designed \mABCD\ to bridge the gap between accessibility and sophistication, making it a versatile tool for researchers across different levels of expertise. This approach ensures that users can generate synthetic multilayer networks that reflect the diverse and sometimes unpredictable patterns observed in real-world systems while still being able to explore a wide range of scenarios, from highly correlated layers to those with minimal interdependence. That kind of adaptability constitutes a crucial step forward in advancing the study and validation of community detection algorithms and other analytical techniques in multilayer network science.

\section{The \mABCD\ Model}\label{subsec:model}

In this section, we introduce a variant of the \textbf{ABCD} model that produces a synthetic collection of graphs that form a multilayer structure, \mABCD. \revision{The subsequent text guides through the details of the network construction process. Its parameters are summarized in Table~\ref{tab:mabcd_global_parameters} and Table~\ref{tab:mabcd_layer_parameters}, while the process itself is is conceptually illustrated in Figure~\ref{fig:schema}.

\subsection{Preliminaries}\label{subsub:preliminaries}

Since the networks produced by the proposed framework follow a power-law distribution in both node degrees and community sizes, and since their generation proceeds sequentially, we first introduce the preliminary concepts that are fundamental to the construction process.}

\subsubsection*{Power-law distribution}

Power-law distributions will be used to generate both the degree sequence and community sizes. For given parameters $\gamma \in (0, \infty)$, $\delta, \Delta \in \N$ with $\delta \leq \Delta$, we define a truncated power-law distribution $\tpl{\gamma}{\delta}{\Delta}$ as follows. For $X \sim \tpl{\gamma}{\delta}{\Delta}$ and for $k \in \N$ with $\delta \leq k \leq \Delta$,
\[
\p{X = k} = \frac{\int_k^{k+1} x^{-\gamma} \, dx}{\int_{\delta}^{\Delta+1} x^{-\gamma} \, dx} \,.
\]

\subsubsection*{The configuration model}

Suppose then that our goal is to create a graph on $n$ nodes with a given degree sequence $\textbf{d} = (d_i, i \in [n])$, where $\textbf{d}$ is a sequence of non-negative integers such that $m = \sum_{i \in [n]} d_i$ is even. We define a random multi-graph $\mathrm{CM}(\textbf{d})$ with a given degree sequence known as the \textbf{configuration model} (sometimes called the \textbf{pairing model}), which was first introduced by \cite{bollobas1980probabilistic}. See: \cite{bender1978asymptotic,wormald1984generating,wormald1999models} for related models and results.

We start by labelling nodes as $[n]$ and, for each $i \in [n]$, endowing node $i$ with $d_i$ half-edges. We then iteratively choose two unpaired half-edges uniformly at random (from the set of pairs of remaining half-edges) and pair them together to form an edge. We iterate until all half-edges have been paired. This process yields $G_n \sim \mathrm{CM}(\textbf{d})$, where $G_n$ is allowed self-loops and multi-edges and thus $G_n$ is a multi-graph.

\subsection{Parameters of the \mABCD\ model}\label{subsec:mabcd_params}

The \mABCD\ model is governed by the following parameters. The first family of parameters is responsible for a few global properties of the model and is listed in Table~\ref{tab:mabcd_global_parameters}.

\begin{table}[ht]
    \caption{Global parameters of \mABCD.}
    \centering
    \begin{tabular}{p{4.5em}|p{9em}|p{23.5em}}
    Parameter & Range & Description \\
    \hline \hline
    $n$ & $\N$ & Number of actors \revision{(recommended $n \geq 1{,}000$)} \\
    $\ell$ & $\N$ & Number of layers \\
    $\Rr$ & $[0,1]^{\ell \times \ell}$ & Inter-\revision{layer edge correlation matrix} \\
    $d$ & $\N$ & Dimension of reference layer \\
    \end{tabular}
    \label{tab:mabcd_global_parameters}
\end{table}

Actors will be associated with \emph{labels} from the set $[n]$. These labels will affect the degrees of actors. Each actor $a \in [n]$ will be associated with $\ell$ nodes, $v_i = (a,i)$ with $i \in [\ell]$, one for each of the $\ell$ layers. Moreover, each actor will be associated with a vector in $\R^d$ representing their features. We will refer to these vectors as vectors in the \emph{reference layer}. This reference layer will affect the process of generating partitions into communities in various layers.

The second family of parameters (Table~\ref{tab:mabcd_layer_parameters}) is responsible for various properties that are specific to each of the $\ell$ layers; subscripts $i \in [\ell]$ indicate that the corresponding parameters shape the $i^{th}$ layer. In particular, the set of parameters $\xi_i$, $i \in [\ell]$, will control the level of noise, that is, the fraction of edges in layer $i$ that are between nodes from two different communities. The suggested range of values for parameters $\gamma$ and $\beta$ (namely, the intervals $(2,3)$ and $(1,2)$, respectively) are chosen according to experimental values commonly observed in complex networks~\parencite{barabasi2016network,orman2009comparison}.

\begin{table}[ht]
    \caption{Layer-specific parameters of \mABCD\ \revision{(the index denotes the $i$-th layer)}.}
    \centering
    \begin{tabular}{p{4.5em}|p{9em}|p{23.5em}}
    Parameter & Range & Description \\
    \hline \hline
    $q_i$ & $(0,1]$ & Fraction of active actors \\
    $\tau_i$ & $[-1,1]$ & Correlation coefficient between degrees and labels \\
    $r_i$ & $[0,1]$ & Correlation strength between communities and the reference layer \\
    \hline
    $\gamma_i$ & $(2,3)$ & Power-law degree distr. with exponent $\gamma_i$ \\
    $\delta_i$ & $\N$ & Min. degree at least $\delta_i$ \revision{(recommended $\delta_i \geq 10$)} \\
    $\Delta_i$ & $\N \; (1 \le \delta_i \le \Delta_i < n)$ & Max. degree at most $\Delta_i$ \\
    \hline
    $\beta_i$ & $(1,2)$ & Power-law community size distr. with exponent $\beta_i$ \\
    $s_i$ & $\N$ & Min community size at least $s_i$ \revision{(recommended $s_i \geq 50$)} \\
    $S_i$ & $\N \; (\delta < s_i \le S_i \le n)$ & Max community size at most $S_i$ \\
    \hline
    $\xi_i$ & $(0,1)$ & Level of noise \\
    \end{tabular}
    \label{tab:mabcd_layer_parameters}
\end{table}

\subsection{The \mABCD\ construction}\label{subsec:abcd_construction}

We will use $\abcdDist$ for the distribution of graphs (layers) generated by the following 6-phase construction process. The model generates $\ell$ graphs; graph $G^i_n = ([n] \times \{i\}, E^i)$, $i \in [\ell]$, is the graph representing the $i$th layer. Once they are generated, we simply take $V = \bigcup_{i \in [\ell]} ([n] \times \{i\})$ and $E = \bigcup_{i \in [\ell]} E^i$.

\begin{figure}[ht!]
    \centering
    \includegraphics[width=\linewidth]{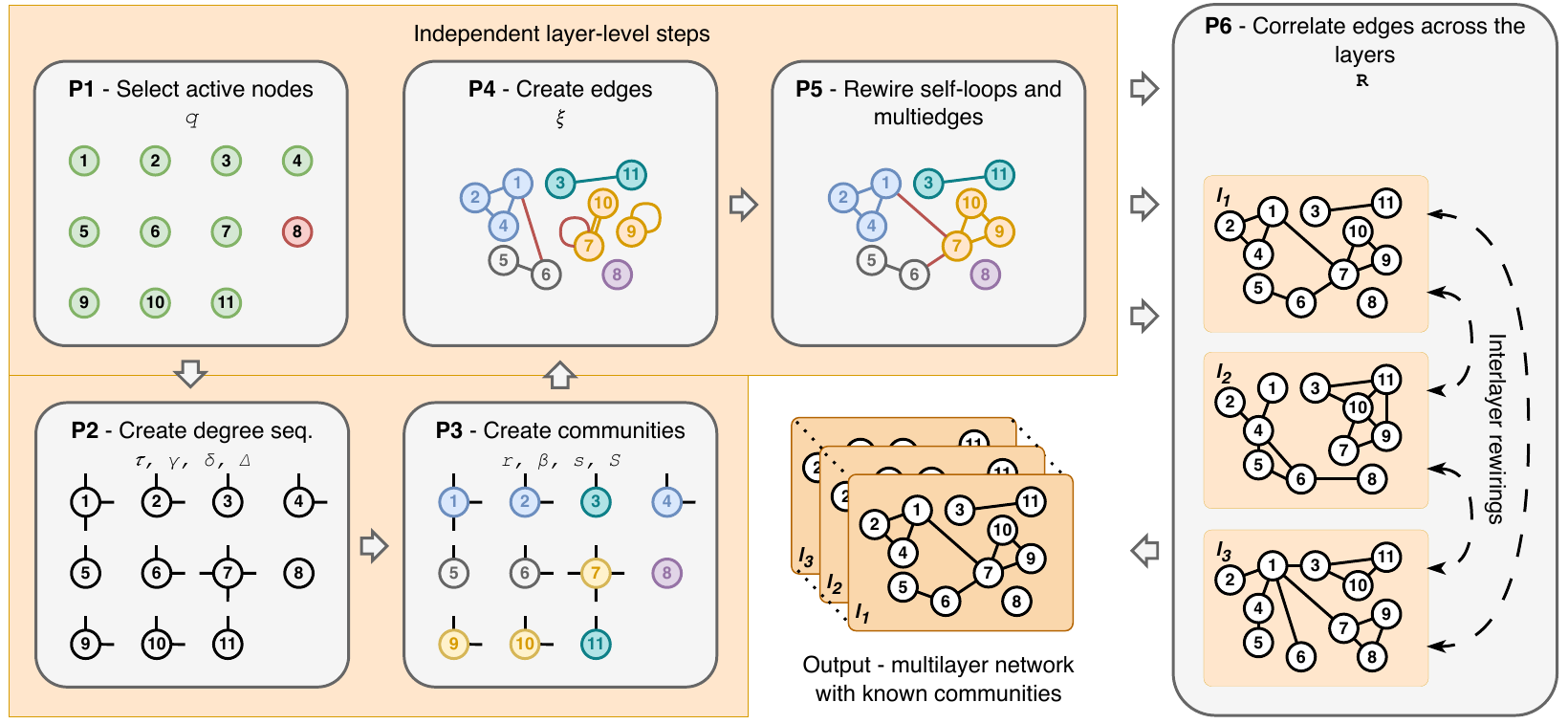}
    \caption{\revision{A schematic illustration of the \mABCD\ model’s operating principle, exemplified by the generation of a three-layer network. The process consists of six steps. The first five are performed independently for each layer, yielding scale-free monoplex graphs with a known community structure drawn from a shared latent space. In the final step, the layers are glued through a sequence of rewirings to achieve the desired cross-layer edge correlations.}}
    \label{fig:schema}
\end{figure}

\revision{Conceptually, the phases of network construction are illustrated in Figure~\ref{fig:schema}. The first five steps are carried out independently for each of the $\ell$ layers, while the sixth binds the generated graphs into a single multilayer network. The process is governed by the parameters listed in Table~\ref{tab:mabcd_global_parameters} and Table~\ref{tab:mabcd_layer_parameters}. In the first step, the so-called active nodes (i.e., those that will not be isolated) are determined. Each active node is then assigned its degree value. During the third step, communities are created and populated with nodes according to the latent, biscuit-like layer. Phase four focuses on connecting nodes within the same community as well as establishing inter-community (i.e. background) links. These graphs are subsequently simplified to remove self-loops and multiedges while preserving the intended structural properties. The final phase performs a series of edge rewirings to achieve the targeted edge correlations between each pair of layers. The illustration is complemented by pseudocode published on GitHub repository.}

\subsubsection*{Phase 1: Selecting active nodes}
As mentioned above, in a multilayer network, not all of the actors are active in all the layers. Actor $a \in [n]$ is \emph{active} in layer $i$ with probability $q_i$, independently for each $i \in [\ell]$ and all other actors. If an actor is not active in a given layer, then it will be represented by an ``artificial'' \emph{inactive} node. We use $N^i$ to denote the number of active nodes (and actors) in layer $i$. (Clearly, $N^i$ is a random variable with expectation $q_i n$.) For convenience, we will keep inactive nodes as a part of the corresponding graphs, but one may think of them as being removed from a given layer. 

\subsubsection*{Phase 2: Creating degree sequences}
The degree sequences for all of the $\ell$ layers are generated independently so we may concentrate on a given layer $i \in [\ell]$. We ensure that the degree sequence satisfies (a) a power-law with parameter $\gamma_i$, (b) a minimum value of at least $\delta_i$, and (c) a maximum value of at most $\Delta_i$ \revision{(see Section~\ref{subsub:preliminaries})}. 

Inactive nodes (representing actors not present in particular layer) are easy to deal with, they simply have degree zero. The remaining $N^i$ degrees are i.i.d.\ samples from the distribution $\tpl{\gamma_i}{\delta_i}{\Delta_i}$. We use $\textbf{d}^i_n= (d^i_v, v \in [n])$ for the generated degree sequence of $G^i_n$ with $d^i_1 \geq d^i_2 \geq \dots \geq d^i_n$; $\textbf{d}^i_{N^i}$ is a degree subsequence of active nodes. Finally, to ensure that $\sum_{v \in [n]} d^i_v$ is even, we decrease $d^i_1$ by 1 if necessary; we relabel as needed to ensure that $d^i_1 \geq d^i_2 \geq \dots \geq d^i_n$. 

Parameter $\tau_i \in [-1,1]$ controls how degrees of the nodes are correlated with labels of the associated actors (recall that node $(a,i)$ in layer $i$ is associated with an actor with label $a$). In one important case, namely, when $\tau_i=0$, there is no correlation at all and the degree sequence $\textbf{d}^i_{N^i}$ is assigned randomly to the $N_i$ active nodes. When $\tau_i=1$, the order of active nodes with respect to their labels is the same as the order with respect to their degrees; the largest degree node is first. In other words, if nodes $(a_1,i), (a_2,i)$ with $1 \le a_1 < a_2 \le n$ are active, then $\deg^i(a_1) \ge \deg^i(a_2)$, where $\deg^i(a_1)$ is the degree of node $(a_1,i)$. Similarly, if $\tau_i=-1$, then the order of active nodes with respect to their labels is also consistent with the order with respect to their degrees but this time the last node is of the largest degree. Since $\tau_i$'s could be different for different layers, one node could have large degrees in some layers but small ones in some other ones.

To achieve the desired property, each active node $(a,i)$ independently generates a normally distributed random variable $X_a = N( a/n, \sigma_i )$, where the variance $\sigma_i$ is a specific function of $\tau_i$. (Recall that we concentrate on a given layer $i \in [\ell]$. For convenience, we simplify the notation and stop referencing to layer $i$ in notation such as $X_a$. Still, there are many independent random variables for each active node $(a,i)$ associated with actor $a$.) We sort active nodes in increasing order of their values of $X_a$ and assign the degree sequence accordingly; that is, node $(a,i)$ gets degree $d^i_r$, where $r \in [N^i]$ is the rank of $X_a$. In particular, the node with the smallest value of $X_a$ gets assigned the largest degree, namely, $d^i_1$. Note that if $\sigma_i=0$, then $X_a = a/n$ (deterministically), and so we recover the perfect correlation between the degrees and the labels ($\tau_i=1$). On the other hand, if $\sigma_i \to \infty$, then the order of nodes is perfectly random (with uniform distribution), so we recover the other desired extreme ($\tau_i=0$). 

Function $\sigma_i : [0,1] \to [0, \infty)$ is empirically approximated so that the variance $\sigma_i = \sigma_i(\rho_i)$ yields the Kendall rank correlation close to $\tau_i \in [0,1]$ between the ordering generated by the ranks of $X_a$ and the labels $a$ associated with corresponding actors that are active in layer $i$. Twenty degree distributions are independently generated (with the same $\sigma$ and different random seeds), and the one with the correlation coefficient that is the closest to the desired value of $\tau_i$ is kept. To deal with negative correlations $\tau_i \in [-1, 0)$, we simply ``flip'' the order generated for $|\tau_i|$. 

\subsubsection*{Phase 3: Creating communities}\label{subsubsec:creating_communities}

Our next goal is to create community structure in each layer of the \mABCD\ model. When we construct a community, we assign a number of nodes to said community equal to its size. Initially, the communities form empty graphs. Then, in later phases we handle the construction of edges using the degree sequence established in Phase~2.

Similarly to the process of generating the degree sequences, the sequence of community sizes are generated independently, ensuring that the distribution for a given layer $i \in [\ell]$, satisfy (a) a power-law with parameter $\beta_i$, (b) a minimum value of $s_i$, and (c) a maximum value of $S_i$  \revision{(see Section~\ref{subsub:preliminaries})}. In addition, we also require that the sum of community sizes is exactly $n$. Specifically, inactive nodes (if there are any) form their own community, namely, $C^i_0$. Other communities are generated with sizes determined independently by the distribution $\tpl{\beta_i}{s_i}{S_i}$. We generate communities until their collective size is at least $n$. If the sum of community sizes at this moment is $n + x$ with $x > 0$, then we perform one of two actions. If the last added community has a size at least $x+s_i$, then we reduce its size by $x$. Otherwise (that is, if its size is $c < x+s_i$), we delete this community, select $c-x$ old communities and increase their sizes by 1. 

Now, given that the sequences of community sizes are already determined (for all layers), it is time to assign nodes to communities. To allow communities to be correlated with each other, we first create a latent \emph{reference} layer that will guide the process of assigning nodes to specific communities across all layers. One may think of this auxiliary layer as properties of actors (such as people's age, education, geographic location, beliefs, etc.) shaping different layers (for example, various social media platforms). This single reference layer will be used for all $\ell$ layers. In this reference layer, each actor $a \in [n]$ gets assigned a random vector in $\R^d$ (by default, $d=2$) that is taken independently and uniformly at random from the ball of radius one centred at $\textbf{0}=(0,0,\ldots,0)$. 

Let us now concentrate on a given layer $i \in [\ell]$. Recall that the community sizes have already been generated. We write $L^i$ for the (random) number of regular communities in layer $i$ partitioning the set of active nodes in this layer and use $\textbf{c}^i = (c^i_j, j \in \{0\} \cup [L^i])$ for the corresponding sequence of community sizes. Recall that inactive nodes (representing actors not present in a particular layer) form their own community (namely, $C^i_0$) so $c^i_0 = n - N^i$ is the number of inactive nodes in layer~$i$. Let $R$ be the set of active nodes. We assign nodes to communities, dealing with one community at a time, in a random order. When community $C^i_j$ is formed (for some $j \in [L^i]$), we first select a node from $R$ that is at the largest distance from the center $\textbf{0}$ (in the reference layer). This node, together with its $c^i_j - 1$ nearest neighbours in $R$, are put to $C^i_j$. We remove $C^i_j$ from $R$ and move on to the next community.

\begin{figure}[ht!]
    \centering
    \includegraphics[scale=0.45]{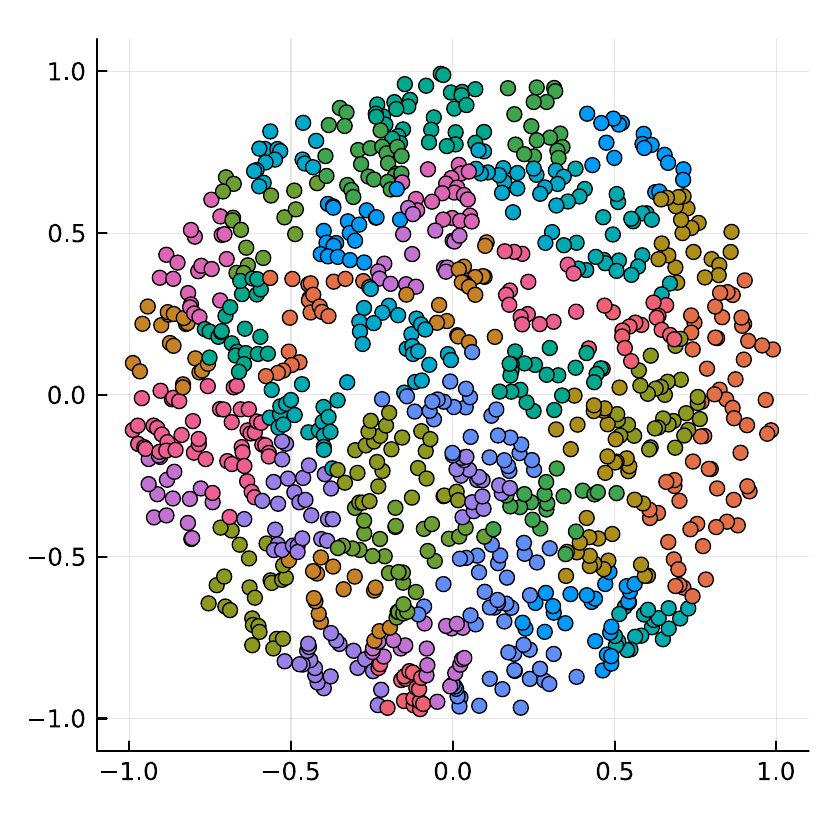}
    \includegraphics[scale=0.45]{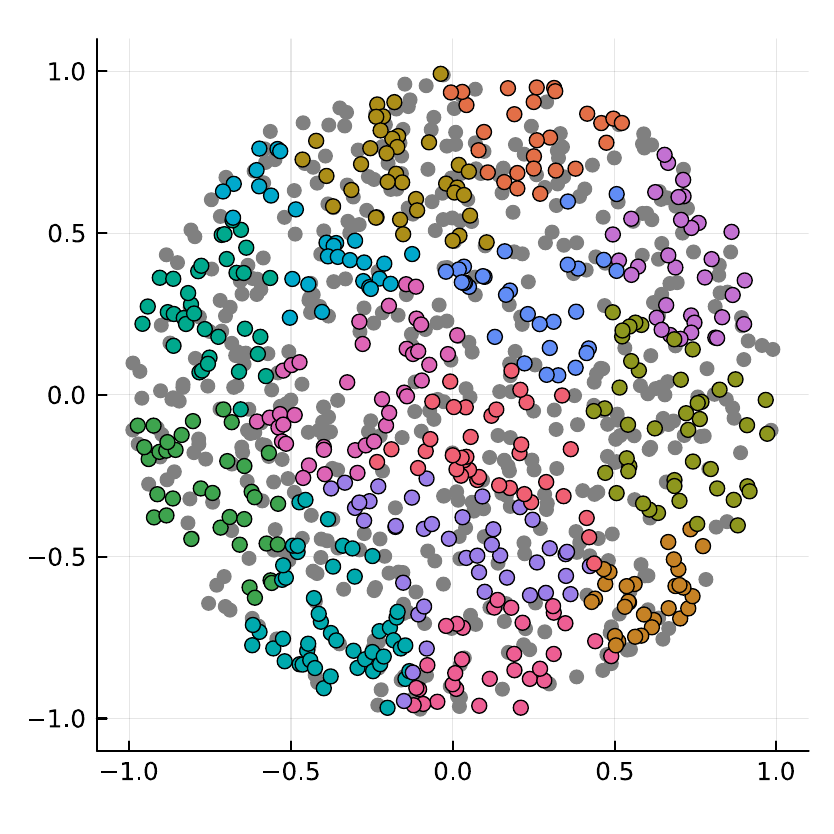}
    \caption{Two partitions generated based on the same reference layer with $n=1{,}000$ nodes: (left) $q_1 = 1$ (all nodes active), $S_1 = 32$, $s_1 = 16$, $\beta_1=1.5$, (right): $q_2 = 0.5$ (50\% nodes active), $S_2 = 50$, $s_2 = 25$, $\beta_2=1.5$ \revision{with inactive nodes shown in grey.}}
    \label{fig:reference_layer}
\end{figure}

The above strategy creates a partition of nodes that is highly correlated with the geometric locations of nodes in the reference layer; nodes that are close to each other in the reference layer are often in the same community. \revision{An example of this process is illustrated in Figure~\ref{fig:reference_layer}, which shows how nodes are assigned to communities in two network layers, both guided by the same reference layer. This mechanism can be intuitively compared to eating a biscuit, bite by bite, from the edge towards the centre. Nodes that are close to each other are more likely to be ``eaten at the same time'' and thus placed in the same group. Moreover, the severed pieces may vary in size, so do communities. Finally, the same biscuit can be enjoyed by two different gourmets, which in this analogy corresponds to reusing the same reference layer for each of the relationships modelled within the generated network.} To reduce the correlation strength (modelled by the parameter $r_i \in [0,1]$), we perform the following procedure. Each active node independently leaves its own community with probability $1-r_i$, freeing a spot in this community. All the nodes that left are then put back randomly to any available spot (which typically is in a~community that this node was \emph{not} originally in). Note that in the extreme case, when $r_i=0$, the resulting partition does not depend on the reference layer at all, so there is no correlation. We write $\textbf{C}^i_n = (C^i_j,j \in \{0\} \cup [L^i])$ for the generated collection of communities in $G^i_n$. Again, let us stress the fact that $\textbf{C}^i_n$ is a random partition of $[n]$ of random size $L^i+1$. \revision{At this point, it is worth mentioning that, due to the shared latent layer and a parameter $r_i$ specific to each relationship, it is possible to model the alignment of communities across the network. Following~\cite{magnani2021community}, where the notion of a \emph{pillar} is introduced to describe a community that persists across at least two layers, layers with $r_i \approx 1$ tend to exhibit this form of cluster alignment. Conversely, if the values of $r_i$ are set to more distinctive numbers, the communities are expected to be largely disjoint across the layers.}

Finally, note that in the above process of assigning nodes to communities, as opposed to the original \textbf{ABCD} model, we ignore the degree of nodes. Indeed, the original \textbf{ABCD} model tries to make sure that large degree nodes are not assigned to small communities. In \mABCD, there are many layers and non-trivial correlations between partitions into communities and degree sequences between layers. In a hypothetical extreme situation, it might happen that each node belongs to some small community in some layer. Hence, the \mABCD\ model does not try to prevent such unavoidable situations and will resolve potential issues later (see Phase~5).

\subsubsection*{Phase 4: Creating edges}

Now, it is time to form edges in \mABCD. It will be done in the next three phases. 
Phases~4 and~5 will independently generate $\ell$ graphs $G^i_n$, $i \in [\ell]$, for each of the $\ell$ layers whereas Phase~6 will make sure that edges across various layers are correlated, if needed. We may then concentrate on a given layer $i \in [\ell]$.

At this point $G^i_n$ contains $n$ nodes labelled as $(a,i)$, $a \in [n]$, partitioned by the communities $\textbf{C}^i_n$, with node $(a,i)$ containing $\deg^i(a)$ unpaired half-edges. Firstly, for each $a \in [n]$, we split the $\deg^i(a)$ half-edges of $(a,i)$ into two distinct groups, which we call \textit{community} half-edges and \textit{background} half-edges. For $a \in \Z$ and $b \in [0,1)$ define the random variable $\round{a+b}$ as
\[
\round{a+b} = \bigg\{
\begin{array}{ll}
a & \text{ with probability } 1-b, \text{ and}\\
a+1 & \text{ with probability } b \,.
\end{array}
\]
(Note that $\E {\round{a+b}} = a(1-b) + (a+1)b = a+b$.) Now define $Y_a = \round{(1-\xi) \deg^i(a)}$ and $Z_a = \deg^i(a) - Y_a$ (note that $Y_a$ and $Z_a$ are random variables with $\E{Y_a} = (1-\xi) \deg^i(a)$ and $\E{Z_a} = \xi \deg^i(a)$ and since we generate each layer separately they are different for each layer) and, for all $a \in [n]$, split the $\deg^i(a)$ half-edges of $(a,i)$ into $Y_a$ community half-edges and $Z_a$ background half-edges. Next, for all $j \in [L^i]$, construct the \textit{community graph} $G^i_{n,j}$ as per the configuration model \revision{(see Section~\ref{subsub:preliminaries})} on node set $C^i_j$ and degree sequence $(Y_a, a \in C^i_j)$. Note that $C_0$ consists of inactive nodes which, by design, have degree zero. Hence, there is no need to do anything with them. Finally, construct the \textit{background graph} $G^i_{n,0}$ as per the configuration model on node set $[n]$ and degree sequence $(Z_a, a \in [n])$. In the event that the sum of degrees in a community is odd, we pick a maximum degree node $(a,i)$ in said community and replace $Y_a$ with $Y_a + 1$ and $Z_a$ with $Z_a - 1$. Note that $G^i_{n,j}$ is a graph, and $C^i_j$ is the set of nodes in this graph; we refer to $C^i_j$ as a \textit{community} and $G^i_{n,j}$ as a \textit{community graph}. Note also that $G^i_n = \bigcup_{0 \leq j \leq L^i} G^i_{n,j}$.

\subsubsection*{Phase 5: Rewiring self-loops and multi-edges}

We continue concentrating on a given layer $i \in [\ell]$. Note that, although we are calling $G^i_{n,j}$ ($j \in \{0\} \cup [L^i]$) \textit{graphs}, they are in fact \textit{multi-graphs} at the end of Phase~4. To ensure that $G^i_n$ is simple, we perform a series of \textit{rewirings} in $G^i_n$. A rewiring takes two edges as input, splits them into four half-edges, and creates two new edges distinct from the input. We first rewire each community graph $G^i_{n,j}$, $j \in [L^i]$, independently as follows.
\begin{enumerate}
\item For each edge $e \in E(G^i_{n,j})$ that is either a loop or contributes to a multi-edge, we add $e$ to a \textit{recycle} list that is assigned to $G^i_{n,j}$.
\item We shuffle the \textit{recycle} list and, for each edge $e$ in the list, we choose another edge $e'$ uniformly from $E(G^i_{n,j}) \setminus \{e\}$ (not necessarily in the \textit{recycle} list) and attempt to rewire these two edges. We save the result only if the rewiring does not lead to any further self-loops or multi-edges, otherwise we give up. In either case, we then move to the next edge in the \textit{recycle} list.
\item After we attempt to rewire every edge in the \textit{recycle} list, we check to see if the new \textit{recycle} list is smaller. If yes, we repeat step 2 with the new list. If no, we give up and move all of the ``bad'' edges from the community graph to the background graph. 
\end{enumerate}
We then rewire the background graph $G^i_{n,0}$ in the same way as the community graphs, with the slight variation that we also add edge $e$ to \textit{recycle} if $e$ forms a multi-edge with an edge in a community graph or, as mentioned previously, if $e$ was moved to the background graph as a result of giving up during the rewiring phase of its community graph. At the end of Phase~5, we have a simple graph $G^i_n$ representing the $i$-th layer of a multilayer network.

\subsubsection*{Phase 6: Correlations between edges in various layers}

During this last phase, we continue performing a series of rewiring (in batches) with the goal of creating a multilayer network with the correlations between edges in various layers (as defined in Subsection~\ref{subsubsec:correlations}) to be as close to the desired matrix $\Rr$ (provided as one of the parameters of the model) as possible. It is important to highlight the fact that during this phase, not only do the degrees of the involved nodes not change, but the community degrees stay the same (as well as the background ones). Hence, in particular, the level of noise stays the same. 

We run $t$ independent \emph{batches} of operations (by default, $t=100$). Before every batch, we re-compute the (empirical) correlation matrix $\hat{\Rr}$ for the current multilayer network $(G^i_n: i \in [\ell])$ and compare it with the desired matrix~$\Rr$. We select an entry $ij$ at random with the probability proportional to the discrepancy between the empirical and the desired values. In other words, we select a pair $(i,j)$ ($1 \le i < j \le \ell$) with probability
$$
p_{ij} = \frac {|\hat{r}_{ij} - r_{ij}|}{\sum_{1 \le r < s \le \ell} |\hat{r}_{rs} - r_{rs}|}.
$$
We attempt to rewire $\lceil \epsilon \min\{ |E_i^j|, |E_j^i| \} \rceil$ of edges in each batch with the goal to bring $\hat{r}_{ij}$ closer to $r_{ij}$ (by default, $\epsilon=0.05$). Recall that $E_i^j$ can be viewed as the set of edges in layer $i$ that are between actors that are active in layer $j$ (and, trivially, also active in layer $i$ since inactive actors form isolated nodes), see Equation~\ref{eq:Eij}.

Suppose first that $\hat{r}_{ij} < r_{ij}$, that is, the correlation between layer $i$ and layer $j$ is smaller than what we wished for. Each of the attempts does the following. Randomly select one of the two graphs, $G^i_n$ or $G^j_n$, and call it \emph{primary}. Then, pick a random edge $uv$ from the primary graph between actors that are active in both layers. Our goal is to try to introduce edge $uv$ in the other graph (call it \emph{secondary}) unless it is already there, in which case we simply finish this attempt prematurely. If $u$ and $v$ belong to one of the communities in the secondary graph (say to the community $C$), then we take $u'$ to be a random neighbour of $u$ in $C$ (if there are any), take $v'$ to be a random neighbour of $v$ in $C$ (again, if there are any). If $u, v, u', v'$ are four distinct nodes and there is no edge $u'v'$ in the secondary graph, then we remove the two edges $uu'$ and $vv'$ and introduce two new edges $uv$ and $u'v'$. If anything goes wrong, then we simply finish prematurely and move on to another attempt. If $u$ and $v$ are from two different communities in the secondary graph, then the procedure is exactly the same, but this time, our goal is to select four nodes, each from a different community. Specifically, we try to pick a random neighbour $u'$ of $u$ outside of the communities $u$ or $v$ belong to. Then, we try to pick a random neighbour $v'$ of $v$ outside of the communities $u$, $v$, or $u'$ belong to. If the four selected nodes are different and there is no edge $u'v'$ in the secondary graph, we do the rewiring.

Suppose now that $\hat{r}_{ij} > r_{ij}$, that is, the correlation between layer $i$ and layer $j$ is larger than what we wished for. As before, during each attempt, we randomly make one of the two graphs, $G^i_n$, $G^j_n$, to be \emph{primary} and the second one to be \emph{secondary}. Then, pick a random edge $uv$ from the intersection of the two graphs. Our goal is to try to remove edge $uv$ from the secondary graph. If $u$ and $v$ belong to one of the communities in the secondary graph (say to the community $C$), then we take a random edge $u'v'$ from $C$. If $u, v, u', v'$ are four distinct nodes and there are no edges $uu'$ nor $vv'$ in the secondary graph, then we remove the two edges $uv$ and $u'v'$ and introduce two new edges $uu'$ and $vv'$. As before, if anything goes wrong, then we simply finish prematurely and move on to another attempt. If $u$ and $v$ are from two different communities in the secondary graph, then we pick a random edge $u'v'$ from the secondary graph with the property that all four nodes belong to different communities. We try to rewire the two edges, making sure that no multi-edges get created. 

The goal of the sequence of $t$ batches is to bring the (empirical) correlation matrix $\hat{\Rr}$ closer to the desired matrix~$\Rr$. Unfortunately, fixing one entry of $\Rr$ may affect the other entries. Hence, it is not guaranteed that the best solution is found after exactly $t$ batches. To take this into account, we track the quality of the multilayer networks $(G^i_n: i \in [\ell])$ at the beginning of each bath (via $L_2$ norm between $\hat{\Rr}$ and $\Rr$) and the final network is the one of the $t$ networks that performed best.

\subsubsection*{Implementation}
The algorithm is implemented in Julia programming language, a high-level, general-purpose dynamic programming language, designed to be fast and effective. Source code and documentation are available on \mABCD\ GitHub repository\footnote{\url{https://github.com/KrainskiL/MLNABCDGraphGenerator.jl}}. The Python ports are available as well\footnote{\url{https://github.com/anty-filidor/spreading-vs-mln-structure}}.

\section{Properties of the \mABCD\ Model}\label{sec:properties}

In this section, to justify \mABCD\ model design and to highlight important features, we provide various experiments to investigate some important properties of the model.

\subsection{Correlations between node degrees}

Recall that in Phase~2 of the \mABCD\ model, independently for all layers, each active node associated with an actor $a \in [n]$ independently generates a normally distributed random variable $X_a = N(a/n,\sigma)$ for some fixed value of the variance $\sigma$. In the first experiment, we independently generate the degree distributions for various values of $\sigma \in [0,20]$. We compute the Kendall rank correlation coefficient between actors' labels $a$ and the corresponding random variables $X_a$. 

\begin{figure}[ht!]
    \centering
    \includegraphics[scale=0.38]{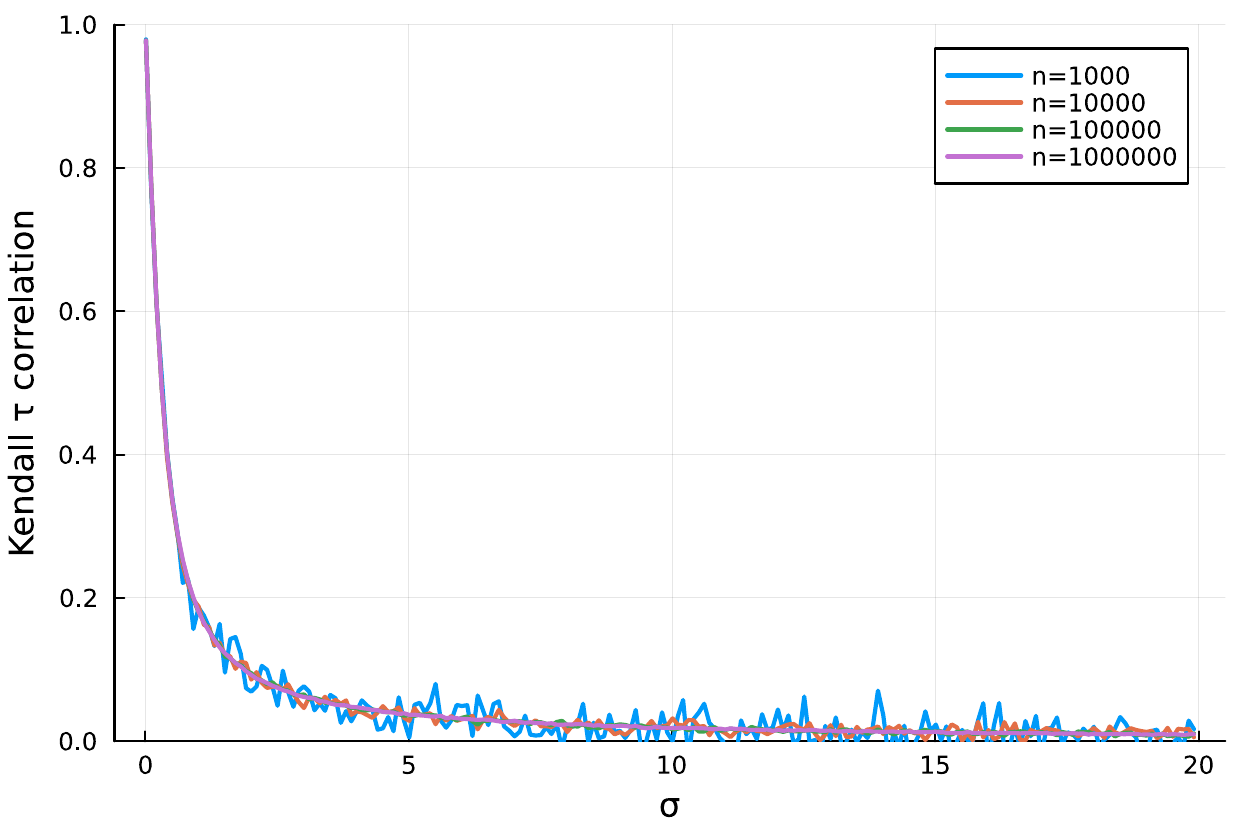}
    \includegraphics[scale=0.38]{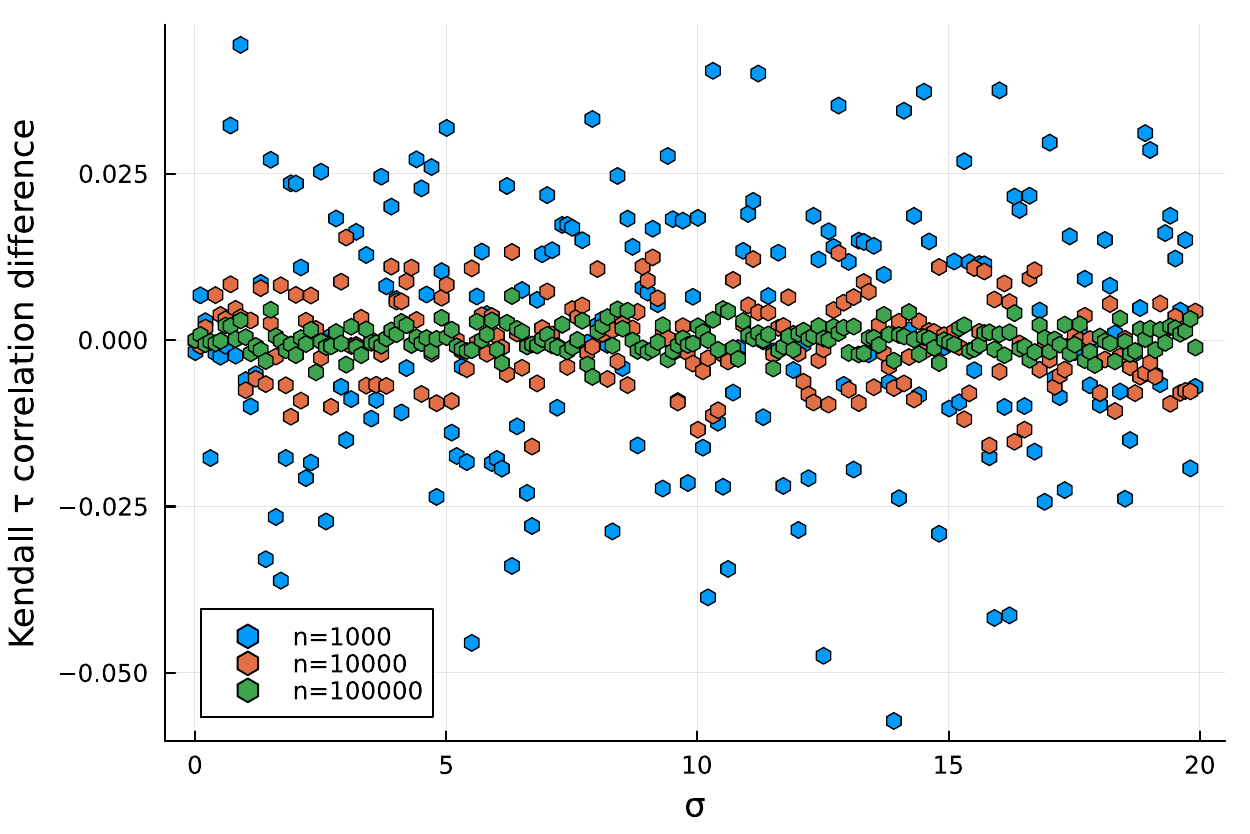}
    \caption{
    Left: Kendall rank correlation coefficients between actors' labels and generated random variables $X_a = N(a/n,\sigma)$ for various values of $n$.
    Right: Difference between Kendall rank correlation coefficients for small values of $n$ and the ones for $n=1{,}000{,}000$.}
    \label{fig:degree_correlation_sigma}
\end{figure}

To see how quickly the process converges, we compare the results for the number nodes $n$ in $\{ 10^3, 10^4, 10^5, 10^6 \}$ and all actors being active---see Figure~\ref{fig:degree_correlation_sigma}~(Left). To see how close the coefficients are to their asymptotic limit (as $n \to \infty$), we also look at the difference between the values for $n \in \{ 10^3, 10^4, 10^5 \}$ and the ones for $n = 10^6$---see Figure~\ref{fig:degree_correlation_sigma}~(Right). The differences are rather small, even for $n$ as small as $1{,}000$. Of course, there are some natural fluctuations, but they are rather insignificant compared to the expectation. It justifies our design of the algorithm. In practice, based on the experiments with $n = 10^6$ (that are saved and accessible by the algorithm), one can select an appropriate value of the parameter $\sigma$ that is expected to produce the desired Kendall rank correlation coefficients $\tau$. With such a choice of $\sigma$, generating 20 independent sequences is enough to make sure that one of them is very close to the desired value of $\tau$.

In the second experiment, we generate the degree sequences for $n=1{,}000$ actors in 5 layers. The desired correlations between the actors'
labels and the order of the corresponding random variables $X_a$ is fixed to be $(\tau_i) = (1.0, 0.5, 0.0, -0.5, -1.0)$ and not all the actors are active in all layers: $(q_i) = (1.0, 0.9, 0.8, 0.7, 0.6)$. The degree sequence for active nodes is independently generated in each layer with the following parameters: $\gamma_i=2.5$, $\delta_i=5$, $\Delta_i=50$.

\begin{figure}[ht!]
    \centering
    \includegraphics[scale=0.38]{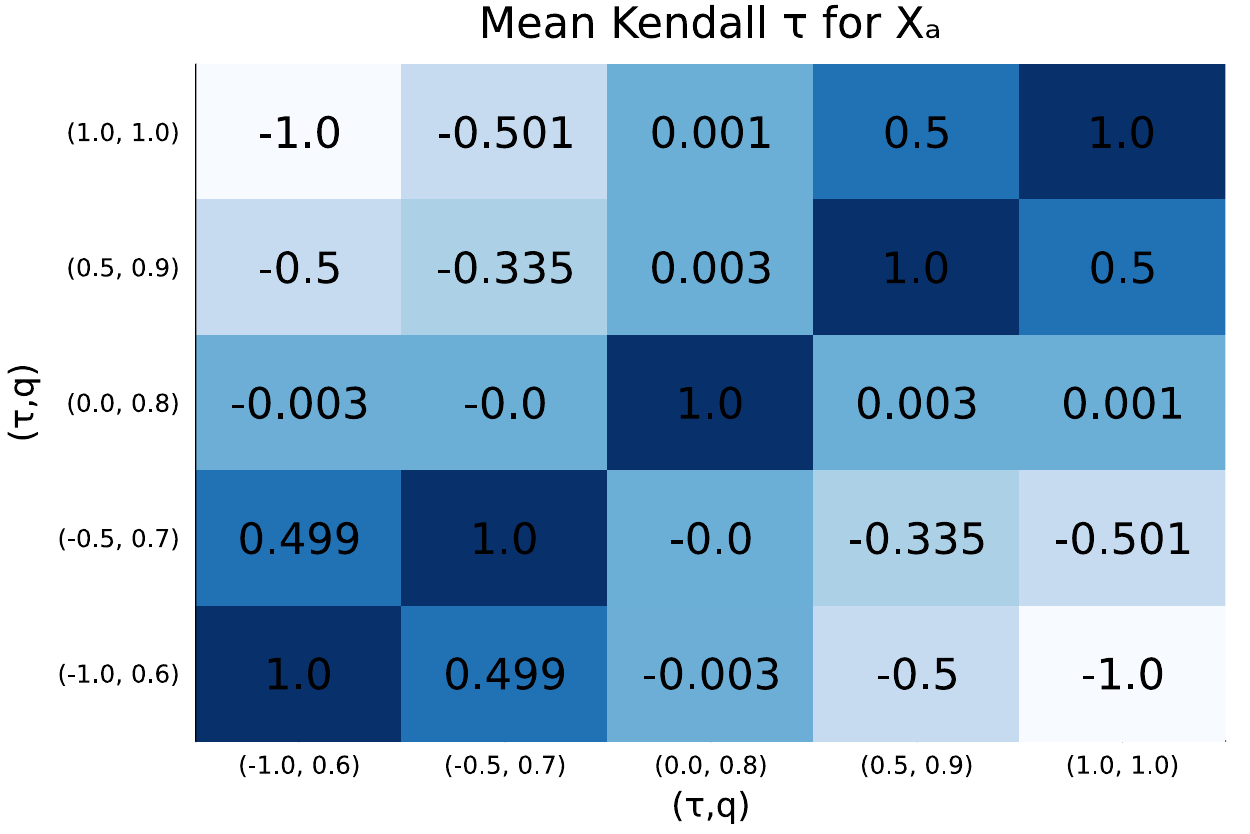}
    \includegraphics[scale=0.38]{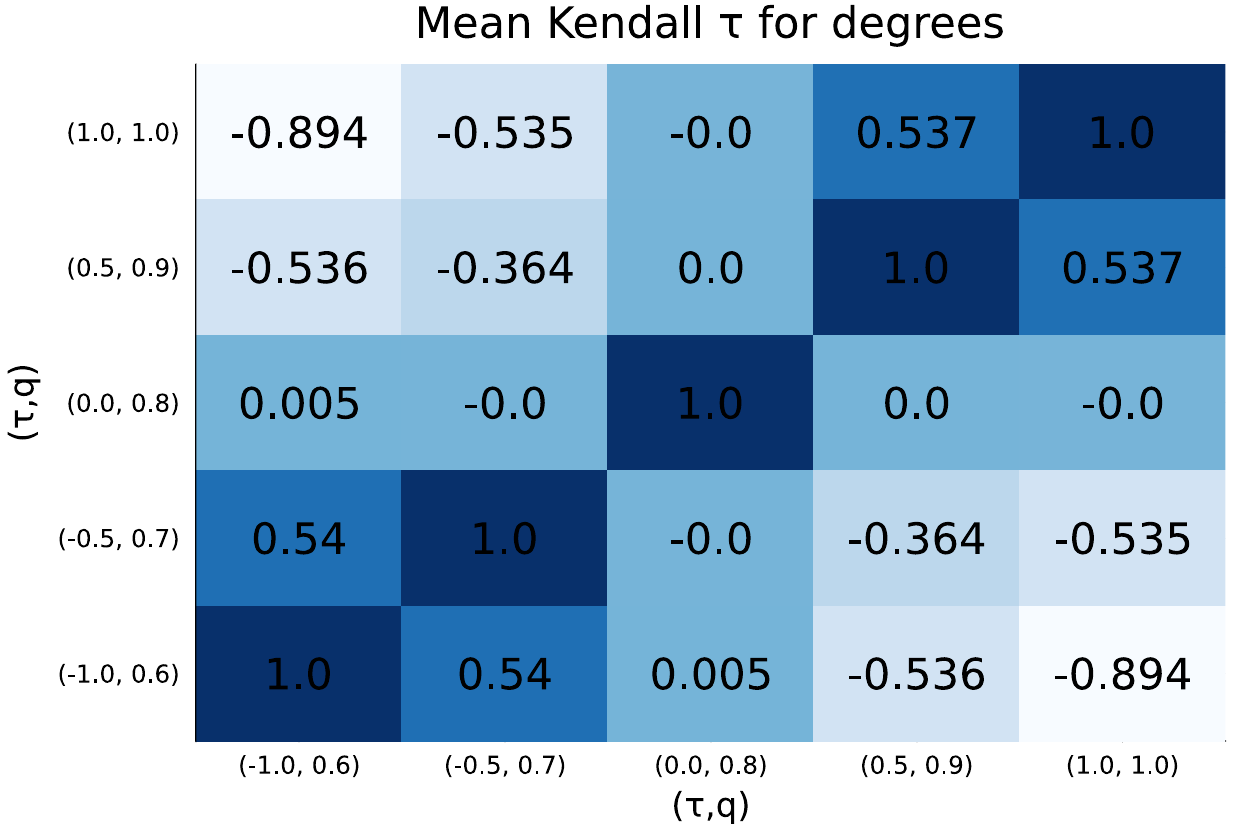}
    \caption{The average value of Kendall $\tau$ rank correlation (over 100 repetitions) between sequences of random variables $X_a$ (Left) and degree sequences (Right) in two layers. Correlations are computed on nodes that are active in both layers.}
    \label{fig:rhos_q_sampled_spearman}
\end{figure}

We repeat the above experiment 100 times. The experimental means and standard deviations of the corresponding Kendall $\tau$ correlations between the hidden labels of active actors and generated random variables $X_a$ are $(1.0, 0.5, 0.0, -0.5, -1.0)$ and, respectively, $(0.0, 0.001, 0.002, 0.001, 0.0)$, very close to the desired sequence $(1.0, 0.5, 0.0, -0.5, -1.0)$. We also computed correlations between sequences of $X_a$ as well as degree sequences in all $\binom{5}{2}$ pairs of layers (on nodes that are active in both layers). The experimental means are reported in Figure~\ref{fig:rhos_q_sampled_spearman} for sequences of $X_a$ (Left) and for degree sequences (Right). As expected, both matrices contain similar values, and correlations behave as expected. For example, layers 4 and 5 are negatively correlated with the labels of actors ($-0.5$ and $-1.0$, respectively) and so are positively correlated with each other ($0.5$ and $0.537$ in the two corresponding correlations).

\begin{figure}[ht!]
    \centering
    \includegraphics[scale=0.38]{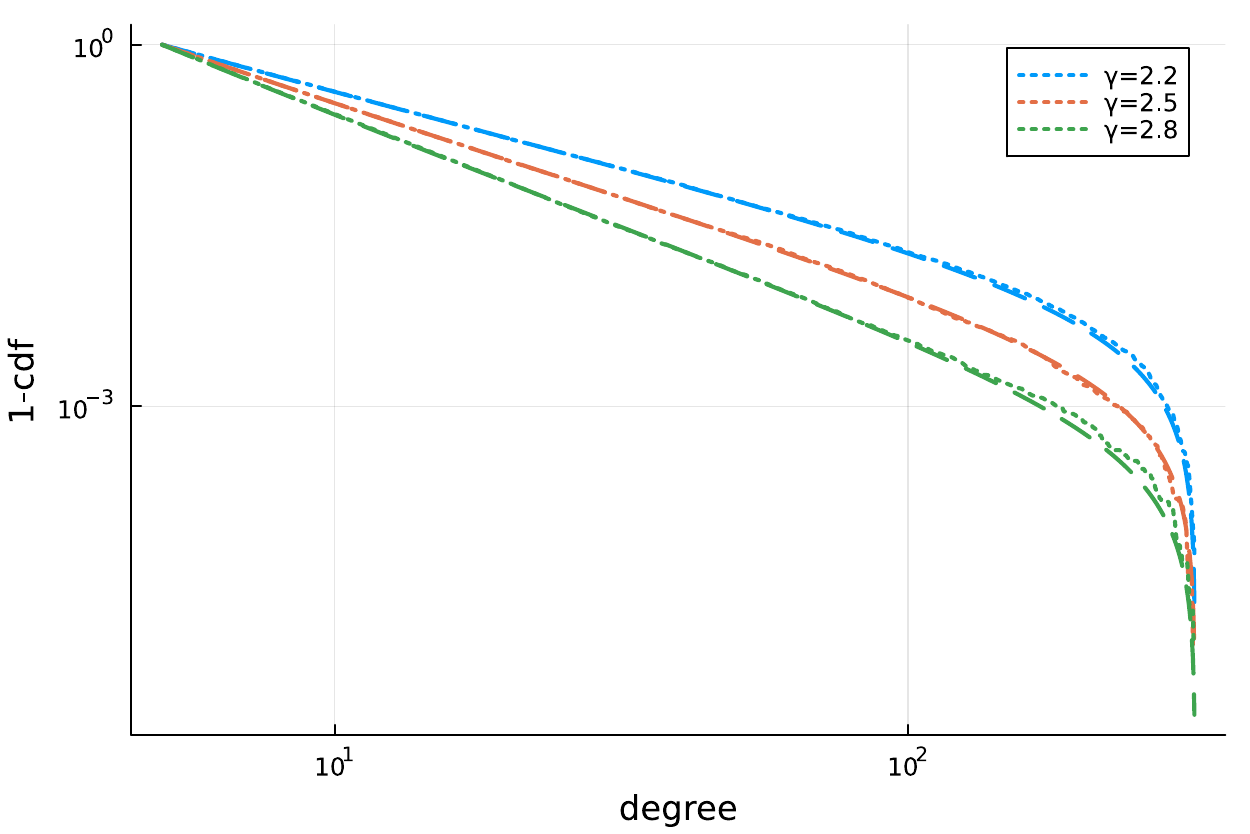}
    \includegraphics[scale=0.38]{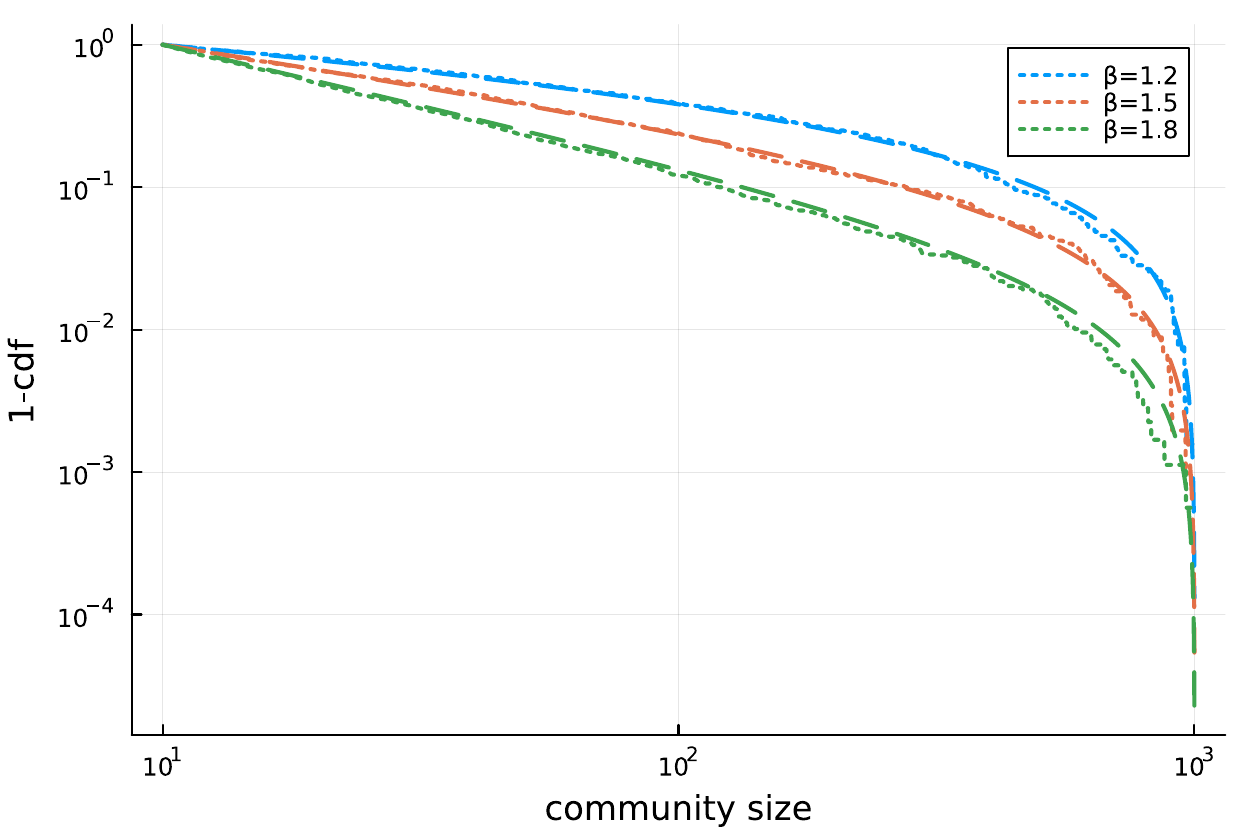}
    \caption{Left: Empirical (dots) and theoretical (dashes) log-log cumulative degree distributions for three degree sequences: $n=100\,000, \delta=5,\Delta=316.$ Right: Empirical (dots) and theoretical (dashes) log-log cumulative community sizes distributions for three sizes sequences: $n=100\,000, s=10, S=1\,000.$}
    \label{fig:powerlaws}
\end{figure}

The third and final experiment in this subsection is merely a sanity check, as this property is enforced in the model by design. We generate degree sequences for three layers with $n=100{,}000$ actors in each, all of them being active ($q=1$). The minimum and the maximum degrees are fixed to be $\delta=5$, $\Delta=316 \approx \sqrt{n}$ across the three layers, but the power-law exponents vary: $(\gamma_i) = (2.2, 2.5, 2.8)$. For a given integer $k$, let $f(k)$ be the experimental cumulative degree distribution, that is, $f(k)$ is the fraction of nodes of degree at least $k$. For a given set of parameters, the theoretical cumulative degree distribution is equal to 
$$
\hat{f}(k) = \frac { \sum_{i = k}^{\Delta} i^{-\gamma} } { \sum_{i = \delta}^{\Delta} i^{-\gamma} } .
$$
We show that the experimental degree distributions are very close to the desired, theoretical, ones --- see Figure~\ref{fig:powerlaws} (Left) for the cumulative degree distributions of the three layers.

\subsection{Correlations between partitions}

Recall that in Phase~4 of the \mABCD\ model, active nodes in each layer are independently partitioned into communities of given sizes. The sequences of community sizes can be substantially different across layers, and the process itself is random. However, the underlying geometry of the reference layer (that is used for all of the layers) ensures that the generated partitions are correlated.

\begin{figure}[ht!]
    \centering
    \includegraphics[scale=0.38]{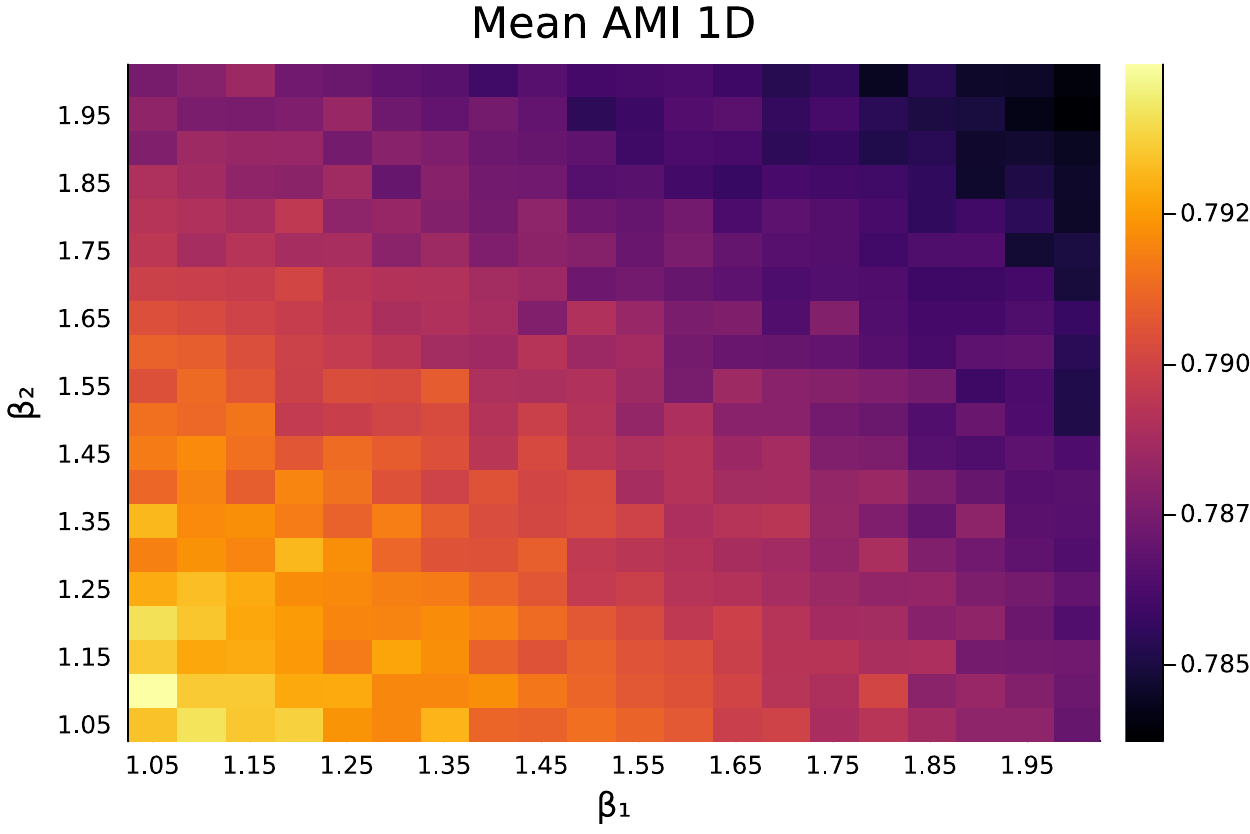}
    \includegraphics[scale=0.38]{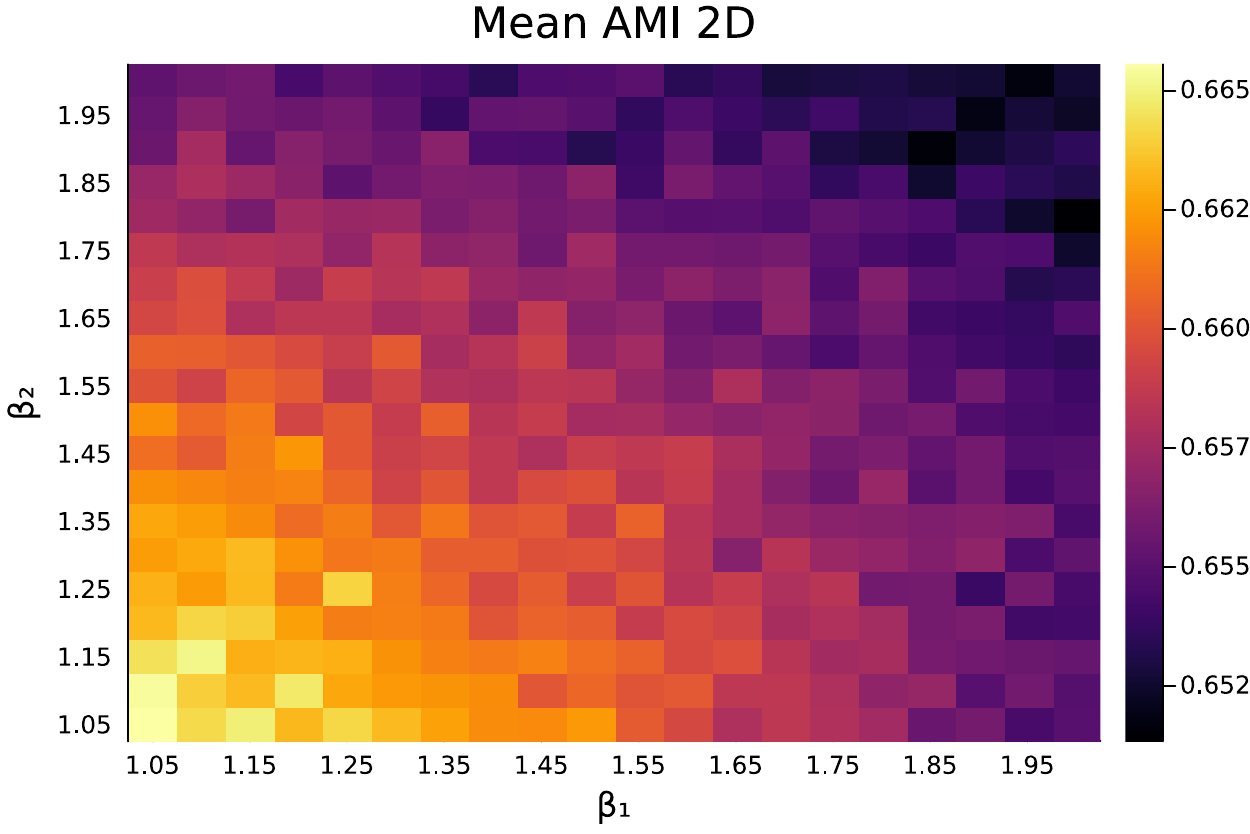}
    \caption{The average value of \textbf{AMI} (over 500 repetitions) between two partitions generated with $\beta_1$ and, respectively, $\beta_2$: $n=1000$, $s_1=s_2=8$, $S_1=S_2=32$; $d$-dimensional reference layer was used: (left) $d=1$, (right) $d=2$.}
    \label{fig:ami_betas_heatmap}
\end{figure}

In the first experiment, we verify how different values of the parameter $\beta$ (power-law exponent for the community sizes) affect the \textbf{AMI} between two partitions, the first one generated with parameter $\beta_1$ and the second one generated with parameter $\beta_2$---see Figure~\ref{fig:ami_betas_heatmap}. The difference is very small, with a slightly larger correlation obtained for small values of $\beta$. We also check how the choice of the dimension $d$ affects the \textbf{AMI}. Not surprisingly, a larger correlation is obtained for dimension $d=1$, but the 2-dimensional reference layer still produces strong correlations.

\begin{figure}[ht!]
    \centering
    \includegraphics[scale=0.38]{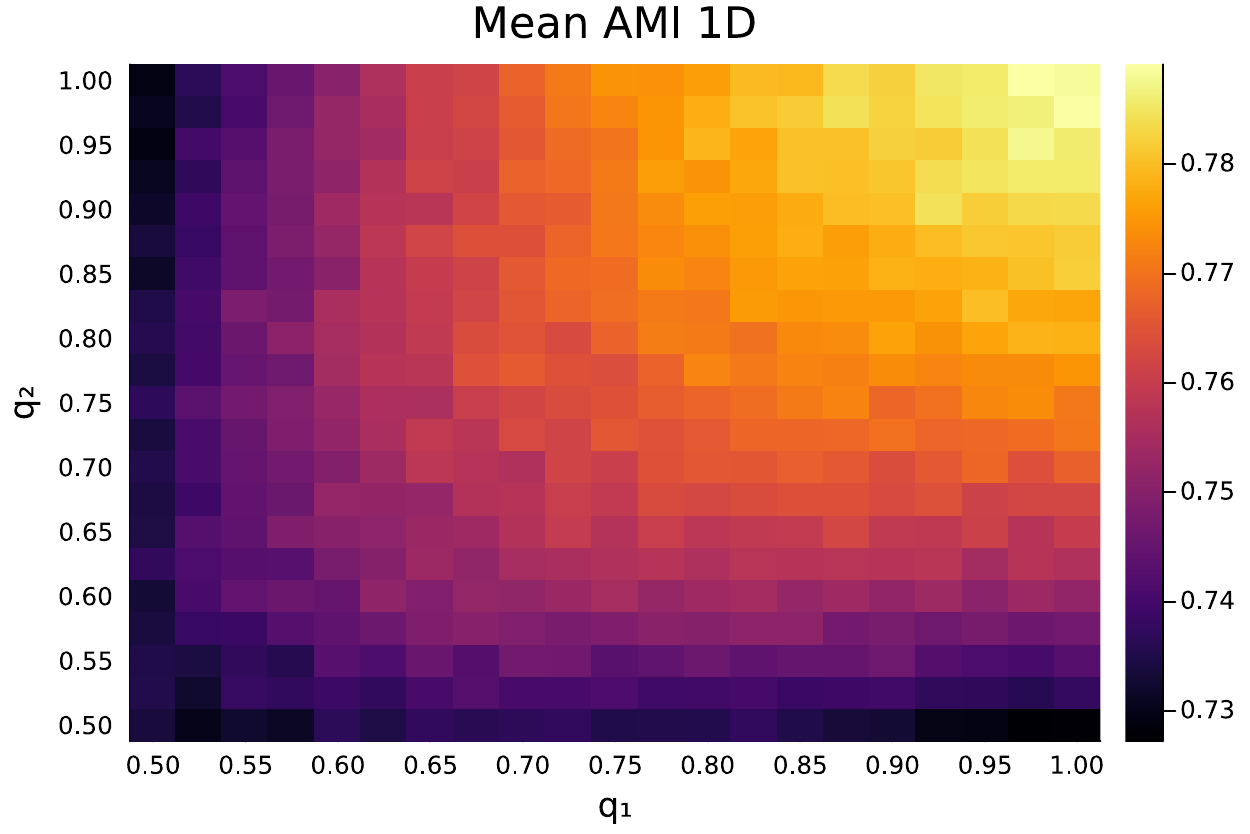}
    \includegraphics[scale=0.38]{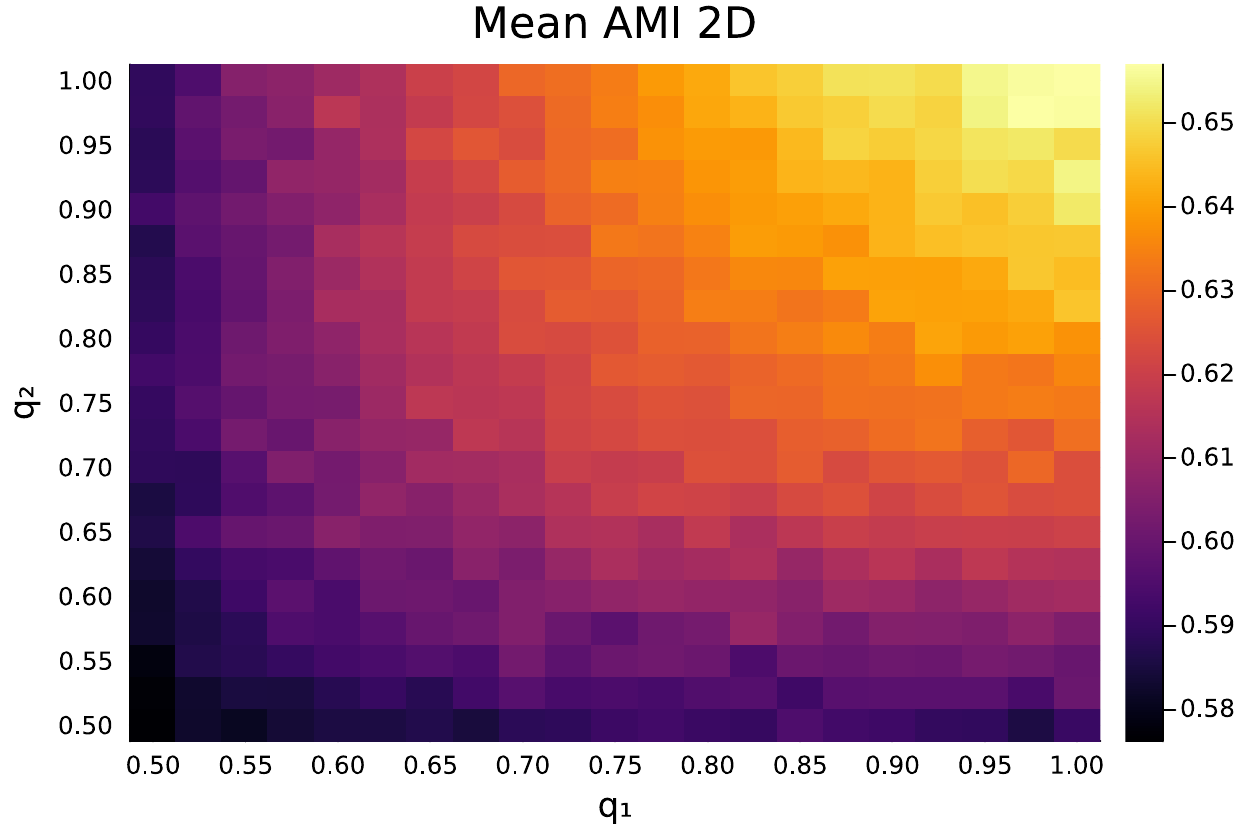}
    \caption{The average value of \textbf{AMI} (over 100 repetitions) between two partitions generated with $q_1$ and, respectively, $q_2$: $n=1{,}000$, $s_1=s_2=8$, $S_1=S_2=32$, $\beta_1=\beta_2=1.5$; $d$-dimensional reference layer was used: (left) $d=1$, (right) $d=2$. Only common active nodes are kept in both layers.}
    \label{fig:ami_S_heatmap_only_active}
\end{figure}

\begin{figure}[ht!]
    \centering
    \includegraphics[scale=0.38]{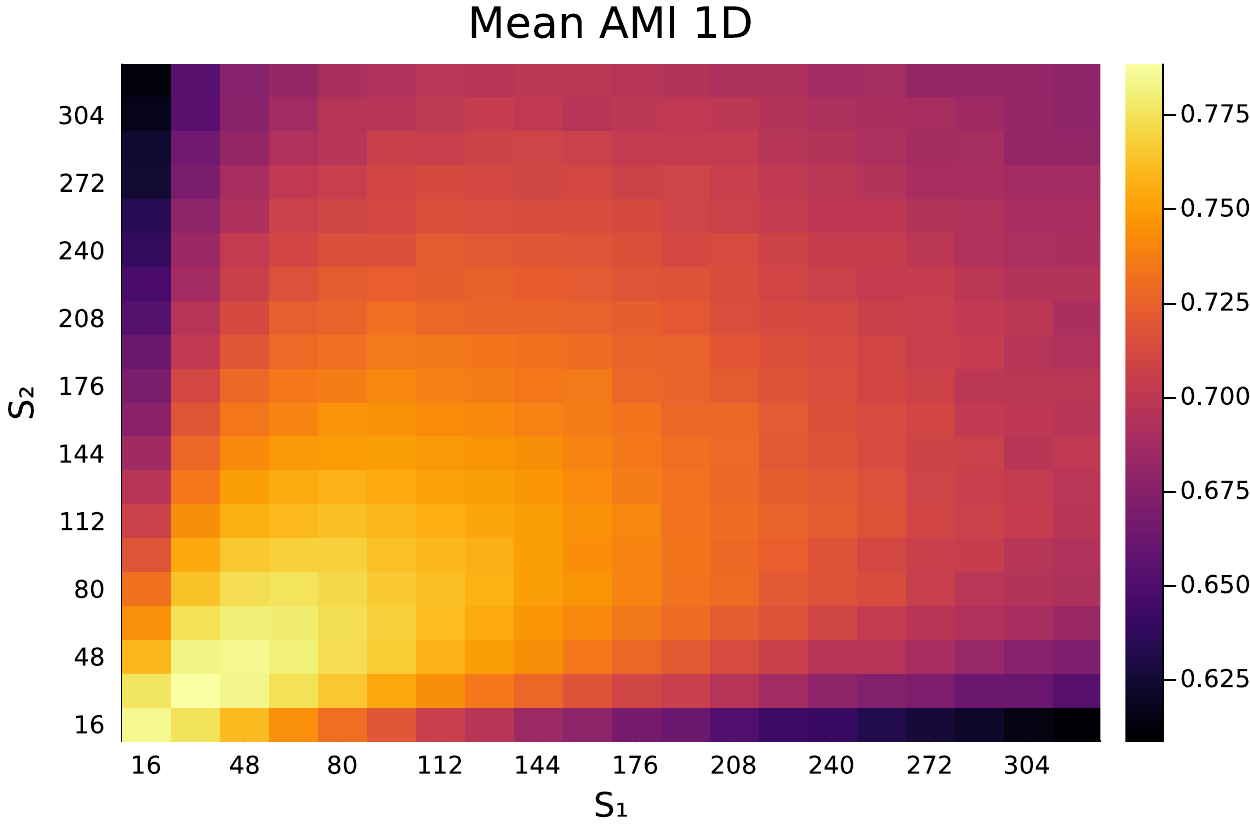}
    \includegraphics[scale=0.38]{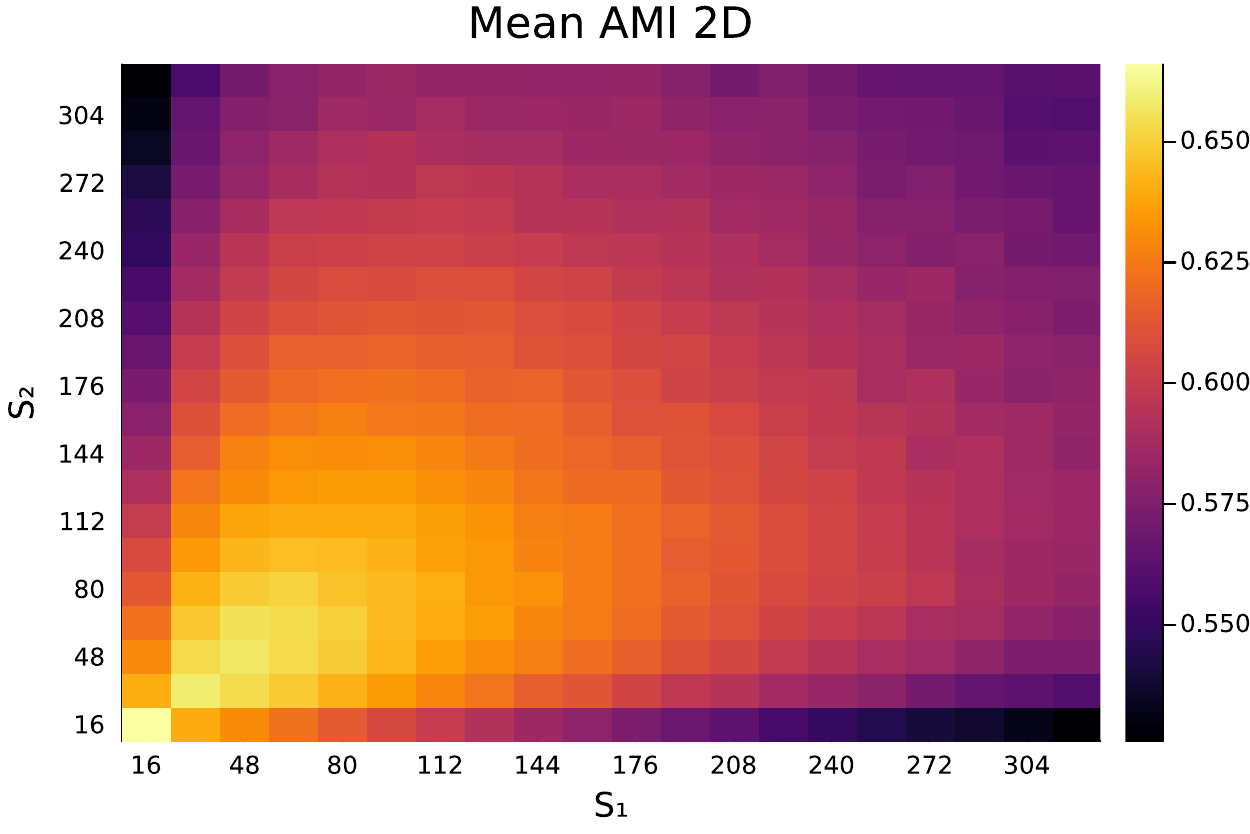}
    \caption{The average value of \textbf{AMI} (over 500 repetitions) between two partitions generated with $S_1$ and, respectively, $S_2$: $n=1{,}000$, $s_1=s_2=8$, $\beta_1=\beta_2=1.5$; $d$-dimensional reference layer was used: (left) $d=1$, (right) $d=2$.}
    \label{fig:ami_S_heatmap}
\end{figure}

We also experiment with the parameter $q$ (fraction of active nodes) and obtain similar conclusions --- see Figure~\ref{fig:ami_S_heatmap_only_active}. It seems that the \textbf{AMI} slightly decreases as $\min \{q_1, q_2\}$ decreases. The \textbf{AMI} is more sensitive with respect to parameter $S$ (the upper bound for community sizes) --- see Figure~\ref{fig:ami_S_heatmap}. As expected, the correlation decreases in highly imbalanced scenarios; one of the partitions has many small communities, whereas the second one has a few large ones.

\begin{figure}[ht]
    \centering
    \includegraphics[scale=0.38]{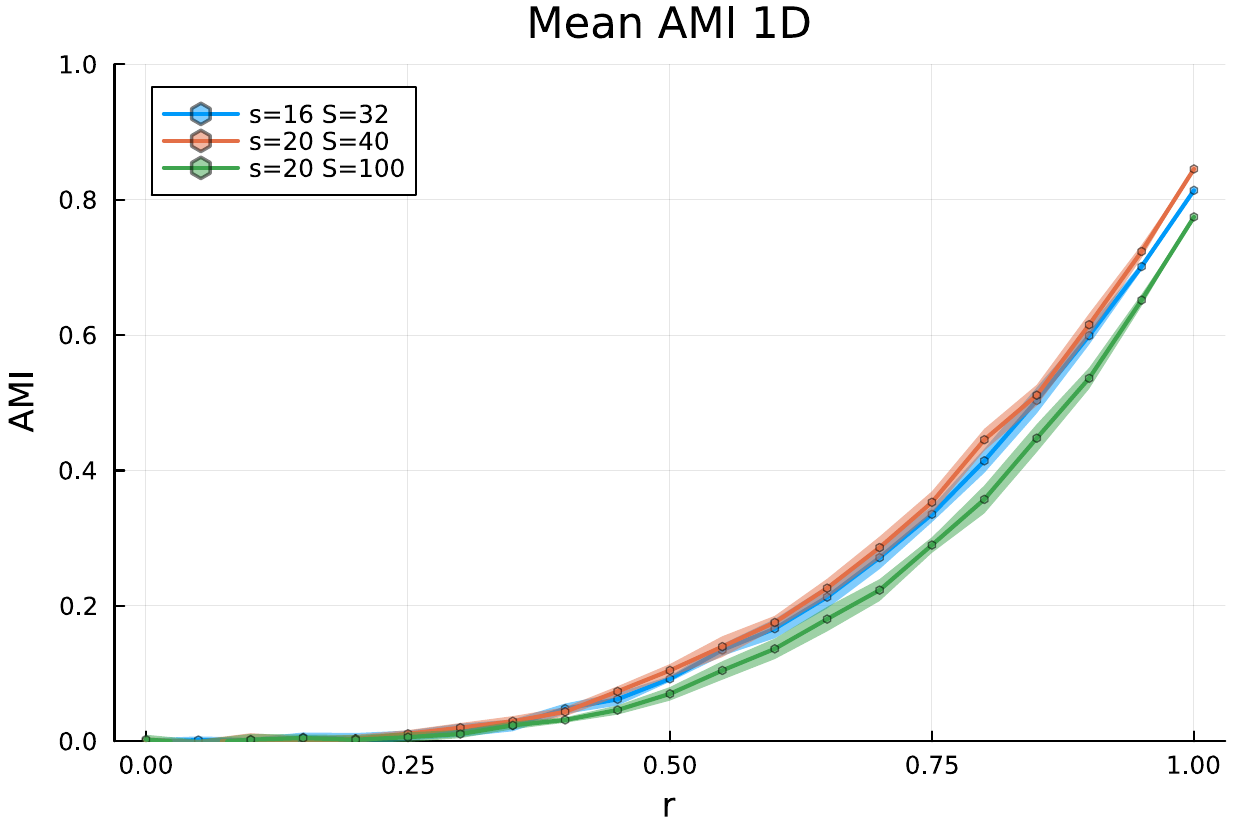}
    \includegraphics[scale=0.38]{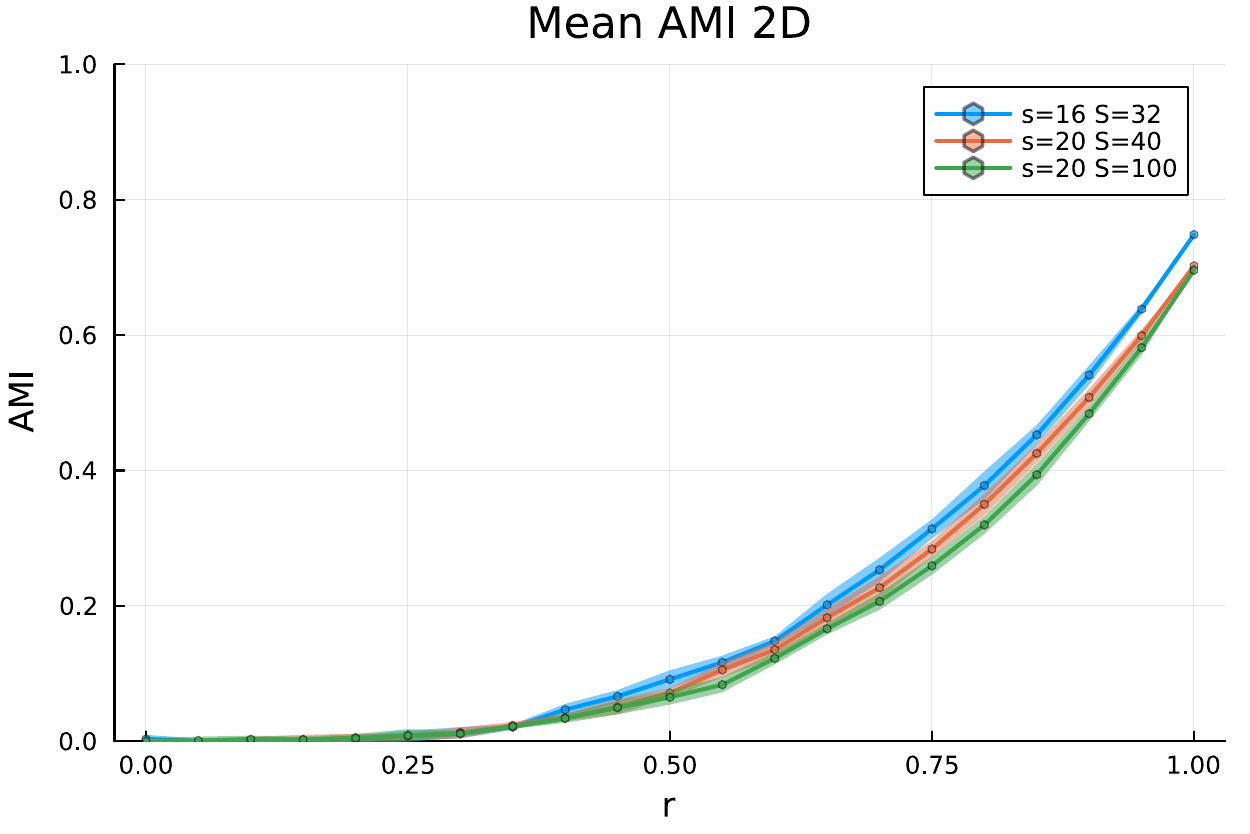}
    \caption{The average value of \textbf{AMI} $\pm$ 1 standard deviation (over 10 repetitions) between two partitions generated with different values of $r$ ($x$-axis): $n=1{,}000$, $\beta_1=\beta_2=1.5$; $d$-dimensional reference layer was used: (left) $d=1$, (right) $d=2$.}
    \label{fig:ami_r_lineplot_compared_to_r}
\end{figure}

In the second experiment, we check how the correlation strength parameter $r$ affects the \textbf{AMI} between the two layers --- see Figure~\ref{fig:ami_r_lineplot_compared_to_r}. It was introduced to provide a smooth transition between the maximum possible correlation (with $r=1$) and completely random partitions (with $r=0$). Our experiments confirm this desired property and show that the \textbf{AMI} is quite stable even for small networks on $n=1{,}000$ nodes.

\begin{figure}[ht!]
    \centering
    \includegraphics[scale=0.38]{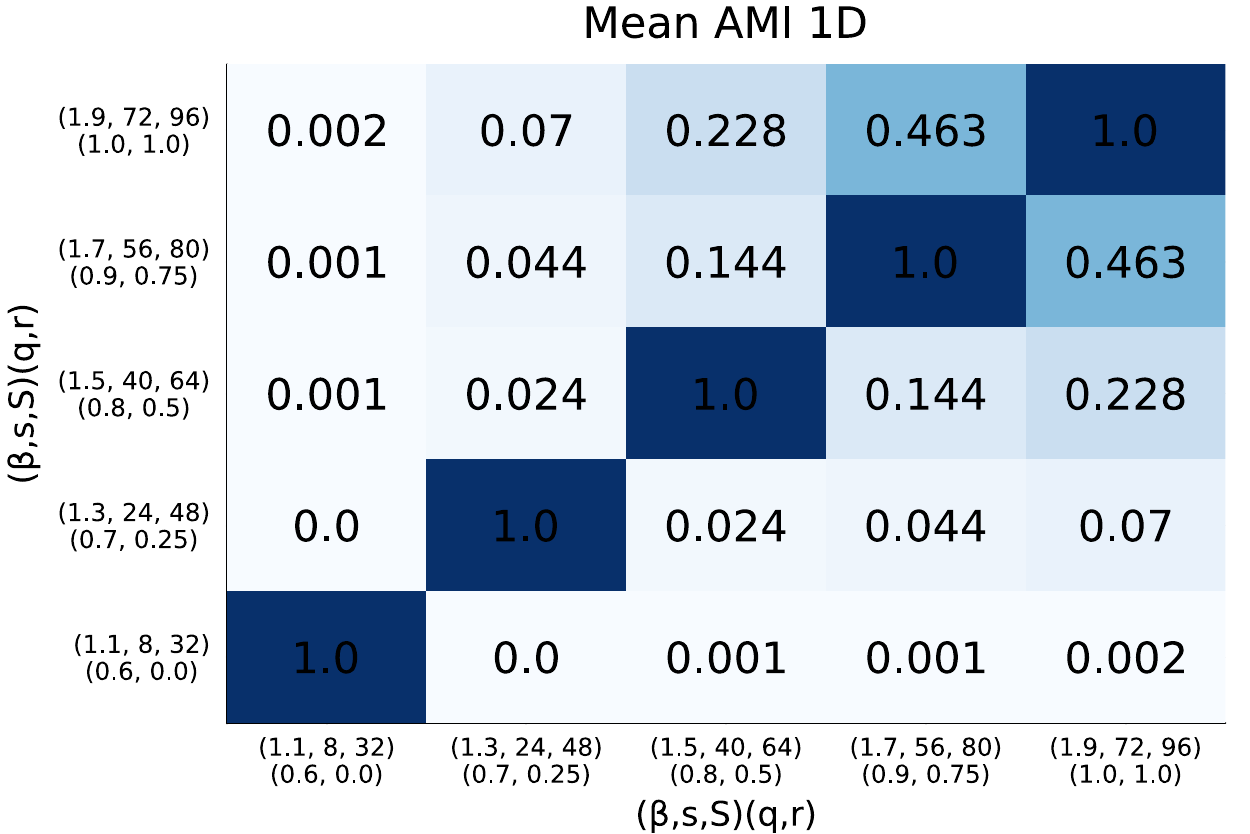}
    \includegraphics[scale=0.38]{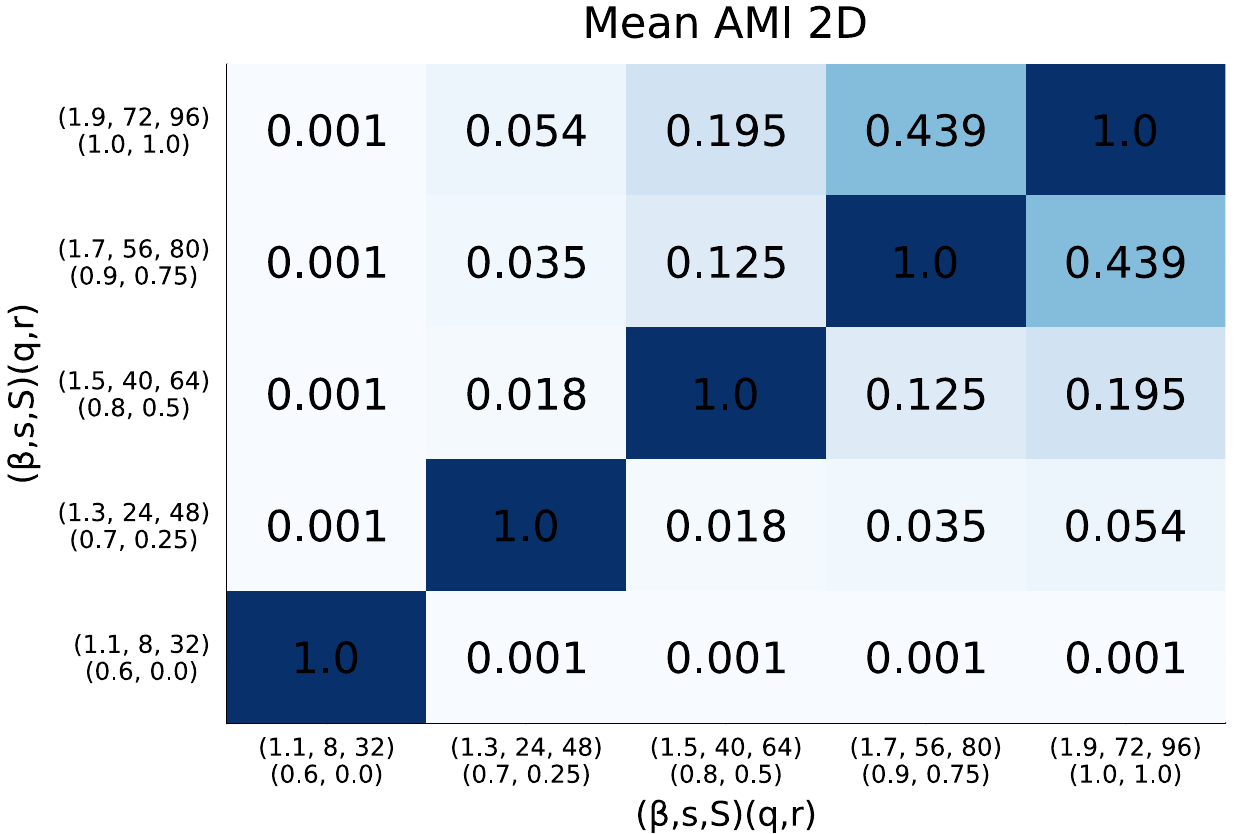}
    \caption{The average \textbf{AMI} value between pairs of partitions and in generated \mABCD\ models as a function of $\beta$, $s$, $S$, $q$, $r$ (over 100 repetitions); $d$-dimensional reference layer was used: (left) $d=1$, (right) $d=2$. Only common active nodes are kept in compared layers.}
    \label{fig:beta_s_S_q_r_heatmap_5_layers}
\end{figure}

In the third experiment, we generate partitions of $n=1{,}000$ actors in 5 layers. As before, not all the actors are active in all layers: $(q_i) = (0.6, 0.7, 0.8, 0.9, 1.0)$. We test different values for the distribution of the community sizes: $(\beta_i) = (1.1, 1.3, 1.5, 1.7, 1.9)$, $(s_i) = (8, 24, 40, 56, 72)$, and $(S_i) = (32, 48, 64, 80, 96)$. More importantly, we test different values of the correlation strength between communities and the reference layer: $(r_i) = (0.0, 0.25, 0.5, 0.75, 1.0)$. The experiment is repeated 100 times, and the experimental means of the \textbf{AMI} between $\binom{5}{2}$ pairs of partitions are reported in Figure~\ref{fig:beta_s_S_q_r_heatmap_5_layers}. As expected, the partition of active actors in the first layer (with $r_1 = 0.0$) is not correlated with any other partition. Moreover, the values of \textbf{AMI} increase together with the corresponding correlation strengths.

In the fourth and final experiment in this subsection, we generate sequences of community sizes with different distributions. There are $n=100{,}000$ actors in each of the three layers, all of them being active ($q=1$). The minimum and the maximum community size is fixed to be $s=10$ and $S=1000$ across the three layers, but the power-law exponents vary: $(\beta_i) = (1.2, 1.5, 1.8)$. We show that the experimental community sizes distributions are very close to the desired ones --- see Figure~\ref{fig:powerlaws} (Right) for the cumulative community sizes distributions of the three layers.

\subsection{Correlations between edges in various layers}

In this subsection, we investigate the performance of Phase~6 of the process generating \mABCD. Recall that in this phase, in each of the $t$ batches, we carefully rewire a random fraction of edges in one of the layers with the goal to bring the empirical correlation matrix $\hat{\Rr}$ closer to the desired matrix $\Rr$. However, because of the very rich dependence structure between layers (associated with various objects: degree sequences, partitions into communities) and the fact that some nodes are inactive, it is not clear that the desired correlations can be achieved.

\begin{figure}[!htbp]
    \centering
    \includegraphics[scale=0.35]{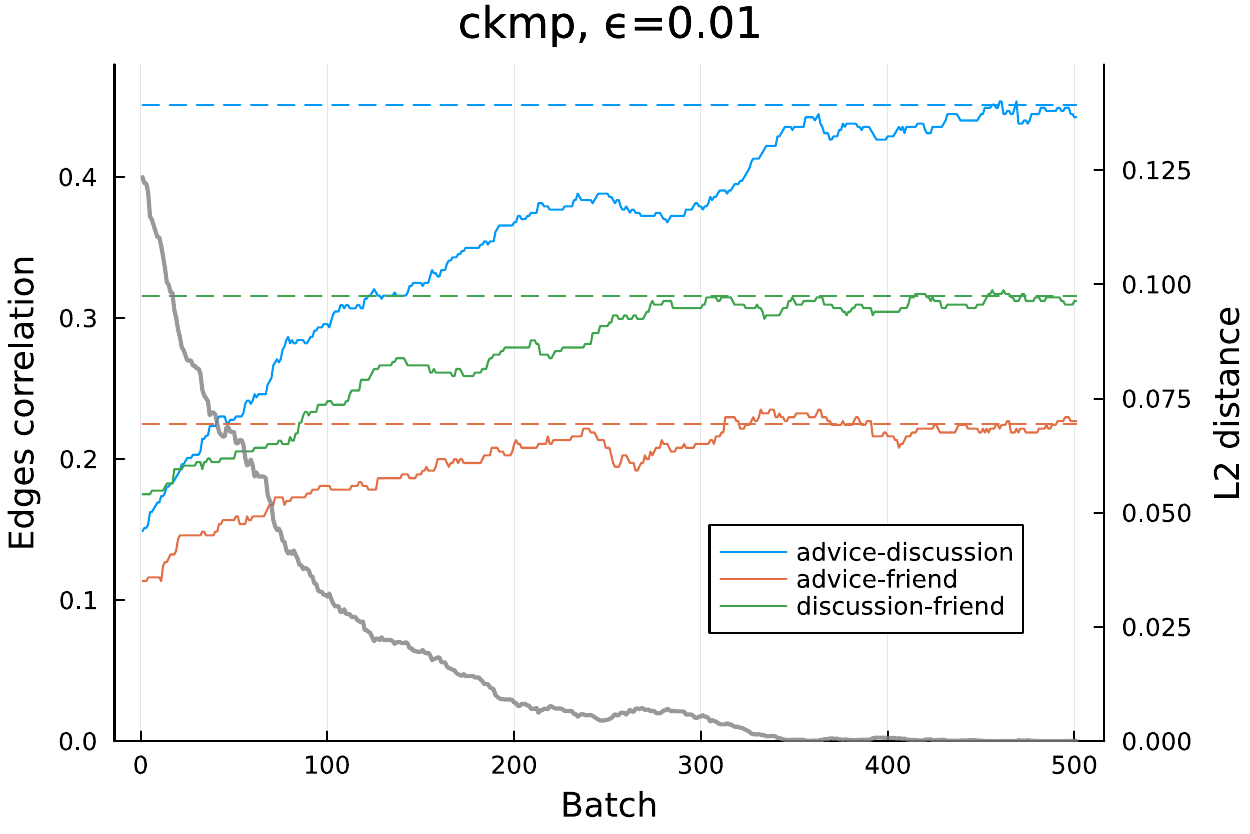}
    \includegraphics[scale=0.35]{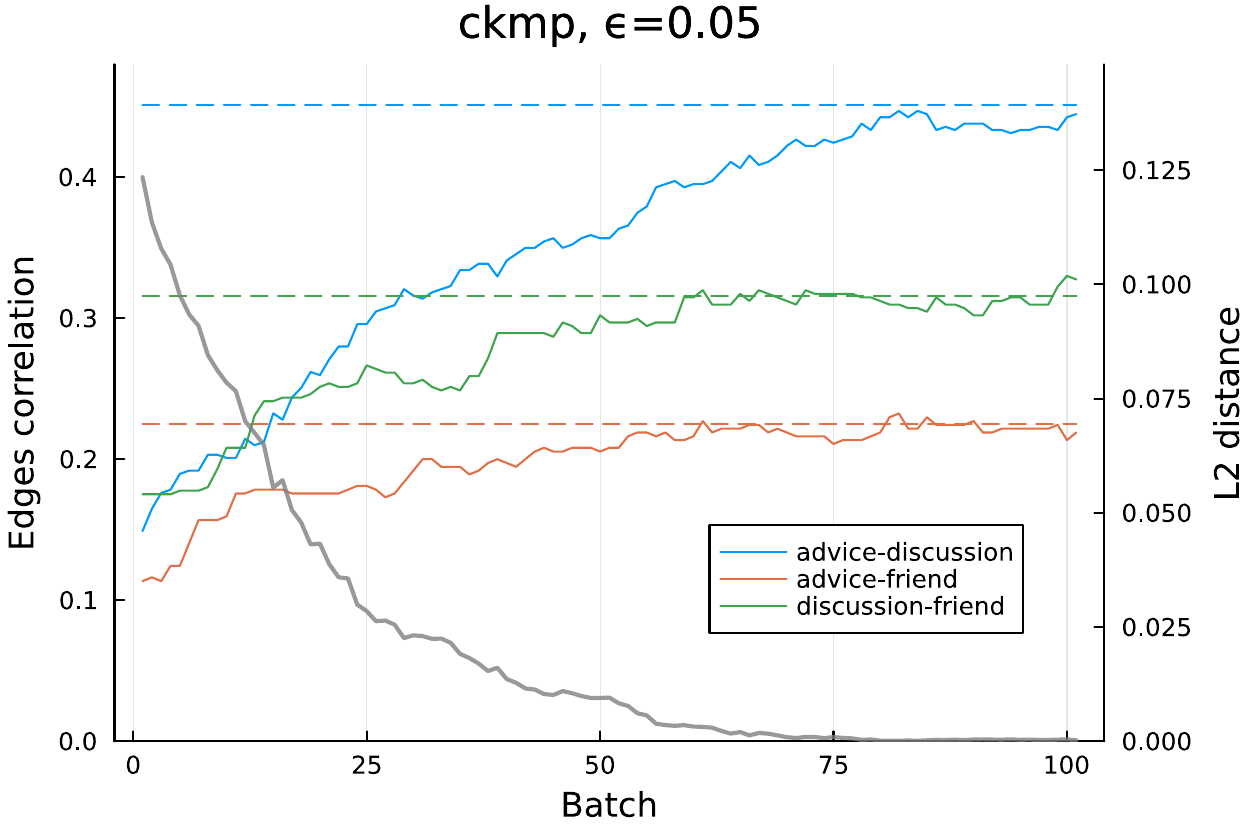}
    \includegraphics[scale=0.35]{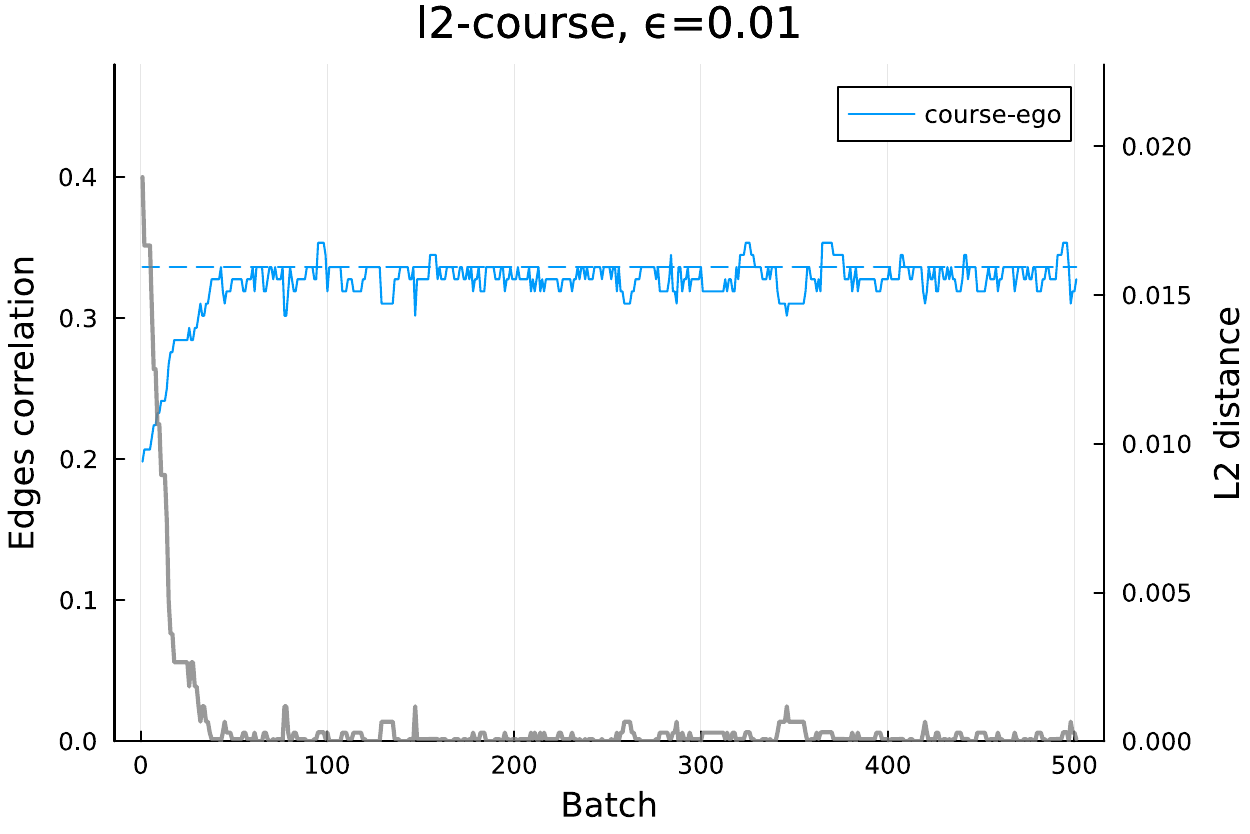}
    \includegraphics[scale=0.35]{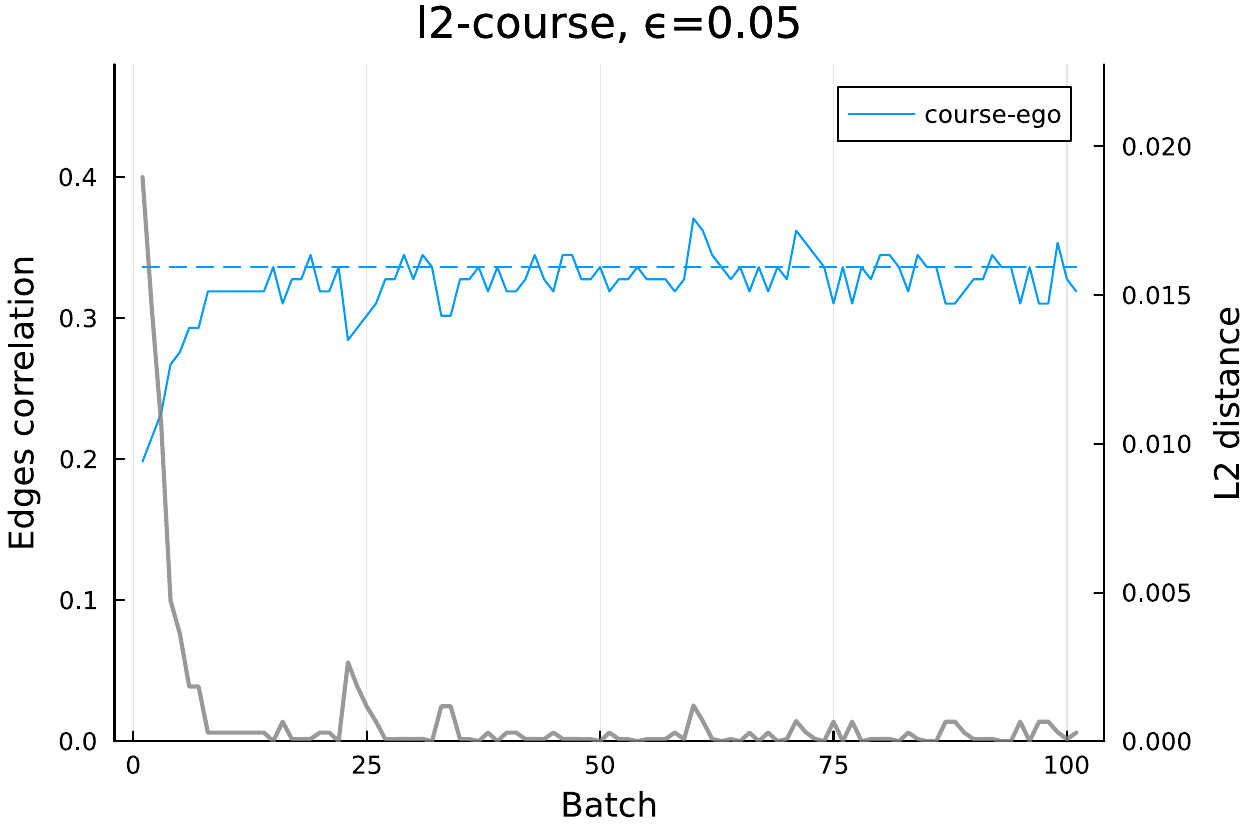}
    \includegraphics[scale=0.35]{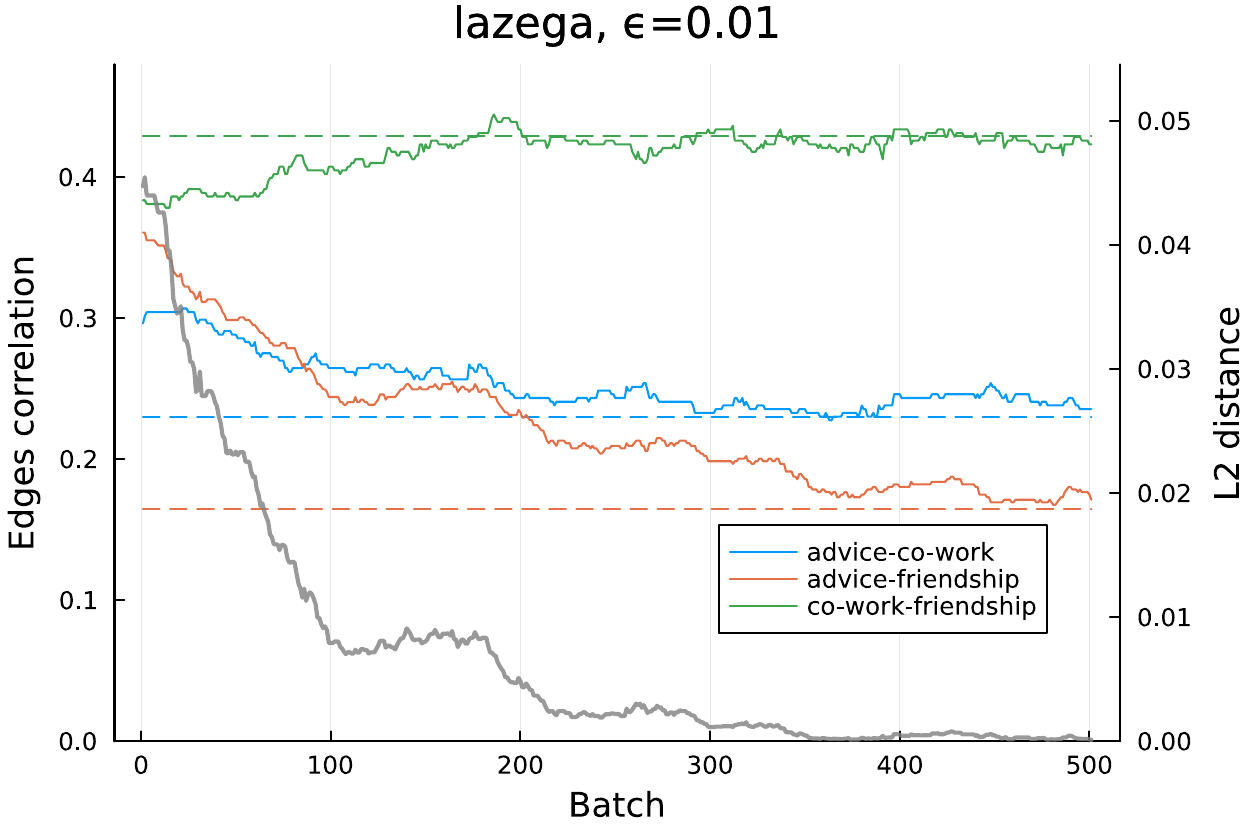}
    \includegraphics[scale=0.35]{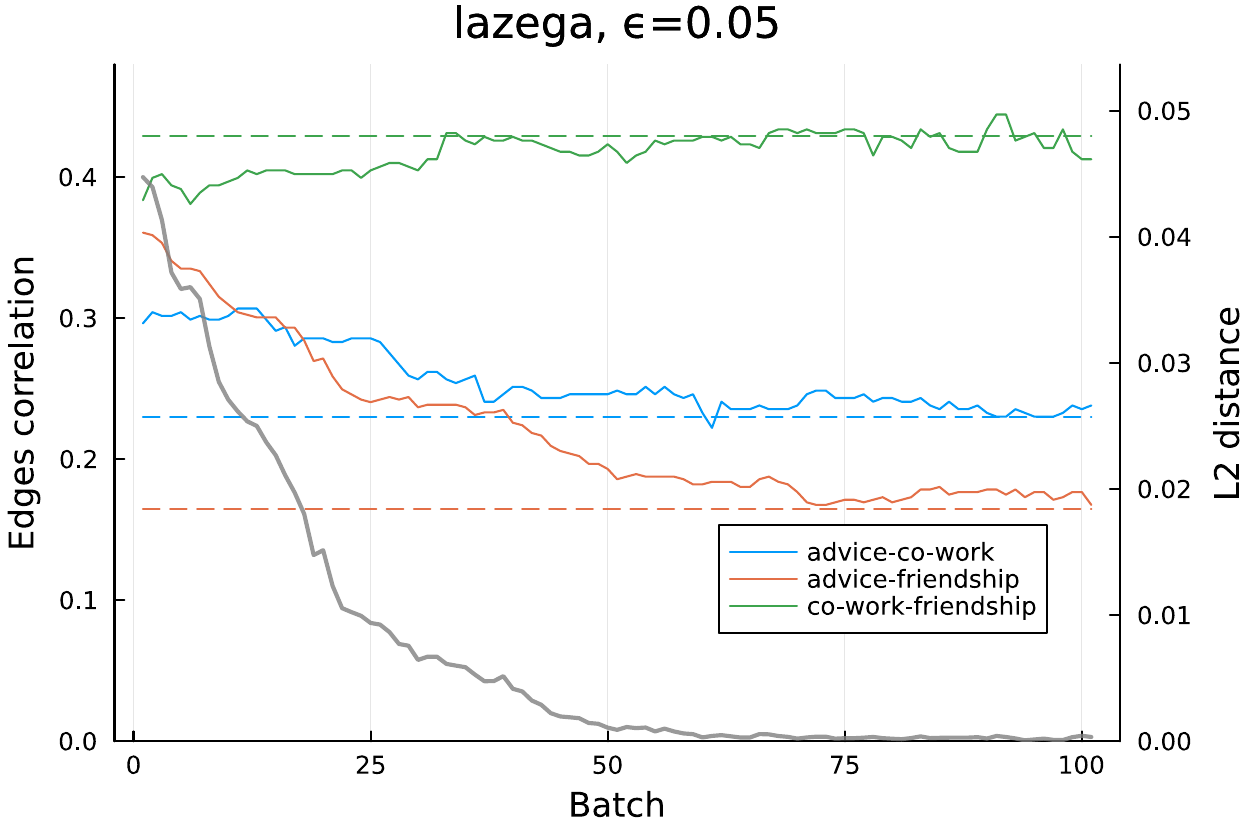}
    \includegraphics[scale=0.35]{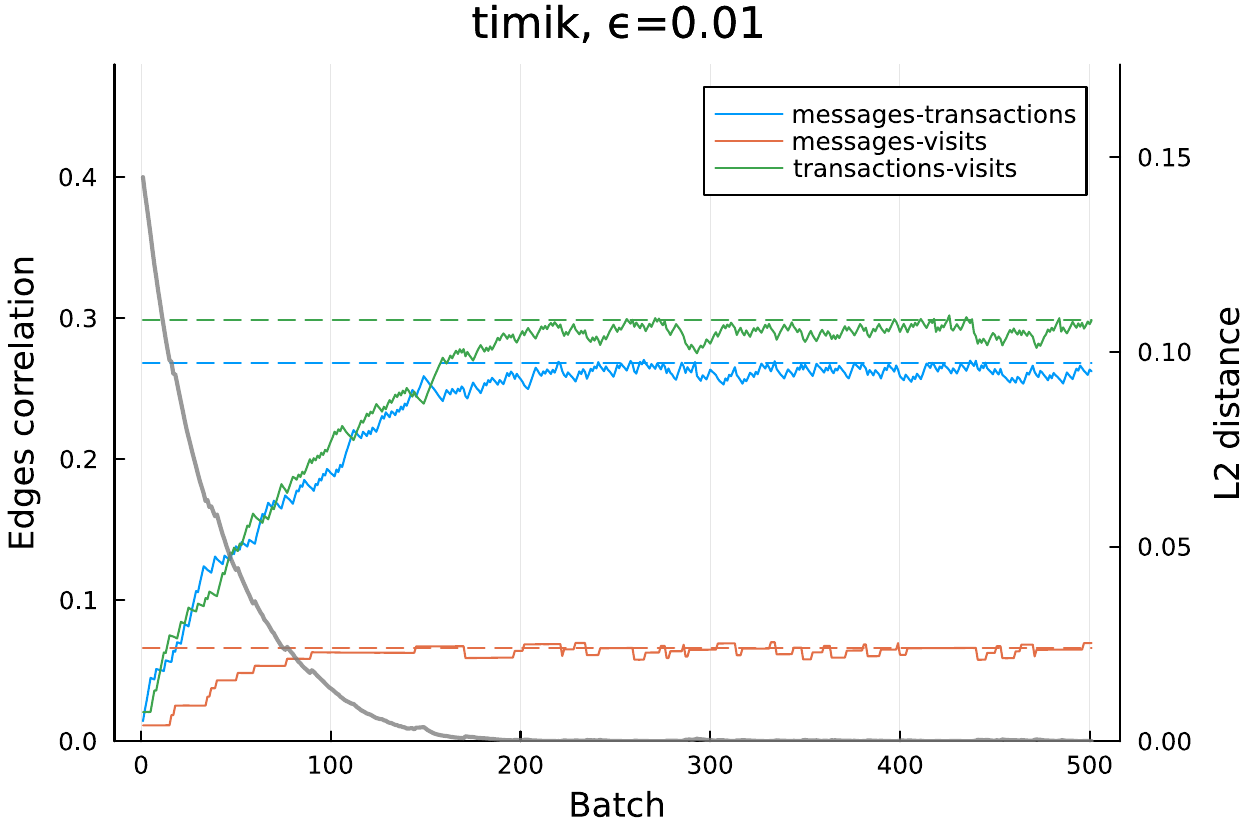}
    \includegraphics[scale=0.35]{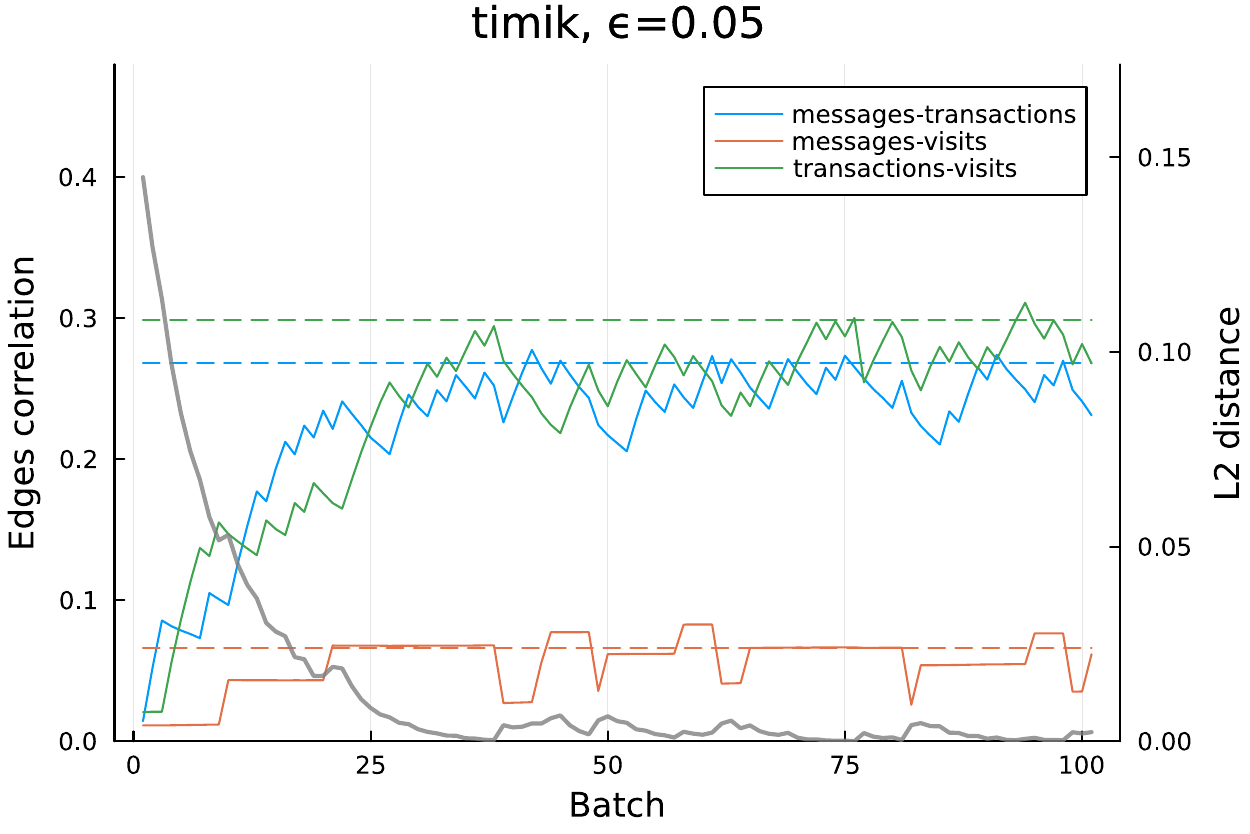}
    \caption{Changes in edge correlation between layers (solid) compared to the desired correlation (dashed). Grey thick line represents the $L_2$ distance between the desired and the empirical correlation matrices ($\Vert \Rr - \hat{\Rr} \Vert_2$).}
    \label{fig:real_graphs_edges_cor_convergence}
\end{figure}

We selected 4 real-world networks (ckmp, l2-course, lazega, and timik), extracted their properties (such as degree sequences, community sizes sequences, noise ratio, correlations; see Section~\ref{sec:parameters_extraction} for more details on how it can be done), and created corresponding \mABCD\ networks mimicking these real-world networks. To provide a good fit to real-world networks we injected exact degree sequences (and therefore active nodes) into the algorithm. As the ground-truth community structure is not known, we used Louvain algorithm to generate community sizes sequence, estimated power law exponent for each layer, and used extrema of community sizes as input to synthetic graph. The convergence to the desired correlation matrix $\Rr$ between edges in various layers is presented in Figure~\ref{fig:real_graphs_edges_cor_convergence}. The default parameters for the number of batches ($t=100$) and the fraction of edges rewired ($\epsilon=0.05$) were tested (right column) but we also checked other set of parameters with the same total number of rewires, namely, $t=500$ and $\epsilon=0.01$.

There is no visible difference between the two scenarios which justifies the default parametrization: the smaller number of batches yields a faster algorithm as the algorithm needs to recompute the empirical correlation matrix $\hat{\Rr}$ at the beginning of each batch. In general, the process converges quite quickly: the $L_2$ norm between $\hat{\Rr}$ and $\Rr$ decreases rapidly. However, there are some fluctuations and the final network is not necessarily the best one. This, again, justifies the design of the algorithm that keeps the best network generated during the process.

\revision{\subsection{Community detection in \mABCD\ networks}

As noted in Section \ref{sec:intro}, the \textbf{mABCD} model supports the validation and assessment of community detection algorithms. Its generation of ground-truth communities enables direct comparison with partitions produced by tested methods. As an extension of \textbf{ABCD}, \textbf{mABCD} incorporates findings from prior analyzes of monoplex community detection algorithms \parencite{kaminski2021artificial,kaminski2022modularity}. An open question remains regarding its utility for validating multiplex partitioning algorithms \parencite{magnani2021community}. Although relevant, this issue lies beyond the scope of this study. We therefore conduct a simple validation experiment using one multiplex partitioning algorithm and the monoplex counterpart, leaving broader analysis for future work.

We generated a network with 1{,}000 actors and three layers. Configuration of layers was identical ($q=1, \tau=0.0, \gamma=2.5, \delta=2, \Delta=25, \beta=1.5, s=16, S=64$) except for the community correlation coefficient $r$ ($r_1=1.0, r_2=0.5, r_3=0.0$). The edges correlation matrix $\Rr$ was not defined, hence no edges rewiring occurred. The noise ratio $\xi$ varied from 0.1 to 0.9 to simulate increasing inter-community connectivity. Due to the stochastic nature of \textbf{mABCD}, ten graphs were produced for each $\xi$. Community detection was performed using the Generalized Louvain algorithm \parencite{Mucha_2010} on the whole multiplex network and the base Louvain algorithm \parencite{blondel2008fast} on each layer independently. The resulting partitions were compared with the ground truth from \textbf{mABCD} via the AMI metric. The experiment aimed to examine how the quality of group detection depends on two parameters: the fraction of inter-community edges within layers ($\xi$) and the community alignment across layers ($r$). The Generalized Louvain algorithm is capable of detecting \emph{pillar communities} \parencite{magnani2021community}. One can think about pillar communities as groups comprising the same nodes that consistently emerge across the network layers. Therefore, \textbf{mABCD} communities in layers with high $r$ (correlated with latent layer, which implies that also between themselves) should create higher-level structures possible to detect with multiplex algorithm. Generalized Louvain highlights pillar communities by producing consistent clusters' labels for nodes in all layers. For reporting purposes, we compare the clustering for each layer separately (Figure \ref{fig:glouvain}).

\begin{figure}[ht!]
    \centering
    \includegraphics[scale=0.38]{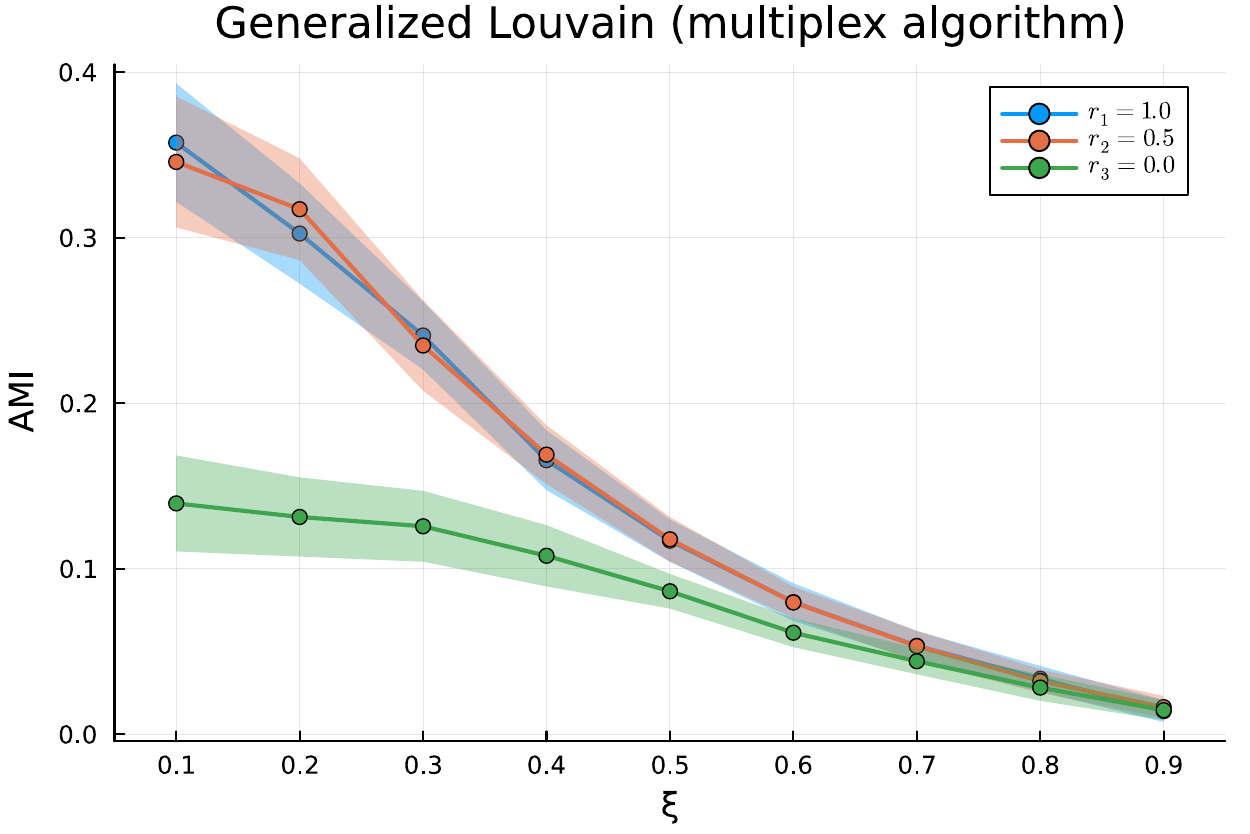}
    \includegraphics[scale=0.38]{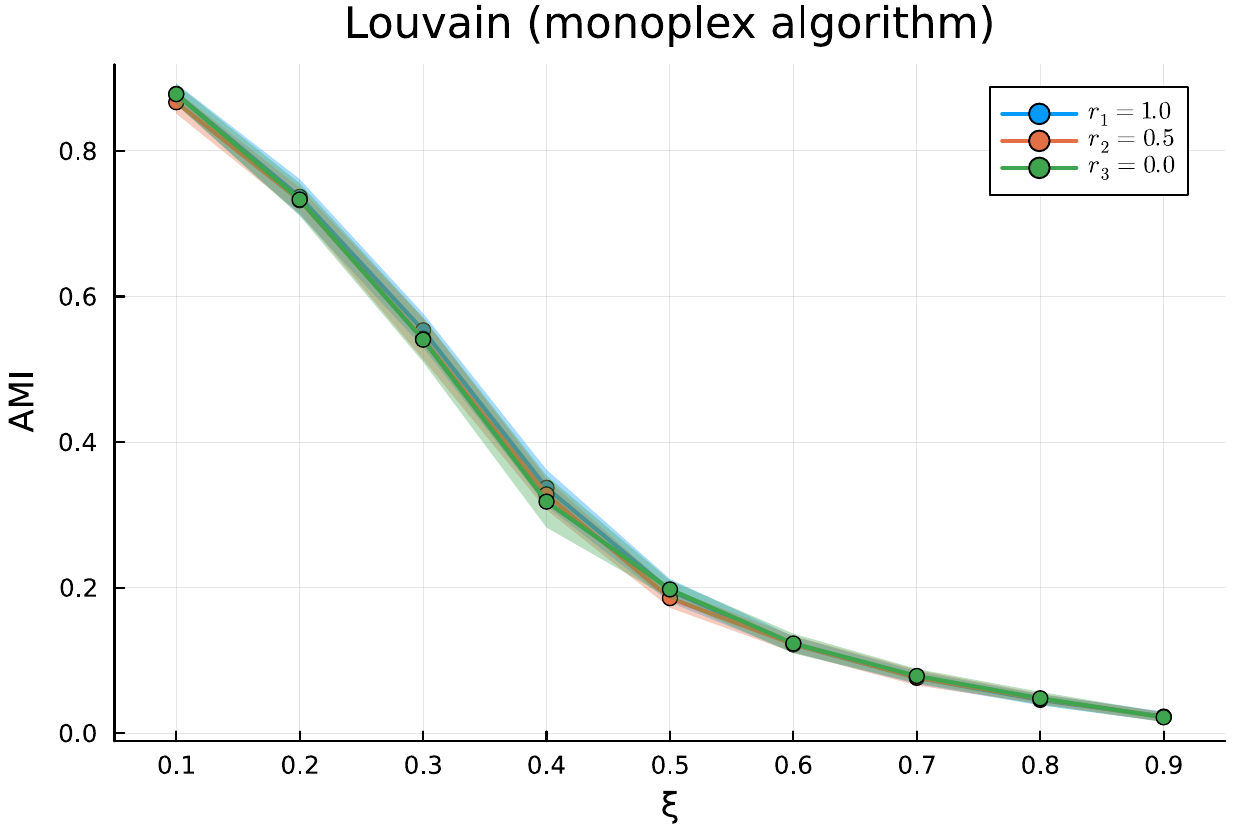}
    \caption{\revision{The average ($\pm$ 1 std) Adjusted Mutual Information between the ground truth and a) Generalized Louvain's communities (left), b) Louvain's communities (right); 50 repetitions for each $\xi$. (Note that for Louvain's communities all three lines overlap; for Generalized Louvain lines for $r_1=1.0$ and $r_2=0.5$ are almost overlapped.)}} 
    \label{fig:glouvain}
\end{figure}

Aligned with expectations, the results show a decline in clustering quality with increasing noise. Clustering within each layer converges to random partitioning for high $\xi$ values. Notably, Generalized Louvain exhibits curious tendency related to the $r$ parameter. The method performed equally well for perfectly correlated ($r_1=1.0$) and moderately correlated ($r_2=0.5$) layers (the reason is that the algorithm does not see the latent layer, so both layers are indistinguishable from the perspective of community detection, since the third layer is uncorrelated). Performance for the uncorrelated layer is significantly lower, which is expected as the assignment of actors to communities is randomly shuffled, so the signal coming from it to the Generalized Louvain algorithm is weaker. Captured similarities may indicate the ability of Generalized Louvain to detect intra-layer patterns that affect the structure of communities. In contrast, the performance of the Louvain algorithm is much better than the multiplex extension. The monoplex variant is insensitive to the value of $r$, which is aligned with \textbf{mABCD} design. The $r$ parameter controls correlation between layers, but does not affect strength of communities within layers --- this property is controlled by the $\xi$ parameter.}

\subsection{Computational efficiency of the model}

One of the reasons why the original \textbf{ABCD} model was developed was the fact (pointed out by the industry partners we worked with back then) that other synthetic random graph models with community structure \revision{do not perform well in terms of computational efficiency}. Keeping this concern in mind, the design of the \textbf{ABCD} model aims to generate networks fast. Indeed, after careful preprocessing (that is very fast), \textbf{ABCD} is essentially a union of independent copies of the configuration model. In two minutes, on a typical laptop, one can generate networks on $n = 1.5 \cdot 2^{21}$ nodes. The \mABCD\ model is clearly much more complicated than the original \textbf{ABCD}, but it builds on fast ingredients of \textbf{ABCD}. \revision{Therefore, in order to validate whether it preserves computational advantages inherited from its monoplex counterpart, we performed experiments aimed to address this problem. The data reported below was obtained during simulations executed on Macbook Air M1 (2020) with macOS v15.5.}

\begin{figure}[ht!]
    \centering
    \includegraphics[scale=0.4]{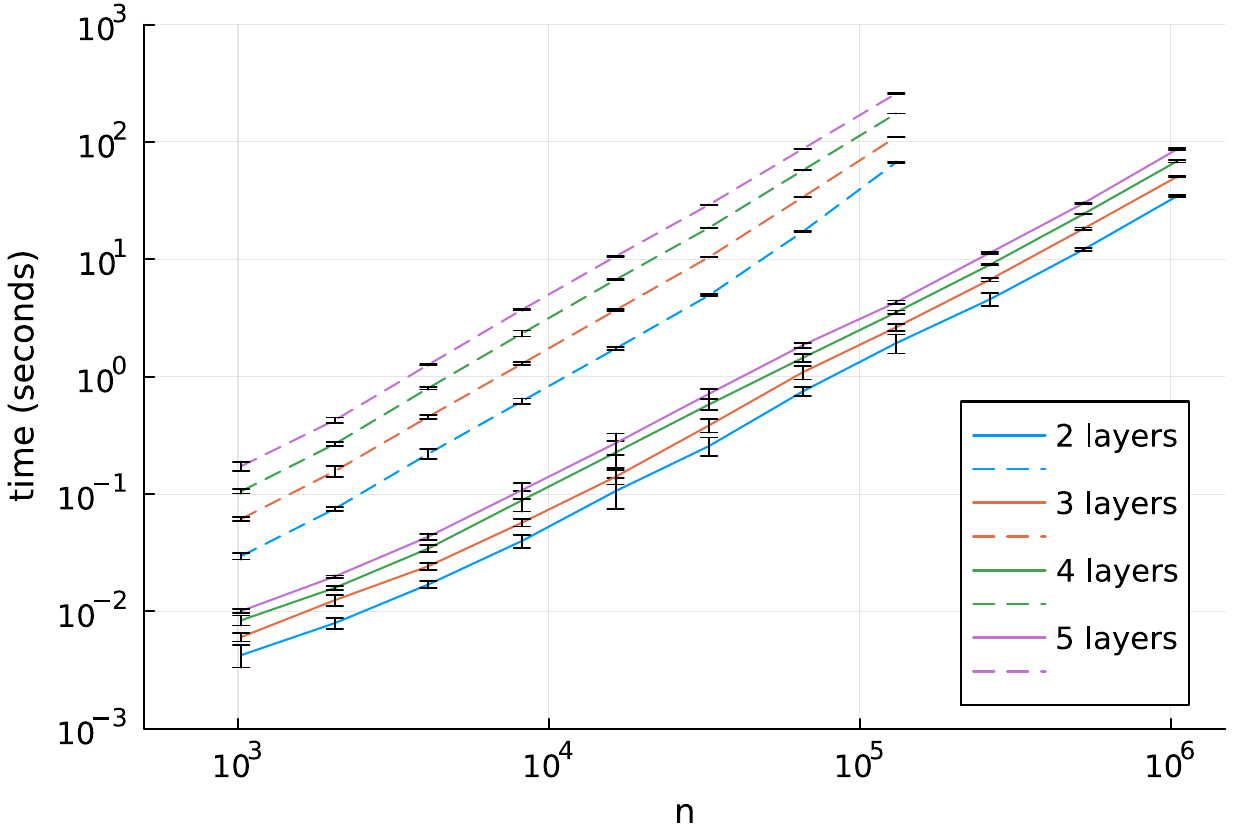}
    \caption{The average ($\pm$ 1 std) execution time of \mABCD\ (in log-log scale) with Phase~6 (dashes) and without it (solid) for $n = 2^k$ nodes, $k\in \{10, 11, \ldots,20\}$, and $\ell \in \{2,3,4,5\}$ layers; 10 repetitions for each combination.} 
    \label{fig:execution_time}
\end{figure}

\begin{figure}[ht!]
    \centering
    \includegraphics[scale=0.3]{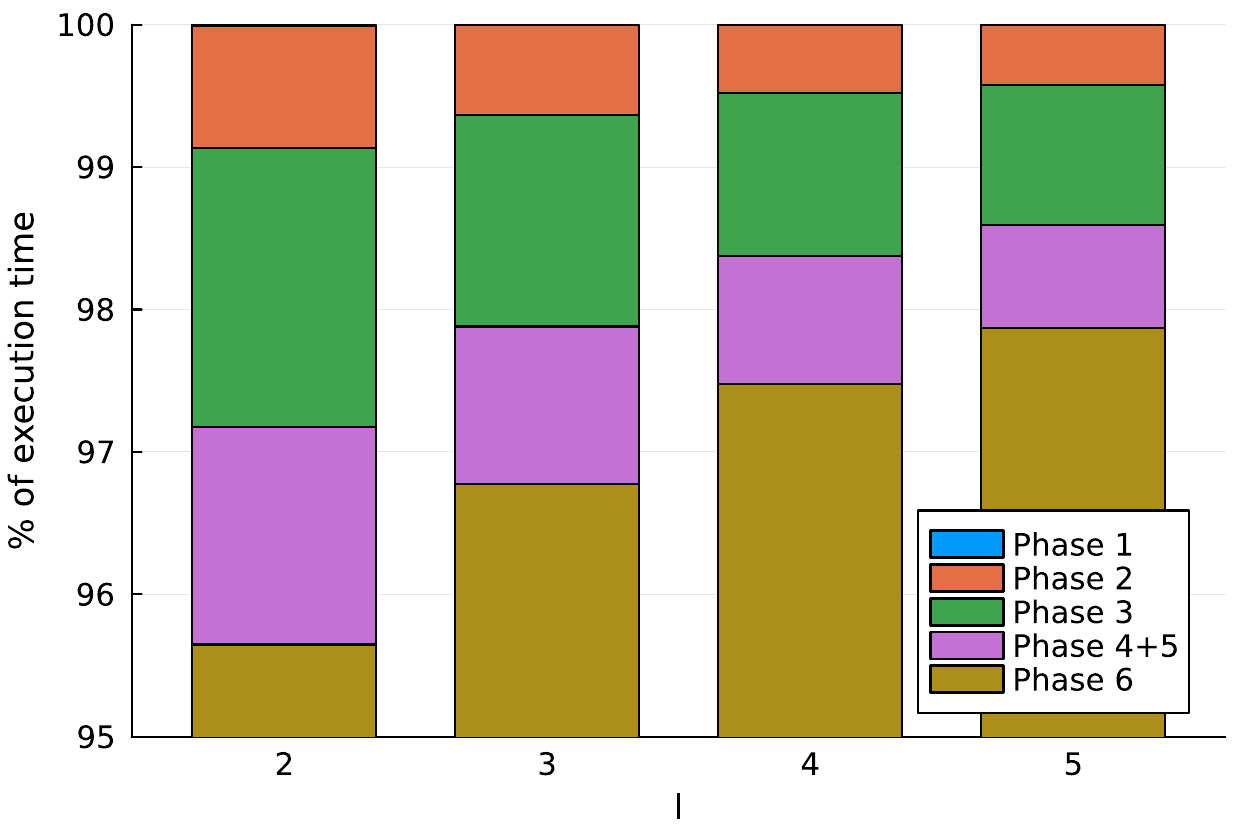}
    \includegraphics[scale=0.3]{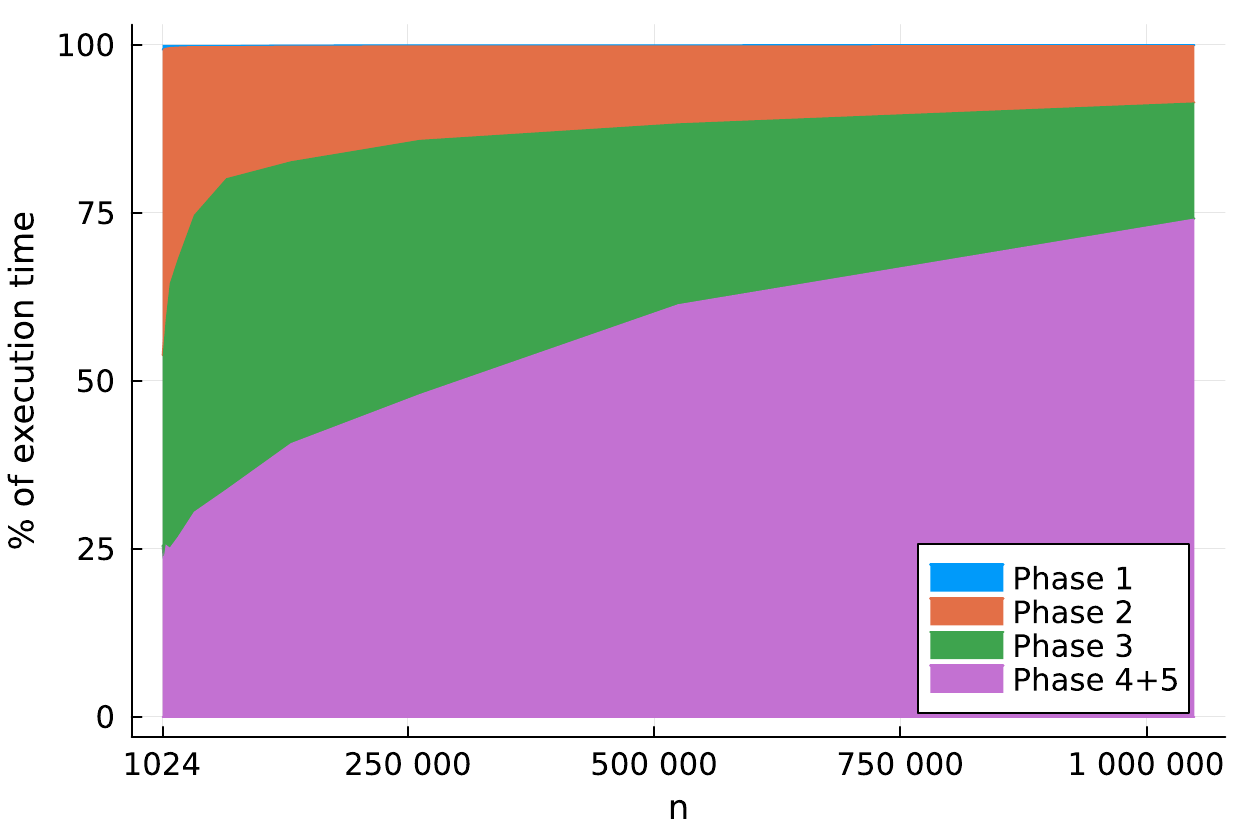}
    \caption{Percentage breakdown of the execution time between phases of the algorithm: with Phase~6 included $n=2^{16}$ (Left) and with Phase~6 excluded, $\ell=5$ (Right).}
    \label{fig:execution_time_breakdown}
\end{figure}

Not surprisingly, Phase 6 (switching edges to get the desired correlations between edges in various layers) is the most time-consuming phase. Moreover, it is clear that the more layers, the longer the process is. In two minutes, one can generate \mABCD\ networks on $n=2^{16} = 65{,}536$ nodes. However, without Phase~6, in the same time one can produce much larger networks on $n=2^{20} = 1{,}048{,}576$ nodes --- see Figure~\ref{fig:execution_time}. The percentage breakdown is presented in Figure~\ref{fig:execution_time_breakdown}.

\begin{figure}[ht!]
    \centering
    \includegraphics[scale=0.4]{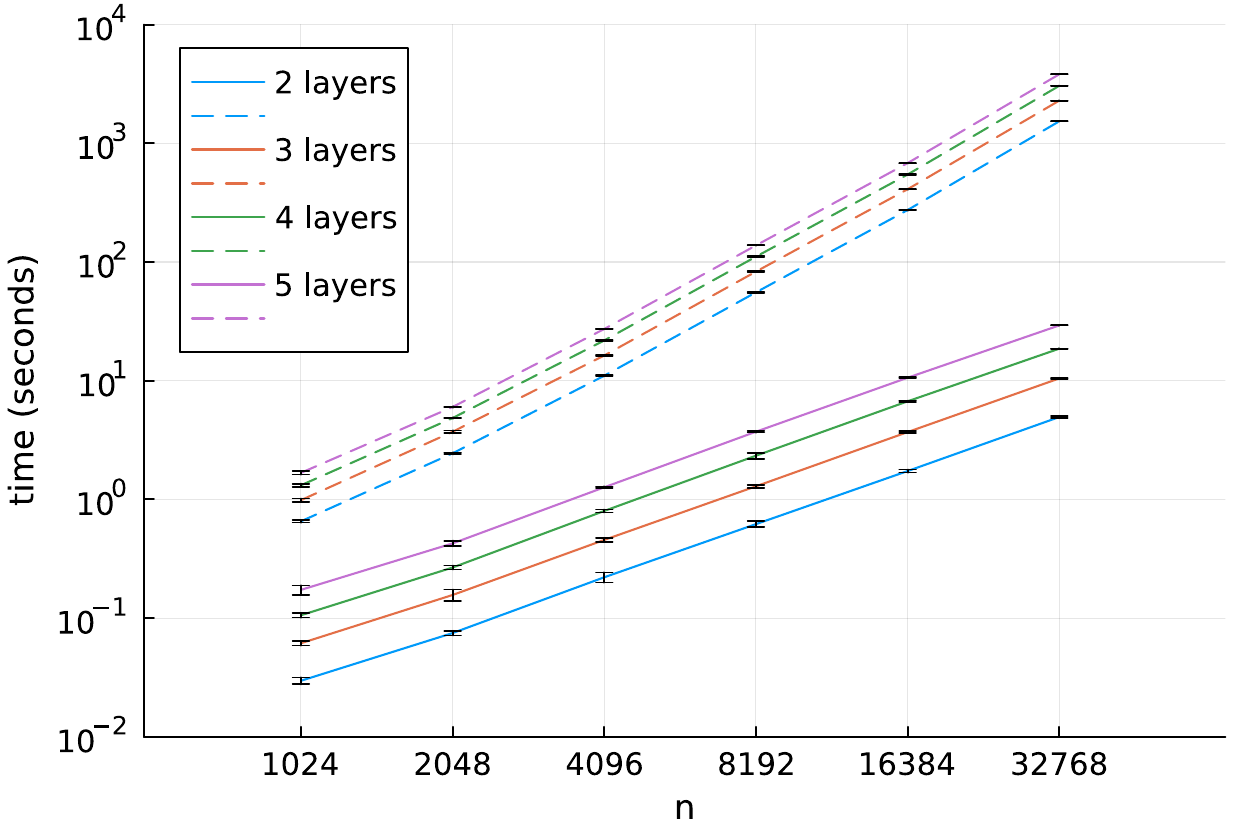}
    \caption{The average ($\pm$ 1 std) execution time (in log-log scale) for \textbf{multilayerGM} framework (dashes) and \mABCD\ model (solid) for $n = 2^k$ nodes, $k\in \{10, 11,...,15\}$, and $\ell \in \{2,3,4,5\}$ layers; 10 repetitions for each combination.} 
    \label{fig:execution_time_multilayerGM_comparison}
\end{figure}

Finally, we have compared the speed of \mABCD\ model to the \textbf{multilayerGM} framework \revision{proposed by~\cite{bazzi2020framework} (see Section~\ref{subsec:literature} for more details)}. The results are presented in Figure~\ref{fig:execution_time_multilayerGM_comparison} and Table~\ref{tab:multilayergm_mABCD_time}, and indicate that for the same setup, \mABCD\ is roughly 10 times faster for small networks (1{,}024 nodes), and the difference in speed quickly increases with the network size (for a two-layer network with 32{,}768 nodes \mABCD\ is 310 times faster). Since generating one five-layer network with 32{,}768 nodes took more than one hour for \textbf{multilayerGM}, and we repeated experiments ten times for each combination of node and layer, we stopped our comparison there.

\begin{table}[ht!]
\centering
\caption{The average execution time (in seconds) of \textbf{multilayerGM} (first) and \mABCD\ (second) algorithms for varying number of nodes $n$ and number of layers $\ell$; ratio of execution times in brackets.}
\begin{tabular}{r|r|r|r|r}
$n$     & $\ell$=2               & $\ell$=3               & $\ell$=4              & $\ell$=5               \\ \hline \hline
1,024  & 0.65/0.03 (22)    & 0.98/0.06 (16)    & 1.31/0.11 (12)   & 1.67/0.17 (10)    \\ \hline
2,048  & 2.44/0.07 (33)    & 3.71/0.16 (24)    & 4.84/0.27 (18)   & 6.00/0.43 (14)       \\ \hline
4,096  & 11.01/0.22 (50)   & 16.30/0.46 (36)    & 21.83/0.80 (27)   & 27.27/1.26 (22)   \\ \hline
8,192  & 55.46/0.62 (90)   & 82.99/1.29 (64)   & 110.85/2.33 (48) & 137.93/3.72 (37)  \\ \hline
16,384 & 274.03/1.73 (158) & 410.23/3.69 (111) & 545.92/6.70 (82)  & 681.25/10.59 (64) \\ \hline
32,768 & 1,533.2/4.9 (310) & 2,290.1/10.5 (219) & 3,051.3/18.6 (164)  & 3,817.6/29.2 (131) \\
\end{tabular}
\label{tab:multilayergm_mABCD_time}
\end{table}

\section{Using mABCD for Spreading Phenomena Analysis}\label{sec:spreading}

To demonstrate the applicability of the \mABCD\ synthetic network in relevant fields, we present three use cases from the area of spreading phenomena analysis. These experiments aim to illustrate the usefulness of the introduced framework and to provide guidance for conducting similar experiments. In general, the problem which we aim to tackle is how the topological features of the network affect the spread of influence in multilayer networks. To do so, we are using \mABCD\ to generate various networks where one feature changes while the remaining ones remain unchanged. This allows us to asses how a single feature affects the spreading process. It is worth noting that the problems depicted below can also be regarded as standalone research outlines within a broader context beyond the examples presented here. Before describing the experiments, we first discuss the employed methodology, as they are conducted in a similar manner.

\subsection{Scope and methodology}\label{subsec:spread_methodology}

The experimental part consists of several steps. First, we select a real-world network to serve as a baseline; specifically, we decided to use the \textit{timik} network (see Table~\ref{tab:networks_eda}), due to its large size which helps minimize sampling errors (typical to any graph generator). Then, a set of configuration parameters for this network are estimated. Having done so, we are able to manipulate specific parameters and generate synthetic networks that can be considered derivatives of the original network. It is worth noting that the idea of utilizing a real network stems from the need to bring the experiments closer to real-world conditions. In the context of spreading phenomena, one can imagine that, as the baseline network represents a real-world system, its artificially generated variants using the \mABCD\ model can correspond to possible configurations of such a system, resembling the impact of certain policies imposed on it. For instance, since the employed network (\textit{timik}) represents interactions on an online social platform, its artificial counterpart generated with an adjusted $\Delta_i$ parameter may reflect the effect of a policy that regulates the maximal number of friends allowed. Thus, by comparing spreading on the original network and its derivative, we assess the consequences of counterfactual scenarios that affect the construction of the evaluated system.

To ensure a fair comparison of spreading effectiveness, only graphs generated by \mABCD\ are evaluated. Consequently, instead of simulating spread on the original real-world network, its ``digital twin'' is employed. Based on the estimated configuration parameters (adjusted for minimum degree and minimum community size to remain within the capabilities of the framework), 20 instances of the network are generated using \mABCD. Spread is then simulated on generated networks under a selected range of spreading regimes that produce non-trivial dynamics; that is, the network neither saturates too rapidly nor inhibits the initiation of spreading.

Each experiment involves manipulating a single parameter of the \mABCD\ model, with the aim of evaluating networks that differ from the baseline by only that feature. Similarly to the baseline graph, \emph{we generated 20 multilayer networks for each set of configuration parameters} to mitigate the randomness inherent in \mABCD. Diffusion is then simulated on these networks using the same spreading conditions as for the ``digital twin'' of the real-world network. Since the employed spreading model is also nondeterministic, \emph{each simulation was repeated 30 times} to reduce the influence of random effects. Finally, the diffusion effectiveness observed in the modified and original networks is compared to assess the influence of the parameter under investigation on the spreading process.

As the influence spreading mechanism, we utilize the Independent Cascade Model~\parencite{goldenberg2001icm} (ICM), extended to the multilayer scenario (MICM). This adaptation draws upon a similar rationale to that used in extending the related Linear Threshold Model (LTM), as described by~\cite{zhong2022mltm}. Consistent with their approach, we regard actors as the principal entities in the diffusion process, while their corresponding nodes across various layers serve a supplementary function. In the original ICM, the probability that an active agent will activate its inactive neighbour is controlled by a single parameter, $\dot{\pi}$, with each activation attempt occurring only once. Our modification introduces an additional parameter, $\dot{\delta}$, termed the \emph{protocol function}, which dictates how activation signals from different layers are combined to establish the actor’s overall state. Activation of the actor (and all its layer-specific nodes) occurs only if the protocol condition is fulfilled; if not, the actor remains inactive even if some layer-specific nodes receive positive activation signals. We examine two extreme cases for $\dot{\delta}$: the $AND$ protocol, where activation requires positive signals from \emph{all} layers representing the actor; and the $OR$ protocol, where activation in \emph{any} single layer suffices.

At this point it is worth noting that, in contrast to the ICM, employing its well-known counterpart in the experimental part, the LTM~\parencite{granovetter1978threshold}, can lead to unstable results. This is due to the way \mABCD\ generates graphs, namely, it aims to capture community structure (i.e., global properties) rather than local ones. As a result, graphs generated with \mABCD\ contain groups of nodes that induce denser subgraphs compared to the global density, but locally neighbourhoods still resemble tree structure, as there are very few short cycles. By contrast, real networks do not exhibit this local behaviour; they contain many short cycles, even triangles. Consequently, diffusion in \mABCD-generated networks under the LTM, which relies on a threshold mechanism requiring a certain number of active neighbours to activate a given node, can either quickly die out or abruptly saturate the network, with no parameter range that produces gradual, stable spreading. On the other hand, the mechanism inherent to the ICM is ``immune'' to this tree-like local structure, as it focuses on individual edges rather than the entire neighbourhood of a given node.

Finally, the metric used to compare the effectiveness of spreading was \textit{gain}~\parencite{czuba2024rankrefininginfmaxmln}. It quantifies the total network coverage achieved by initiating the diffusion from a seed set $S_0$. Specifically, it is computed as the proportion of actors activated during the diffusion process, excluding the initial seeds, relative to the total number of activatable actors (i.e., those not in $S_0$): 
\[
\Gamma = \frac{|S_{\infty} - S_{0}|}{|A - S_{0}|}, \quad \Gamma \in [0, 1].
\]

In summary, the experimental methodology adopted in this study comprises the following steps: (1) selecting a real-world network; (2) obtaining a set of configuration parameters of the model for the network; (3) generating 20 network instances serving as ``digital twins'' of the real-world network; (4) simulating diffusion under a predefined regime; (5) manipulating a specific parameter of the configuration model; (6) generating 20 network instances based on the modified configuration; (7) simulating diffusion under the same predefined regime; (8) comparing the diffusion dynamics to assess how the evaluated parameter influences spreading efficiency.

The simulations described in this chapter were conducted on a workstation running Ubuntu 20.04.4 LTS with kernel version 6.5.3-arch1-1, equipped with 376~GB of RAM and an Intel(R) Xeon(R) Gold 6238 CPU @ 2.10~GHz (x86\_64 architecture). The source code was written in Python 3.12 utilizing \mABCD\ v1.0, \texttt{networkx} v3.3~\parencite{hagberg2008networkx}, and \texttt{network-diffusion} v0.18~\parencite{czuba2024networkdiffusion}, with particular attention to reproducibility. All scripts, data, and results are publicly available at Github repository\footnote{\url{https://github.com/anty-filidor/spreading-vs-mln-structure}}.

\subsection{Extracting configuration parameters from real networks}\label{sec:parameters_extraction}

To describe how the configuration parameters are retrieved from real-world networks, we first refer the reader to Section~\ref{subsec:mabcd_params}, as the notation used below follows that section closely. Furthermore, we do not elaborate here on how the values of $n$, $\ell$, $q$, $\delta$, and $\Delta$ are obtained, as they are straightforward and can be directly extracted from the processed network.

As for the more complicated parts of the \mABCD\ framework, we begin with $\Rr$. This parameter is denoted in a form of matrix of pairwise edge correlations. We estimate it using directly an approach described in Section~\ref{subsubsec:correlations}, especially with equations~(\ref{eq:Eij}) and~(\ref{eq:rij}).

The \mABCD\ model assumes that the labels of actors and nodes representing them follow the natural numbers. However, in real networks, labels may take various other forms. This aspect affects the estimation of $\tau$, which relies on the natural ordering of actor labels. To address this in retrieving $\tau$, we first convert the node labels in the first layer (layers are sorted alphabetically) so as to maximize the correlation between node degrees and their assigned labels; that is, the node with the highest degree is assigned the highest identifier. These re-organized labels are then used to compute the correlations in the remaining layers. Moreover, only nodes with a positive degree are taken into account to obtain the correlation value in order to reduce the noise.

The parameter $r$ is the most challenging to retrieve, as we do not have access to the latent biscuit-like reference layer used during the network generation process. To address this, we apply a coarse estimation of this parameter. Specifically, as it was done before, we select the first layer in alphabetical order and apply the Louvain algorithm to identify its community structure, which is then treated as the reference partition. Subsequently, we detect the community structure in each layer of the network and compute the layer-specific value $r_i$ as the \textbf{AMI} between the layer's partition and the reference partition. Since the alphabetically first layer is always used as the reference, its $r$ value will be large. Due to the non-deterministic nature of the Louvain algorithm~\parencite{blondel2008fast}, this value will not be exactly $1$, but it should be close to it.

Among the degree-oriented properties, only the retrieval of $\gamma$ is non-trivial, as it involves the general challenge of fitting an observed distribution to a specific model. To estimate this parameter, we employ the \texttt{powerlaw} library~\parencite{alstott2014powerlaw}, implicitly assuming that the degree distribution follows a power-law and is not better described by any other distribution type.

The next group of parameters ($\beta$, $s$, and $S$) concerns layer-specific community structural properties. At this stage, we again employ the Louvain algorithm~\parencite{blondel2008fast} to obtain a partition of the nodes into communities on each layer of the network. Then, a value of $\beta$ is estimated by applying the \texttt{powerlaw} library~\parencite{alstott2014powerlaw} to the sequence of community sizes. The values of $s$ and $S$ are likewise derived from this sequence.

Finally, to estimate the layer-wise noise level between communities, we again analyze the community structure on each network layer. For each $\xi_i$, we compute the fraction of inter-community edges relative to the total number of edges in the given layer.

The implementation of the \mABCD\ framework also includes parameters that do not directly reflect properties of the generated network, but instead serve as variables controlling the generation process. In the subsequent experiments, we use a fixed set of such parameters. Specifically, the dimensionality of the latent reference layer is set to $d = 2$. The maximum number of iterations for sampling both degrees and community sizes is set to $1{,}000$. For the rewiring step, we use $0.05$ as the proportion of edges to be rewired in each batch, and $100$ as the number of rewiring batches.

It is worth noting that, like any artificial model, \mABCD\ accepts only a feasible range of configuration values. Therefore, after retrieving the parameters for \textit{timik}, we tailored them to fit within the constraints of the framework \revision{according to Section~\ref{subsec:mabcd_params}}. The code to extract all parameters mentioned above from any network is also present in the previously mentioned GitHub repository, so even inexperienced user can generate a ``digital twin'' of any real network.

\subsection{Experiments}

Having introduced the outline of the experimental methodology, we now turn to the description of the experiments. To demonstrate the capabilities of \mABCD, we selected three use cases, each designed to shed light on a different aspect of the relationship between network geometry and spreading phenomena. Specifically, we evaluate the impact of the noise level between communities, network compression, and inter-layer community correlation.

\subsubsection*{Effect of noise level between communities}\label{subsubsec:experiment_a}

The first experiment conducted aimed to answer the following question: how does the inter-community noise level ($\xi$) impact information spreading? Intuitively, we expected higher $\xi$ values to facilitate the information dissemination and lower ones to suppress it. Indeed, the lower values of $\xi$ could create ``echo chambers'', closed environments where actors are primarily exposed to information and opinions that reinforce their existing beliefs, potentially leading to increased fragmentization. On the other hand, dissemination might still occur through other layers. Hence, the situation for multilayered networks is more complex and interesting compared to single-layer ones.

Following this hypothesis, we generated four types of networks with modified values of $\xi$ as follows: \textit{series 2} with $\xi_i=1.00$, \textit{series 3} with $\xi_i=2\xi_{i_{\textit{timik}}}$, \textit{series 4} with $\xi_i=0.5\xi_{i_{\textit{timik}}}$, and \textit{series 5} with $\xi_i=0.01$. In other words, two types of networks exhibit extreme inter-community noise levels across all layers, while the remaining two have intermediate values, determined proportionally to those of the baseline. We then selected a spreading regime for the baseline networks (i.e., ``digital twins'' of \textit{timik}, denoted as \textit{series 1}). Specifically, we used the Neighbourhood Size Discount~\parencite{czuba2024rankrefininginfmaxmln} to select seeds, and considered two sets of parameters for each protocol function: $(\dot{\delta}, \dot{\pi}, s) \in \{AND\} \times \{0.15, 0.2, 0.25, 0.3\} \times \{1, 3, 5, 7\}$ and  $(\dot{\delta}, \dot{\pi}, s) \in \{OR\} \times \{0.03, 0.05, 0.07, 0.09\} \times \{1, 5, 10, 15\}$, where $\dot{\delta}$ and $\dot{\pi}$ denote the parameters of the MICM (see Section~\ref{subsec:spread_methodology}), while $s$ refers to the seed set budget, expressed as a percentage of the number of actors constituting a given network.

The results indicate that the impact of the noise level between communities on diffusion effectiveness depends on the spreading parameters. Specifically, all three variables --- $\dot{\delta}$, $\dot{\pi}$, and $s$ --- contribute in distinct ways. Therefore, we first briefly discuss the influence of each parameter and subsequently present detailed results for the subset of configurations in which manipulating $\xi$ yields the most pronounced effect. The complete set of results is available in the code repository.

The impact of the protocol function is significant: for the less restrictive one ($OR$), the noise level barely affects the diffusion characteristics (up to $0.0052$ of $\Gamma$ obtained for baseline). The trend is contrary to the initial assumptions --- namely, reducing $\xi$ improves spreading. One possible explanation could be that low values of $\xi$ help with spreading information within one community in one layer and multilayer structure is responsible for spreading to other communities. By contrast, the differences observed for $\dot{\delta} = AND$ were more pronounced: up to an order of magnitude larger than for $OR$ (particularly in networks with decreased inter-community noise levels). Nonetheless, at this level of analysis, there was no clear trend regarding the impact of $\xi$ on spreading effectiveness for this protocol function. Results obtained for both protocol functions indicate that higher activation probability ($\dot{\pi}$) corresponds to a reduced impact of $\xi$, i.e. the results converge towards those of the baseline. Regarding the seed set budget size, for $\dot{\delta}=AND$, lower budgets lead to larger differences with the baseline results, and the results begin to align with the noise level: higher noise yields greater gain. For $OR$, however, different trends were observed. In this case, for the diffusion triggered with higher budgets, the impact of $\xi$ was more visible, although the relative differences with \textit{series 1} remained still small (up to $0.0062$ of baseline $Gain$).

Finally, we turn to the set of parameters for which manipulating $\xi$ produces the most pronounced effects. To this end, we selected following spreading conditions: the lowest evaluated budget for the $AND$ protocol and the highest for the $OR$ protocol. These configurations are listed in Table~\ref{tab:experiment_xi}. As can be observed, the impact of $\xi$ on spreading effectiveness differs significantly between the two diffusion regimes. When the protocol function is more restrictive, increasing the inter-community noise enhances information spread. The trend is even more pronounced when noise is suppressed: in such cases, the diffusion coverage may decrease by up to $25\%$. This outcome aligns with our initial intuition. Since the $AND$ protocol imposes stricter conditions, it becomes easier to activate a given actor when it is more strongly connected at the global level. Consequently, the process can ``reach'' more distant nodes and has a greater chance of persisting. In contrast, when the $OR$ protocol is applied, the effect of modifying $\xi$ is reversed, albeit considerably weaker. A possible explanation is that, when communities are more clearly separated, the diffusion tends to remain within them initially and only later extends outward. Furthermore, due to the lenient nature of the $OR$ protocol, the process is less likely to die out prematurely, which may account for this behaviour.

\begin{table}[ht!]
    \caption{Change in $\Gamma$ achieved through MICM-driven spreading in networks generated with \mABCD\ with \emph{varying inter-community noise levels}, relative to the results obtained on the baseline network (resembling \textit{timik}).}
    \centering
    \begin{tabular}{cc|>{\raggedleft\arraybackslash}p{1.5cm}>{\raggedleft\arraybackslash}p{1.5cm}>{\raggedleft\arraybackslash}p{1.5cm}>{\raggedleft\arraybackslash}p{1.5cm}>{\raggedleft\arraybackslash}p{1.5cm}}
    \multirow{2}*{$\dot{\delta}$} & \multirow{2}*{$\dot{\pi}$} & \multicolumn{5}{c}{$\xi_i$} \\
    & & $0.01$ & $0.50\xi_{i_{\textit{timik}}}$ & $\xi_{i_{\textit{timik}}}$ & $2.00\xi_{i_{\textit{timik}}}$ & $1.00$ \\ \hline \hline
    \multirow{4}*{\rotatebox[origin=c]{90}{$AND$}} & 0.15 & -0.1668 & -0.0392 & 0.0000 & 0.0352 & 0.0397 \\
    & 0.20 & -0.2440 & -0.0450 & 0.0000 & 0.0254 & 0.0290 \\
    & 0.25 & -0.2339 & -0.0183 & 0.0000 & 0.0049 & 0.0041 \\
    & 0.30 & -0.1706 & -0.0066 & 0.0000 & 0.0017 & 0.0018 \\ \hline
    \multirow{4}*{\rotatebox[origin=c]{90}{$OR$}} & 0.03 & 0.0104 & 0.0088 & 0.0000 & -0.0120 & -0.0154 \\
    & 0.05 & 0.0046 & 0.0033 & 0.0000 & -0.0049 & -0.0059 \\
    & 0.07 & 0.0020 & 0.0013 & 0.0000 & -0.0022 & -0.0026 \\
    & 0.09 & 0.0009 & 0.0008 & 0.0000 & -0.0009 & -0.0010 \\
    \end{tabular}
    \label{tab:experiment_xi}
\end{table}

\subsubsection*{Effect of changes in network size}\label{subsubsec:experiment_b}

With the second experiment, we aimed to address one of the fundamental questions in the domain of social network analysis: is it possible to compress a network in such a way that it preserves its topology while being smaller? Our goal was to examine this issue in the context of spreading phenomena. Specifically, we were interested in whether manipulating the number of actors, while preserving all other structural properties of the network (as dictated by the configuration model underlying \mABCD), would result in different diffusion characteristics. It is worth noting that the approach adopted in this study can be considered novel, as most existing work reduces graphs using dedicated compression techniques rather than by explicitly employing standalone graph generators (for a comprehensive review, see~\cite{sun2024gcbench}).

In order to address this question, we generated four series of networks, each differing only in the number of actors, while preserving the remaining configuration model parameters from the baseline. The series were defined as follows: $n = 1.5n_{\textit{timik}}$ (\textit{series 6}), $n = 1.25n_{\textit{timik}}$ (\textit{series 7}), $n = 0.75n_{\textit{timik}}$ (\textit{series 8}), and $n = 0.50n_{\textit{timik}}$ (\textit{series 9}). Hence, two of the network types can be considered as upsampled, while the other two represent downsampled versions of the baseline. Regarding the spreading model, we applied the same parameter settings as in the first experiment (see Section~\ref{subsubsec:experiment_a}).

In contrast to the first experiment, the trends observed in the results are consistent and clear when analyzed with respect to the protocol function, activation probability, and seeding budget. Therefore we move directly to Table~\ref{tab:experiment_n}, which presents these results aggregated over the MICM parameters; nevertheless, grouping by other variables produces similar patterns. The results are shown in the same format as in Section~\ref{subsubsec:experiment_a}, i.e., as the average relative difference in $\Gamma$ obtained by each network series with respect to the baseline. It is noteworthy that, since $\Gamma$ accounts for network size, it is reasonable to compare this metric across diffusion processes conducted on networks with different numbers of actors.

\begin{table}[ht!]
    \caption{Change in $\Gamma$ achieved through MICM-driven spreading in networks generated with \mABCD\ with \emph{varying number of actors}, relative to the results obtained on the baseline network (resembling \textit{timik}).}
    \centering
    \begin{tabular}{cc|>{\raggedleft\arraybackslash}p{1.5cm}>{\raggedleft\arraybackslash}p{1.5cm}>{\raggedleft\arraybackslash}p{1.5cm}>{\raggedleft\arraybackslash}p{1.5cm}>{\raggedleft\arraybackslash}p{1.5cm}}
    \multirow{2}*{$\dot{\delta}$} & \multirow{2}*{$\dot{\pi}$} & \multicolumn{5}{c}{$n$} \\
    & & $0.50n_{\textit{timik}}$ & $0.75n_{\textit{timik}}$ & $n_{\textit{timik}}$ & $1.25n_{\textit{timik}}$ & $1.50n_{\textit{timik}}$ \\ \hline \hline
    \multirow{4}*{\rotatebox[origin=c]{90}{$AND$}} & 0.15 & -0.0011 & 0.0016 & 0.0000 & 0.0911 & 0.0949 \\
    & 0.20 & -0.0003 & 0.0005 & 0.0000 & 0.0793 & 0.0808 \\
    & 0.25 & -0.0001 & -0.0000 & 0.0000 & 0.0428 & 0.0432 \\
    & 0.30 & 0.0002 & 0.0001 & 0.0000 & 0.0202 & 0.0202 \\ \hline
    \multirow{4}*{\rotatebox[origin=c]{90}{$OR$}} & 0.03 & -0.0006 & -0.0011 & 0.0000 & 0.0522 & 0.0542 \\
    & 0.05 & -0.0002 & -0.0003 & 0.0000 & 0.0341 & 0.0353 \\
    & 0.07 & -0.0001 & -0.0001 & 0.0000 & 0.0231 & 0.0237 \\
    & 0.09 & 0.0001 & 0.0000 & 0.0000 & 0.0163 & 0.0168 \\
    \end{tabular}
    \label{tab:experiment_n}
\end{table}

As one can note, for both protocol functions and across all evaluated activation probabilities, downsampling the network does not disturb spreading effectiveness (the differences are minimal). On the other hand, when the network gets upsampled, we observe slight, but non-negligible changes, in the diffusion --- the more upsampled the network is, the higher $\Gamma$ change we observe. However, the disparities are not as big as in the results from Section~\ref{subsubsec:experiment_a}. At this point, one might wonder why network downsampling does not alter the spreading trend, while upsampling does. A possible explanation lies in the average degree of the evaluated networks, as this parameter has a substantial impact on the effectiveness of diffusion under the MICM. We observe that downsampling has little effect on the average degree, whereas upsampling increases it by approximately $0.9$ (from around $20.2$ to $21.1$). Consequently, in the upsampled graphs, the diffusion process may be more effective due to this slight increase in connectivity, especially when the seed set is selected using a heuristic that prioritizes central actors rather than selects them at random.

\subsubsection*{Effect of inter-layer community correlation}

The final experiment aimed to evaluate whether spreading is more effective in networks where the communities are strongly correlated. The \mABCD\ framework allows control over this property, albeit indirectly, via the $r_i$ parameter. (Recall that this parameter governs the extent to which the communities in layer $i$ are correlated with the latent reference layer, where spatial proximity influences community assignment, i.e., actors located closer together are more likely to be placed in the same group; see Section~\ref{subsubsec:creating_communities} for more details). To this end, we intended to simulate a case in which information begins to diffuse from actors who are central in only one of the several types of relationships represented in the network, while manipulating community correlations on the remaining layers. For instance, consider a situation in which a group of department heads returns from an overseas business trip, having contracted a disease. As leaders, they occupy central positions within the company network and communities formed by their teams. Subsequently, the disease begins to spread not only through work-related contacts but also through encounters in other spheres of life. By manipulating the $r$ parameter, we wanted to reflect the extent to which these relationships overlap with the professional one, and to assess whether it influences the effectiveness of spreading.

As in the previous experiments, the networks were also generated according to the configuration derived from \textit{timik}. All of them had the $r$ value on the first layer set to $1.000$ in order to comply with the assumed seed selection scenario. However, they differed in the $r$ values on the remaining layers, as follows. Two series were generated to represent stronger community correlations: \textit{series 10} with $r_2 = r_3 = 1.000$, and \textit{series 11} with $r_2 = r_3 = 0.667$. Additionally, networks with weaker correlations were generated: \textit{series 12} with $r_2 = r_3 = 0.333$, and \textit{series 13} with $r_2 = r_3 = 0.001$. The spreading conditions applied were the same as in the previous experiments, with one difference in the seed selection method. Specifically, instead of employing a heuristic that considers actors' representations across all network layers, we modified it to consider only the first layer (which is maximally correlated with the latent layer used in network generation). Consequently, the Neighbour Size Discount heuristic was reduced to its baseline version designed for single-layer networks: Degree Centrality Discount~\parencite{Chen2009DegreeDiscount} and with that method a seed set was selected.

After executing the simulations, we obtained the results. As in the previous experiments, we initially analyzed them separately with respect to $\dot{\delta}$, $\dot{\pi}$, and $s$. Across all of these parameters, no significant differences in spreading were observed, even though the propagation process could be initiated and sustained for several simulation steps. To maintain clarity, we present the results aggregated over $s$ in Table~\ref{tab:experiment_r}. As can be seen, the relative difference in $\Gamma$ with respect to the \textit{timik}’s ``digital twin'' (\textit{series 1}) remains consistently very small across the evaluated network series. This suggests that, under the considered spreading model, manipulating $r$ does not substantially influence propagation effectiveness.

\begin{table}[ht!]
    \caption{Change in $\Gamma$ achieved through MICM-driven spreading in networks generated with \mABCD\ with \emph{varying correlation strength between communities and the reference layer}, relative to the results obtained on the baseline network (resembling \textit{timik}).}
    \centering
    \begin{tabular}{cc|>{\raggedleft\arraybackslash}p{1.5cm}>{\raggedleft\arraybackslash}p{1.5cm}>{\raggedleft\arraybackslash}p{1.5cm}>{\raggedleft\arraybackslash}p{1.5cm}>{\raggedleft\arraybackslash}p{1.5cm}}
    \multirow{2}*{$\dot{\delta}$} & \multirow{2}*{$\dot{\pi}$} & \multicolumn{5}{c}{$r_2, r_3$} \\
    & & $0.001$ & $0.333$ & $r_{i_{\textit{timik}}}$ & $0.667$ & $1.000$ \\ \hline \hline
    \multirow{4}*{\rotatebox[origin=c]{90}{$AND$}} & 0.15 & 0.0007 & 0.0007 & 0.0000 & 0.0009 & 0.0015 \\
    & 0.20 & 0.0016 & 0.0001 & 0.0000 & 0.0019 & 0.0022 \\
    & 0.25 & 0.0010 & -0.0003 & 0.0000 & 0.0003 & 0.0006 \\
    & 0.30 & 0.0002 & -0.0006 & 0.0000 & -0.0004 & -0.0002 \\ \hline
    \multirow{4}*{\rotatebox[origin=c]{90}{$OR$}} & 0.03 & -0.0002 & -0.0006 & 0.0000 & -0.0010 & -0.0011 \\
    & 0.05 & -0.0002 & -0.0002 & 0.0000 & -0.0004 & -0.0002 \\
    & 0.07 & -0.0001 & -0.0001 & 0.0000 & -0.0002 & -0.0002 \\
    & 0.09 & -0.0000 & 0.0001 & 0.0000 & -0.0000 & -0.0000 \\
    \end{tabular}
    \label{tab:experiment_r}
\end{table}

The reason for this may lie in the operational principles of the MICM and the corresponding geometric characteristics of the generated networks. Given that the effectiveness of spreading under this model depends primarily on local structures (such as degree or clustering coefficient) networks with similar values for these properties are expected to exhibit comparable diffusion dynamics. This appears to be the case in the considered example: manipulating $r$ does not significantly affect either the degree distribution and the clustering coefficient (for reference, see the detailed analysis available in the GitHub repository). Consequently, the propagation proceeds in a similar manner across the evaluated networks.

\section{Conclusions}

In this work, we have presented \mABCD, a framework for generating multilayer networks with community structure and inter-layer relationships. We hope it addresses the existing gap in the field through its highly customizable configuration options and efficient algorithmic design. Furthermore, we release its implementation in Julia, along with Python ports, to make \mABCD\ accessible to the broader research community. As \mABCD\ builds upon the well-established generator introduced by~\cite{kaminski2021artificial}, it can be regarded as a new member of the \textbf{ABCD} family. In addition to outlining the essential components of the model, we have validated its properties, demonstrating that \mABCD\ produces stable results and outperforms its closest competitors in both efficiency and usage flexibility.

The paper also shows potential applications of the framework, illustrated in the context of studying spreading phenomena. Through three experiments, we examined whether and how selected parameters of the \mABCD\ configuration model contribute to diffusion effectiveness. The results are nuanced rather than clear-cut. While there is a consistent trend with respect to the number of actors, the level of inter-community noise affects diffusion depending on the imposed regime. Finally, the strength of correlation between groups appears to have little influence on information spread. Altogether, these findings suggest that the proposed model is capable of supporting the analysis of spreading processes in multilayer networks, albeit up to certain extend.

\revision{As for the limitations of the proposed approach, the most significant one arises from the design assumptions underlying the model’s development. Specifically, \mABCD\ is capable of generating scale-free networks, and therefore its applications are restricted to systems that conform to this property (e.g., social ones). Another constraint concerns the size of the generated network. Although it should not be regarded as the major shortcoming, it is important to note that \mABCD\ fails to properly generate structures when certain parameters are assigned excessively low values, such as the minimum degree ($\delta_i$), the minimum community size ($s_i$), or the number of actors ($n$). The next limitation, which became particularly apparent during the experiments described in Section~\ref{sec:spreading}, concerns the difficulty of representing the properties of a real-world network within the parameter space of \mABCD, especially the correlation strength between communities and the latent reference layer ($r_i$) which constitutes a core of the third phase in the network construction process (see Section~\ref{subsec:abcd_construction}). Although we have proposed a methodology for the coarse estimation of this value, a follow-up study is required to address this issue comprehensively.

Apart from the above, f}uture work may proceed in two directions. The first concerns the \textbf{ABCD} model itself, which could be extended to other types of networks, such as temporal ones \revision{or multilayer ones without a lower bound imposed on the network size. The second way} relates to the experiments involving spreading phenomena. In this context, it would be valuable to evaluate how other parameters of \mABCD\ influence the dynamics of spreading processes. However, a more important and promising direction is the development of a metric for quantifying the difference between a real-world network and its “digital twin” generated by \mABCD. Such a method could open a new avenue for influence maximization studies, as it would allow results and conclusions to be more directly related to real-world systems.

\bigskip

\paragraph{Data availability} The datasets generated during and/or analyzed during the current study and source code are available in following GitHub repositories: \url{https://github.com/KrainskiL/MLNABCDGraphGenerator.jl}, \url{https://github.com/anty-filidor/spreading-vs-mln-structure}. Access to the DVC remote will be granted upon reasonable request.

\paragraph{Funding} This research was partially supported by the National Science Centre, Poland, under Grant no. 2022/45/B/ST6/04145\footnote{\url{https://multispread.pwr.edu.pl/}}, the Polish Ministry of Education and Science within the programme ``International Projects Co-Funded'', and the EU under the Horizon Europe, grant no. 101086321 OMINO. Views and opinions expressed are, however, those of the authors only and do not necessarily reflect those of the National Science Centre, Polish Ministry of Education and Science, the EU or the European Research Executive Agency. 




\printbibliography

\end{document}